\documentclass[manuscript]{aastex}

\slugcomment{}

\shorttitle{Spectral optical monitoring of Ark 564}
\shortauthors{Shapovalova et al.}

\begin{document}


\title{Spectral Optical Monitoring of the Narrow Line Seyfert 1 \\
galaxy Ark 564}


\author{A.I. Shapovalova\altaffilmark{1}, L.\v C. Popovi\'c\altaffilmark{2,3}, A.N. Burenkov\altaffilmark{1}, 
V.H. Chavushyan\altaffilmark{4}, D. Ili\'c\altaffilmark{5,3}, \\
A. Kova\v cevi\'c\altaffilmark{5,3},
W. Kollatschny\altaffilmark{6}, J. Kova\v cevi\'c\altaffilmark{2,3}, 
N.G. Bochkarev\altaffilmark{7}, J. R. Valdes\altaffilmark{4}, \\
J. Torrealba\altaffilmark{4},
J. Le\'on-Tavares\altaffilmark{8}, A. Mercado\altaffilmark{9}, E. Ben\'itez\altaffilmark{10},
L. Carrasco\altaffilmark{4}, D. Dultzin\altaffilmark{10} \\
and E. de la Fuente\altaffilmark{11}}

\email{ashap@sao.ru}

\altaffiltext{1}{Special Astrophysical Observatory of the Russian AS,
Nizhnij Arkhyz, Karachaevo-Cherkesia 369167, Russia}
\altaffiltext{2}{Astronomical Observatory, Volgina 7, 11160 Belgrade 74, Serbia}
\altaffiltext{3}{Isaac Newton Institute of Chile, Yugoslavia Branch, Belgrade, Serbia}
\altaffiltext{4}{Instituto Nacional de Astrof\'{\i}sica, \'{O}ptica y
Electr\'onica, Apartado Postal 51-216, 72000 Puebla, Puebla, M\'exico}
\altaffiltext{5}{Department of Astronomy, Faculty of Mathematics, University
of Belgrade, Studentski trg 16, 11000 Belgrade, Serbia}
\altaffiltext{6}{Institut f\"ur Astrophysik, Georg-August-Universit\"at G\"ottingen, Germany}
\altaffiltext{7}{Sternberg Astronomical Institute, Moscow, Russia}
\altaffiltext{8}{Aalto University Mets\"ahovi Radio Observatory,  Mets\"ahovintie 114,
FIN-02540 Kylm\"al\"a, Finland}
\altaffiltext{9}{Universidad Polit\'ecnica de Baja California, Av. de la Industria
291, 21010 Mexicali, B.C., M\'exico}
\altaffiltext{10}{Instituto de Astronom\'ia, Universidad Nacional Aut\'onoma de 
M\'exico, Apartado Postal 70-264, M\'exico, D.F.  04510, M\'exico}
\altaffiltext{11}{Instituto de Astronom\'ia y Meteorolog\'ia, Dpto. de F\'isica CUCEI,
Universidad de Guadalajara, Av. Vallarta 2602, 44130 Guadalajara,
Jalisco, M\'exico}


\begin{abstract}
We present the results of a long-term (1999--2010) spectral optical monitoring campaign of the 
active galactic nucleus (AGN) Ark 564, which shows a strong \ion{Fe}{2} line emission in {\bf the optical.  
This AGN is a} narrow line Seyfert 1 (NLS1) galaxies, a group of AGNs with specific spectral characteristics. 
We analyze the light curves of the {\bf permitted} H$\alpha$,  H$\beta$, optical \ion{Fe}{2} line fluxes, and 
the continuum flux in order to search for a time lag between them. Additionally, in order to estimate the 
contribution of iron lines from different multiplets, we fit the H$\beta$ and \ion{Fe}{2} lines with a 
sum of Gaussian components. We found that during the monitoring period the spectral variation 
(F$_{\rm max}$/F$_{\rm min}$) of Ark 564 was between 1.5 for H$\alpha$ to 1.8 for the \ion{Fe}{2} lines. 
The correlation between the \ion{Fe}{2} and H$\beta$ flux variations is of higher significance than 
that of H$\alpha$ and H$\beta$ (whose correlation is almost absent). 
The {\bf permitted-line} profiles are 
Lorentzian-like, and did not change shape during the monitoring period.  
We investigated, in detail, the optical \ion{Fe}{2} emission and found different degrees of correlation 
between the \ion{Fe}{2} emission arising from different spectral multiplets and the continuum flux. 
The relatively weak and different degrees of correlations between {\bf permitted} lines and continuum fluxes 
indicate a rather complex source of ionization of the broad line emission region. 
\end{abstract}


\keywords{galaxies: active -- galaxies: quasar -- galaxies: individual
(Ark 564) -- line: profiles}

\section{Introduction}

Narrow-line Seyfert 1 (NLS1) galaxies were first introduced as a class 
of active galactic nuclei (AGNs) by \cite{ost85}. Their optical spectra 
show relatively narrow (FWHM $\leq$ 2000  km s$^{-1}$) permitted lines, 
which are narrower than in a typical Seyfert 1 galaxy. In particular, \cite{ost85}  
did show that, the permitted lines are only slightly broader than the forbidden ones, 
and that a strong \ion{Fe}{2} emission is present in the optical region of the spectrum. 
In addition, the [O III] $\lambda$5007/H$\beta$ ratio, emitted in the narrow line 
region (NLR), varies from 1 to 5 {\bf \citep[][]{rod00}}, instead of the universally 
adopted observed value for Seyfert 1s of around 10 \citep[][]{rod00}, 
indicative of the presence of high-density gas. \cite{ost85} pointed out that 
the H$\beta$ equivalent widths in NLS1s are smaller than typical values for normal 
Seyfert 1s, suggesting that they are not just normal Seyfert 1s seen at a particular 
viewing angle. Renewed interest in NLS1s arises from the discovery of their 
distinctive X-ray properties: they show a steep X-ray excess with a photon index of 3
below 100 keV, a steep hard X-ray continuum, and a rapid large-amplitude X-ray variability on
timescales of minutes to hours \citep[see][and references therein]{le99a,le99b,le00,pa11}. 
Moreover, optical studies have established that NLS1s lie at one end of 
the \cite{bo92} eigenvector 1 (EV1) and that they show a relatively strong 
\ion{Fe}{2} emission and a weak [O III] emission \citep[][]{bol96}. They also represent 
the "extreme Population A" objects (FWHM H$\beta <$ 4000 km/s) as defined by the four 
dimensional eigenvector 1 (4DE1) in \cite{su07,mar10}. 4DE1 involves
four parameters and NLS1 are "extreme" in all of them: they have the narrowest 
broad H$\beta$, strongest \ion{Fe}{2} emission, strongest X-ray excess and 
largest \ion{C}{4} blueshifts.

Arakelian 564 (Ark 564, IRAS 22403+2927, MGC +05-53-012) is a bright $V=14.6$ mag. \citep{dev91},
nearby narrow-line Seyfert 1 galaxy ($z=0.02467$), with an X-ray luminosity 
$L_{2-10 \rm keV}=2.4\times 10^{43},\rm{erg \ s^{-1}}$ \citep[][]{tu01}. This AGN is one of the brightest 
NLS1s in the X-ray band \citep{bol96,co01,sm08}, and it shows a soft excess 
below $\sim$1.5 keV and a peculiar emission-line-like feature at 0.712 keV in the source rest 
frame \citep[][]{ch04}. The variations of the X-ray amplitude in the short-timescale light curve is 
very similar to those in the long-timescale light curve \citep[][]{po01}, that is in contrast to 
the stronger amplitude variability on longer timescales, which is a characteristic of broad-line Seyfert 1 (BLS1) galaxies.  
In the UV part of the spectrum, this galaxy shows intrinsic UV absorption lines  \citep[][]{cr99}.
In order to explore the variability characteristics of a NLS1 in different wavelength bands, 
a multiwavelength monitoring campaign of Ark 564 was conducted \citep{sh01}. {\bf The optical campaign covered the periods
1998 November -- 1999 November and 2000 May -- 2001 January, where the object was observed both photometrically 
(UBVRI filters) and spectrophotometrically (spectral coverage 4800--7300 \AA) \citep{sh01}. 
The data set and analysis is described in details and compared with the simultaneous X-ray and UV campaigns \citep{sh01}.}
The results of this intensive variability multiwavelength campaign
show that the optical continuum is not significantly correlated with the X-ray emission \citep[][]{sh01}.  The UV
campaign, carried out with the HST on 2000 May 9 and 2000 July 8, is described in \cite{co01}. These authors found 
a small fractional variability amplitude of the continuum between 1365 \AA\  and 3000 \AA\  (around ~6\%), 
but reported that large-amplitude short-timescale flaring behavior is present, with trough-to-peak flux changes 
of about 18\% in approximately 3 days \citep[][]{co01}. The wavelength-dependent continuum time delays in Ark 564 
have been detected and these delays may indicate a stratified continuum reprocessing region \citep[][]{co01}.

Here we present the long-term monitoring of Ark 564 in the optical part of the spectrum. 
We analyzed the variability in the {\bf permitted} emission lines and continuum in order to determin the 
size and structure of emitting regions of the {\bf permitted} Balmer and \ion{Fe}{2} lines. We placed particular
interest in the strong \ion{Fe}{2} lines of the H$\beta$ spectral region whose behavior is investigated 
in details and discussed in this paper.

The paper is organized as follow: in section 2 we describe the observations and data reduction procedures, in 
section 3 we give an analysis of the spectral data, in section 4 we explore the correlations between 
different lines and the continuum, as well as between different lines, in section 5 we investigate
in more details the \ion{Fe}{2} variation, in section 6 we discuss our results, and in section 7 
we provide our conclusions.

\section{Observations and data reduction}

Spectral monitoring  of Ark 564 was carried out with the 6 m and
1 m telescopes of the SAO RAS (Russia, 1999--2010), the INAOE's
2.1 m telescope of the ''Guillermo Haro Observatory'' (GHO) at
Cananea, Sonora, M\'exico (1999--2007), and the 2.1 m telescope of
the Observatorio Astron\'omico Nacional at San Pedro Martir (OAN-SPM), Baja
California, M\'exico (2005--2007). Spectra were taken with long--slit
spectrographs equipped with CCDs. The typical {\bf observed} wavelength range was
4000--7500 \AA , the spectral resolution was R=5--15  \AA , and with S/N ratio
$>$ 50 in the continuum near  H$\alpha$ and H$\beta$.
In total 100 blue and 55 red spectra were obtained during 120 nights.
In the analysis, {\bf about 10\% of spectra were discarded for several different reasons: 
e.g. a) large noise (S/N$<$15\AA\ - 2001 Aug 29 (blue, red), 2001 Oct 08 (blue),
2001 Oct 09 (blue, red), taken with Zeiss(1m)+CCD(1k$\times$1k);
b) large noise and badly corrected spectral sensitivity in the blue part - 
2006 Jun 28 (blue, red), 2006 Aug 29(blue), 2006 Aug 30 (blue), 2009 Aug 14 (blue), 
2009 Oct 11 (blue), taken with Zeiss(1m)+CCD(2k$\times$2k). We note here that the 
CCD(2k$\times$2k) sensitivity in the blue part is not good enough, since
it is a red CCD; c) poor spectral resolution (R$>$20\AA, - 2003 Nov 18, 2004 Oct 18, taken with 2.1m GHO).}
Thus our final data set consisted of 91 blue and 50 red spectra, which 
were used in further analysis.

From 1999 to 2003 spectral observations with 1 m Zeiss telescope of the SAO
were carried out with two different CCDs (formats used were 1k$\times$1k or 530$\times$580) 
and the H$\alpha$ and H$\beta$ spectral regions were observed separately. From 2004 to 2010 a CCD (2k$\times$2k,
EEV CCD42-40) was used, allowing us to observe the entire wavelength range (4000--8000) \AA\,
with a spectral resolution of 8-10 \AA. However, this CCD in the blue part of some spectra 
presents large sensitivity variations (i.e. bad S/N), which are badly corrected, thus the blue 
region of these spectra was not used in our analysis.

From 2004 to 2007, the spectral observations with two Mexican 
2.1 m telescopes were carried out with two observational setups. 
In the case of GHO observations we used the following configuration:
1) with a grating of 150 l/mm (spectral resolution of R=15 \AA,
a resolution similar to the observations of 1999--2003);
2) with a grating of 300 l/mm (moderate spectral resolution of R=7.5\AA).
The similar spectral characteristics at the OAN-SPM were, respectively, 
obtained with the following configuration:
1) with a grating of 300 l/mm (spectral resolution of R=15 \AA);
2) with a grating of 600 l/mm (moderate spectral resolution of R=7.5\AA).

As a rule, observations were carried out with the moderate resolution in the blue or 
red bands during the first night of each run. {\bf In order to cover H$\alpha$ and 
H$\beta$ at the same time},  we used the lower resolution mode and observed 
the entire spectral range 4000-7500 \AA; 
and then the moderate resolution was adopted again for the following night.
Since the shape of the continuum of active galaxies practically
does not change during adjacent nights, it was easy to match, the blue and red bands 
obtained with the moderate resolution in different nights. To this aim we used the data 
obtained for the continuum from the low-dispersion spectra for the entire wavelength range.
By this procedure, the photometric accuracy is thus considerably improved with respect
to a match obtained by overlapping the extremes of the blue and red continuum 
(3-5\% instead of 5-10\%).

Spectrophotometric standard stars were observed every night.

Information on the source of spectroscopic observations is listed
in Table~\ref{tab1}. The log of the spectroscopic observations is
given in Table~\ref{tab2}. Taking into account all observations, the mean 
sampling rate is 33.20, and the median rate is 2.95 days.
{\bf The big difference between the mean and median sampling rate is due to
the big gaps in the variability campaign.}

The spectrophotometric data reduction was carried out either with
the software developed at the SAO RAS or with the IRAF package for the
spectra obtained in M\'exico. The image reduction process included
bias and flat-field corrections, cosmic ray removal, 2D wavelength
linearization, sky spectrum subtraction, addition of the spectra for
every night, and relative flux calibration based on observations of 
standard star.

\subsection{Absolute calibration (scaling) of the spectra}

The standard technique of flux calibration of the spectra (i.e.
comparison with stars of known spectral energy distribution) is not
precise enough for the study of AGN variability, since even under
good photometric conditions, the accuracy of spectrophotometry is
usually not better than 10\%. Therefore we used standard stars only 
to provide a relative flux calibration. Instead, for the absolute 
calibration, the observed fluxes of the forbidden, narrow emission
lines are adopted for the scaling procedure the AGN spectra, since 
these fluxes are expected to be constant \citep[][]{pe93}. 
From HST observations \citep[][]{cr02} 
it was shown that the NLR in Ark 564 is about $0.2$'' (95 pc), and
this facts implies a constant [OIII]$\lambda$5007 flux intensity during 
several hundred years. Consequently, the flux of this forbidden line
should not have changed during our monitoring period.
The scaling of the blue spectra was performed using the method of
\citet{vg92} modified by \citet{sh04}\footnote{see Appendix A 
\citet{sh04}}. {\bf We will not repeat the scaling procedure here,
we only note that the flux in the lines was determined after subtraction of a
linear continuum determined by the beginning and the end of
a given spectral interval.} This method allowed us to obtain a homogeneous set of spectra with the
same wavelength calibration and the same [OIII]$\lambda$5007 flux.
The [OIII]$\lambda$5007 flux in absolute units was taken from \citet{sh01}: 
F([OIII]$\lambda$5007)=(2.4$\pm$0.1)$\times 10^{-13} \rm erg \ s^{-1} cm^{-2}$.
The spectra, obtained with 2.1 m telescopes in Mexico
with a resolution of 12--15 \AA, containing both H$\alpha$
and H$\beta$ regions were scaled using the [O III]$\lambda$5007 line.
However, some spectra of Ark 564 were
obtained separately in the blue (H$\beta$) and red (H$\alpha$) wavelength
bands, with a resolution of 8--10 \AA. Usually, the red edge of
the blue spectra and the blue edge of the red spectra overlap in
an interval of 300 \AA. Therefore, first the red spectra (17)
were scaled using the overlapping continuum region with the
blue ones. The latter were scaled with the [OIII]$\lambda$5007 line. In these cases
the scaling uncertainty was about 5\%--10\%.  Then scaling of the red spectra
was refined using the mean flux in [OI]$\lambda$6300 
(mean F[OI]$\lambda$6300$\sim$(1.93$\pm$0.24)$\times 10^{-14}$),
determined from low-dispersion spectra (R$\sim$12--15 \AA). For 3 red spectra
(JD:2452886.9; 2455058.5 and 245116.4) we have no blue spectrum
in {\bf adjacent} nights, and they was scaled using only the mean
flux of the [OIII]$\lambda$6300\AA\ line.

\subsection{Unification of the spectral data}

In order to investigate the long term spectral variability of an
AGN, it is necessary to conform  a consistent, uniformed data set.
Since observations were carried out with 4 different instruments, we must correct 
the line and continuum fluxes for aperture effects \citep[][]{pet83}.
To this effect, we determined a point-source correction factor $\varphi$ given by 
the following expression \citep[see][for a detailed discussion]{pe95}:
$$   F({\rm H}\beta)_{\rm true}= \varphi\cdot F({\rm H}\beta)_{\rm obs}$$
where $F({\rm H}\beta)_{\rm obs}$ is the observed H$\beta$ flux;
$F({\rm H}\beta)_{\rm true}$ is the aperture corrected H$\beta$ flux.
The contribution of the host galaxy to the continuum flux depends
also on the aperture size. The continuum fluxes $F(5235$\AA) {\bf (in the observed-frame)}
were corrected for different amounts of host-galaxy contamination,
according to the following expression (see \citet{pe95}):
$$F(5235\AA)_{\rm true}=\varphi\cdot F(5235\AA)_{\rm obs} - G(g),$$
where $F(5235$\AA$)_{\rm obs}$ is the continuum flux at {\bf 5235 \AA \, in the observed-frame};
 $G(g)$ is an aperture-dependent correction factor to
account for the host galaxy contribution. The GHO observing scheme (Table~\ref{tab1}),
which correspond to a projected aperture ($2.5''\times 6''$)  of the 2.1 m
telescope, were adopted as standard (i.e. $\varphi=1.0$, $G(g)$=0
by definition). 
{\bf The correction factors  $\varphi$ and $G(g)$ are determined empirically 
by simulated aperture photometry of suitable images of the narrow line emission  
and the starlight of the host galaxy in the same way as it is given in \cite{pe95}. 
This  procedure is accomplished empirically by comparing pairs of 
simultaneous observations from each of given telescope data sets to that 
of the standard data set \cite[as it used in AGN Watch, e.g.][]{pe94, pe99, pe02}. 
As noted in these papers, even after scaling of the spectra to a common value of 
the [OIII] 5007 flux, there are systematic differences
between the light curves produced from the data obtained with different telescopes.
Therefore, it is proposed to correct for small offsets between the light curves 
from different sources in a simple, but effective fashion \cite[e.g.][and references therein]{pe02}, 
attributing these small relative offsets to aperture effects \citep{pe95}. 
The procedure also corrects for  other unidentified systematic differences
between data sets (for example, miscentering of the AGN nucleus in 
spectrograph aperture, etc.). In our paper we took the GHO-data as standard, 
because this data set contains the largest number of observed spectra. 
The correction factors  $\varphi$ and $G(g)$ are determined 
empirically  by comparing pairs of nearly simultaneous observations from 
each of the given telescope data sets (L(U), SPM, Z1K, Z2K) to that of the GHO data 
set. In practice, intervals which we defined as "nearly simultaneous"
are typically of 1-2 days. Therefore, the variability on short time 
scales ($<2$ days) is suppressed.} The point-source correction factors
$\varphi$ and $G(g)$ values for different samples are listed in
Table~\ref{tab3}. Using these factors, we re-calibrated the
observed fluxes of H$\alpha$, H$\beta$, \ion{Fe}{2} 48,49 and
continuum to a common scale corresponding to our standard aperture 
$2.5''\times6''$ (Table~\ref{tab4}).

\subsection{Measurements of the spectra and errors}

\begin{figure}
\includegraphics[width=13cm]{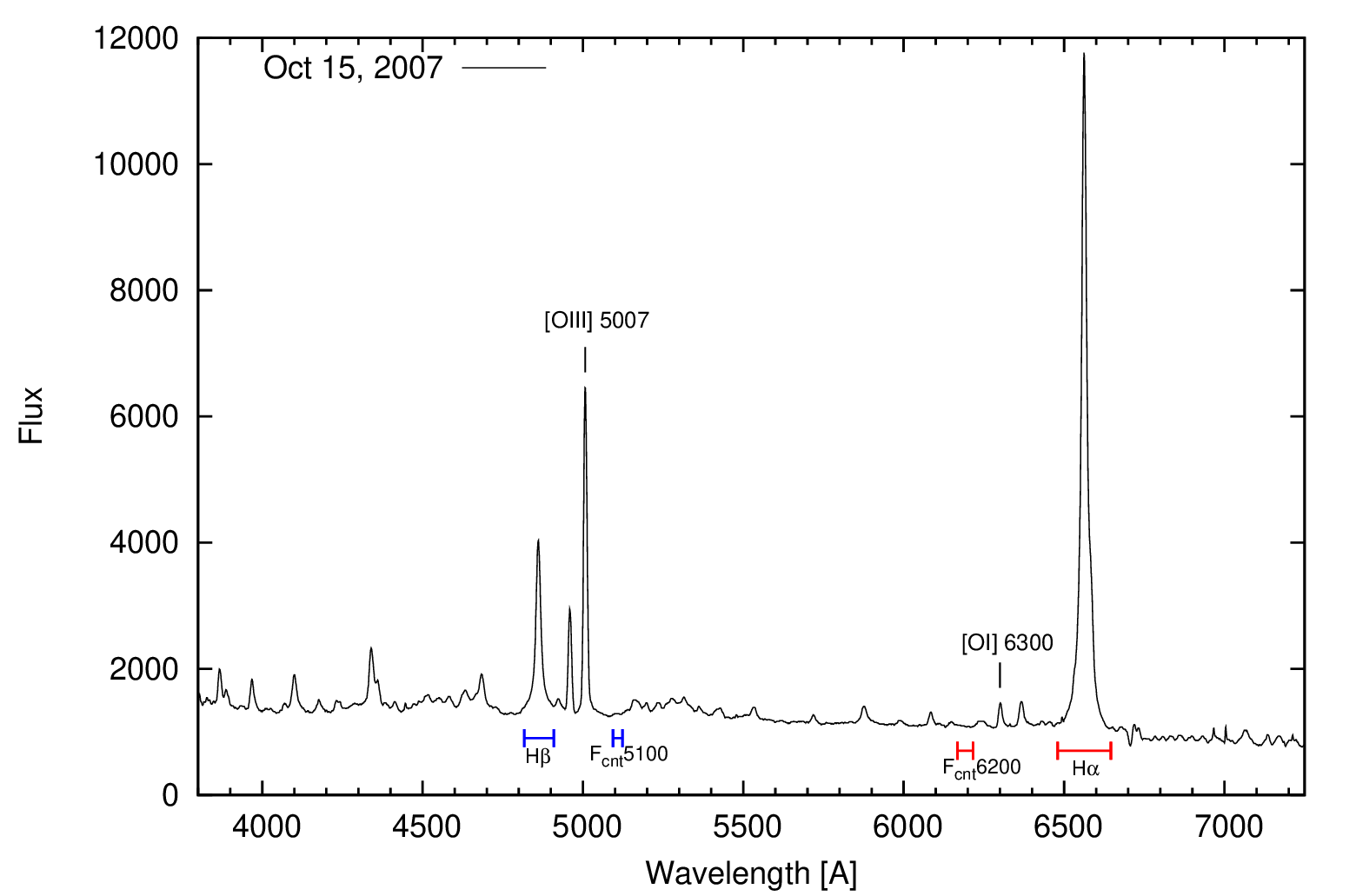}
\caption{An example of the total optical spectrum of Ark 564. The windows for
H$\beta$, H$\alpha$, blue and red continuum measurements are marked.\label{fig01}} 
\end{figure}

From the scaled spectra  we determined the average flux in the
blue continuum at the rest-frame wavelength $\sim 5100$\,\AA, by means of flux
averages in the spectral interval 5094--5123 \AA\ {\bf in the rest-frame} (Table~\ref{tab5}). 
We also calculated the average flux in the red continuum at the rest-frame wavelength $\sim 6200$ \AA, 
by averaging the flux in the spectral interval 6178--6216\,\AA\ 
{\bf in the rest-frame} (Table~\ref{tab5}).  These intervals wavelength were selected because 
they do not contain noticeable emission lines (\ion{Fe}{2} or any other lines, see Fig.~\ref{fig01}).

\begin{figure}
\includegraphics[width=13cm]{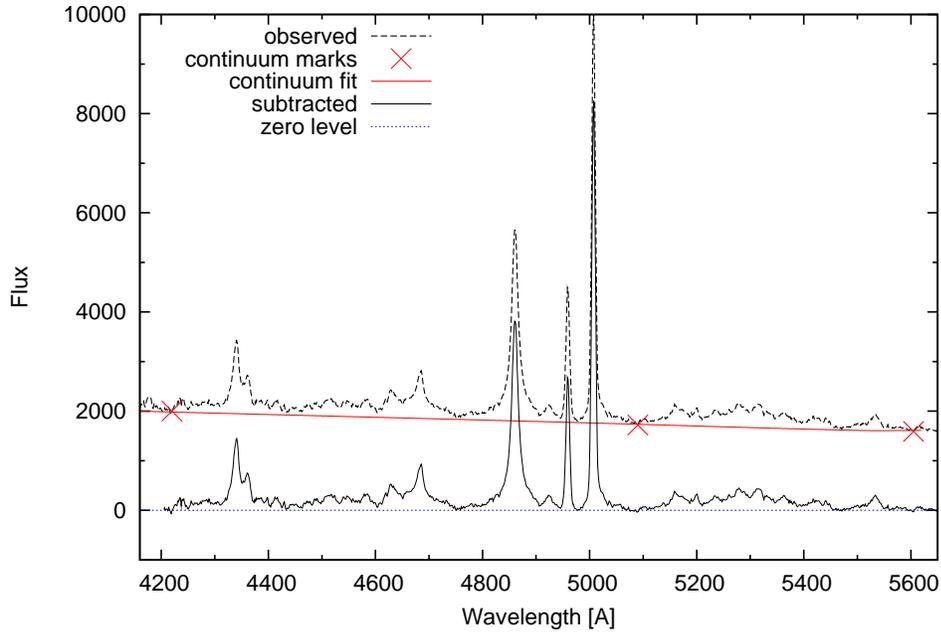}
\caption{Underlying continuum subtraction in the H$\beta$ region needed for accurate \ion{Fe}{2} measurements.\label{fig02}} 
\end{figure}

In order to determine the observed fluxes of the H$\alpha$, H$\beta$ and 
\ion{Fe}{2} lines we need to subtract the underlying continuum, thus, a linear 
continuum was defined through 20\,\AA\ windows, located at {\bf rest-frame wavelengths} 4762\,\AA\, 
(4880\,\AA\ {\bf in the observed-frame}) and 5123\,\AA\, (5250\,\AA\ {\bf in the observed-frame}) for the H$\beta$ line, 
and at {\bf rest-frame wavelengths} 6334\,\AA\ (6490\,\AA\ {\bf in the observed-frame}) and 6656\,\AA\ (6820\,\AA\ {\bf in the observed-frame}) for the H$\alpha$ line (Fig.~\ref{fig01}). In the case of \ion{Fe}{2} emission, 
a precise subtraction of the underlying continuum for a larger wavelength range is required. 
Hence, a polynomial fit for the continuum was drawn through continuum windows 
(Fig.~\ref{fig02}) located at {\bf the rest-frame wavelength intervals} 
4210--4230 \AA, 5080--5100 \AA, 5600--5630 \AA, \citep[see e.g.][]{ku02,ko10}.

After the continuum subtraction, we measured the observed fluxes of the emission 
lines in the following {\bf rest-frame} wavelength intervals: 4817--4909\,\AA\, for H$\beta$, 
6480--6646\,\AA\, for H$\alpha$, and 5100--5470\,\AA\, for the \ion{Fe}{2} emission 
(hereafter \ion{Fe}{2} red shelf). The measurements are given in Table~\ref{tab5}. 
In this \ion{Fe}{2} wavelength range, mainly the 48 and 49 \ion{Fe}{2} multiplets are 
located (Fig.~\ref{fig02}), yet there is also a contribution of the 42 multiplet 
around 5170\,\AA\ {\bf in the rest-frame} \citep[see][]{ko10}.
This spectral interval was chosen because the \ion{Fe}{2} lines there included are not
blended with other either strong broad and narrow emission lines (e.g. He II 4686\,\AA).
This allows to determine the \ion{Fe}{2} line fluxes in a straightforward manner (Fig.~\ref{fig02}).
Further in the text, we discuss a more detailed analysis
of the \ion{Fe}{2} emission in a wider spectral interval 4100--5600\,\AA\ 
{\bf in the rest-frame} (see Section 5).

Worth noting, is the fact that the H$\beta$ and H$\alpha$ fluxes here reported,
include the corresponding narrow component fluxes: in the case of H$\beta$ -- only the narrow H$\beta$ 
is included (the [OIII]$\lambda\lambda$4959,5007 lines are out from the H$\beta$ spectral interval); 
while for H$\alpha$ case -- lines of [NII]$\lambda\lambda$6548, 6584 
and narrow H$\alpha$ are included. As fluxes of narrow lines are assumed to be constant, they have 
no influence on the broad line component variability. The line and continuum 
fluxes were corrected for aperture-effect using the listed correction factors in 
Table~\ref{tab3} (see Subsection 2.2).

\begin{figure*}
\includegraphics[width=8cm]{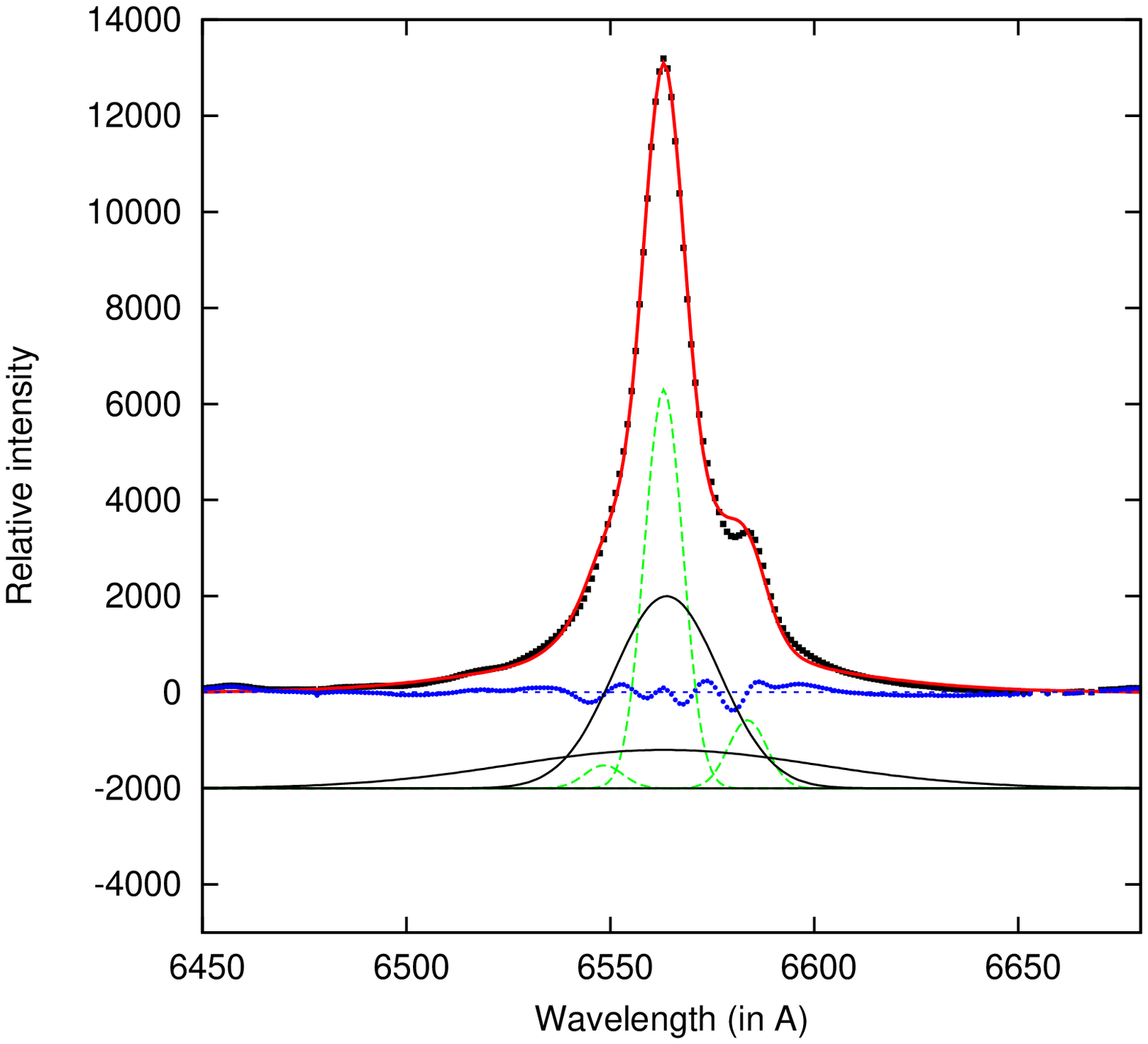}
\includegraphics[width=8cm]{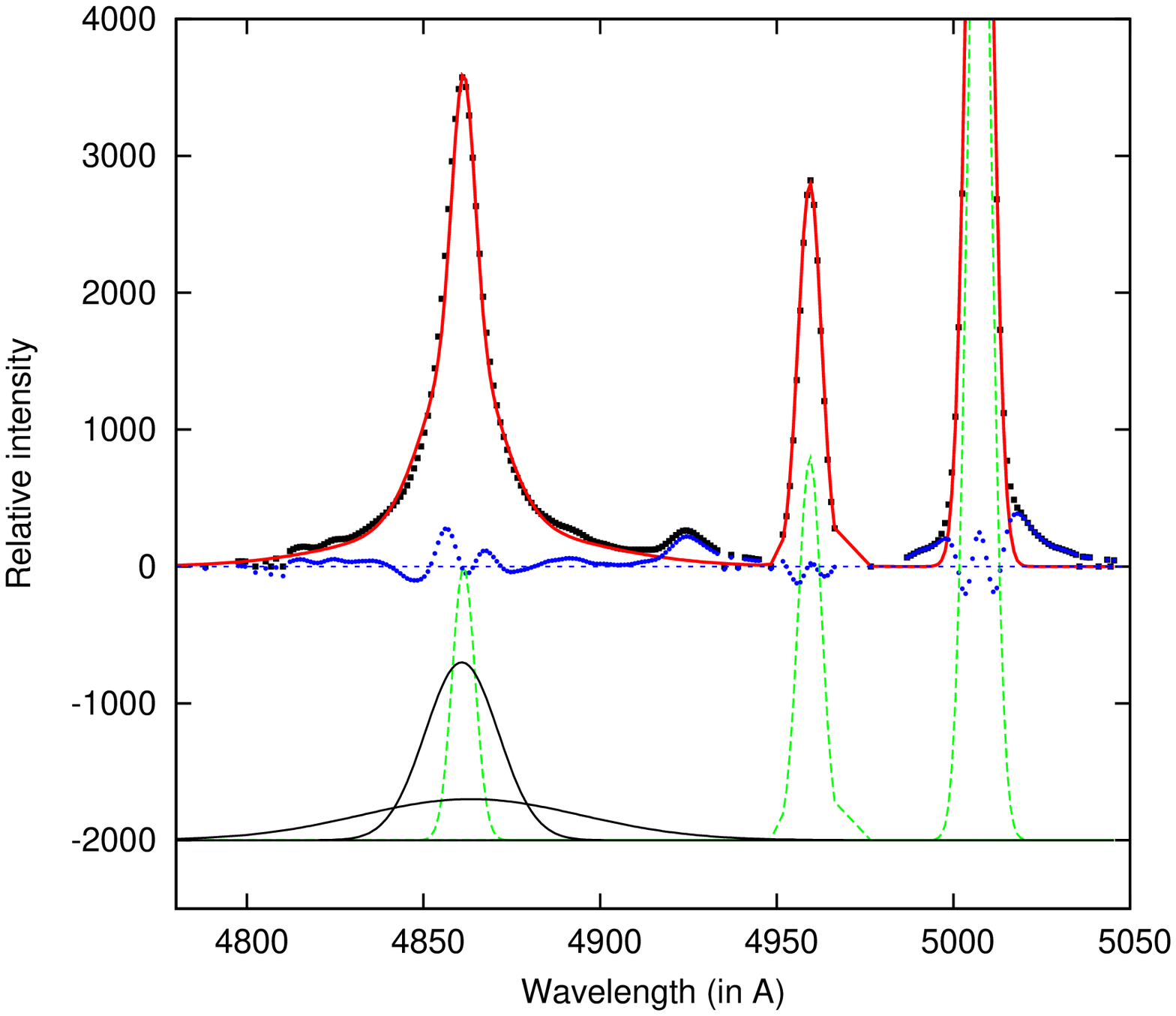}
\caption{The best Gaussian fitting (solid line) of the mean H$\alpha$ (left) and H$\beta$ (right) line profiles 
(dotted line) with a sum of Gaussians. The broad {\bf components} are fitted with two Gaussian (solid lines) and 
the narrow lines with one (dashed lines). The line residuals are also given below the observed spectra. 
In the region of the H$\beta$ the \ion{Fe}{2} contribution is not subtracted.\label{fig03}}
\end{figure*}

In Table~\ref{tab4} the fluxes for the blue continuum (at 5100 \AA), H$\alpha$, 
H$\beta$ and \ion{Fe}{2} lines are listed.
We have also estimated the flux contribution from the H$\beta$ and H$\alpha$ 
narrow components and [NII]$\lambda\lambda$6548, 6584,
from multi Gaussian fit to the blends (H$\beta$+[OIII]$\lambda\lambda$4959,5007
and H$\alpha$+[NII]$\lambda\lambda$6548,6584) of the mean profiles. The best fits are 
plotted in Fig.~\ref{fig03}.
From the mean spectra, the estimated contribution of F(H$\beta$) narrow component 
to the total line flux is $\sim$20\%. The narrow  F(H$\alpha$) has a contribution 
of about 30\% while the fluxes of the [NII]$\lambda\lambda$6548,6584 a 7\% one. 
A similar result (an averaged contribution of $\sim$18\%) was obtained for the F(H$\beta$) 
narrow component from the gaussian fit to every blue spectrum (see section 5).

Additionally, we measured line-segment fluxes. In doing this, we divided the H$\alpha$
and H$\beta $ line profiles into three parts: a blue wing, a core and a red wing.
The adopted intervals in wavelength and velocity are listed in Table~\ref{tab5}.

The mean uncertainties (errorbars) for the fluxes of continuum, H$\alpha$ and H$\beta$ lines, 
and their line segments (wings and core) are listed in Table~\ref{tab5}.
These quantities were estimated from the comparison of the results of spectra obtained 
within a time interval shorter than 3 days. The details of evaluation techniques of these
uncertainties (errorbars) are given in \cite{sh08}. As can be notices, in Table~\ref{tab5} 
the mean error of the continuum flux, total H$\alpha$, H$\beta$ lines
and their cores, is $\sim$4\%. While due to their relatively weaker flux, the errors in 
the determination of the fluxes of the \ion{Fe}{2} and line wings are larger, about $\sim$(7--9)\%.

\section{Results of the data analysis}

\subsection{Variability of the emission lines and of the optical continuum}

\begin{figure}
\includegraphics[width=13cm]{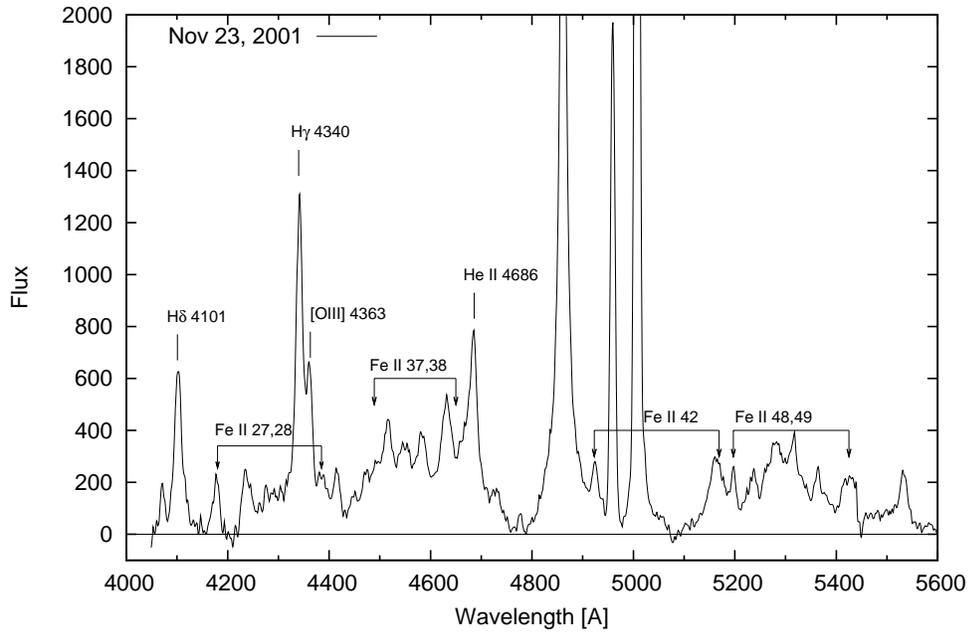}
\caption{The  \ion{Fe}{2} emission around the H$\beta$ line for Ark 564.\label{fig04}} 
\end{figure}

We analyzed flux variations in the continuum and emission lines from a total of 91 spectra 
covering the H$\beta$ wavelength region, and 50 spectra covering the H$\alpha$ 
line vicinity. In Fig.~\ref{fig04} the blue continuum subtracted spectrum of Ark 564, obtained with the 6 m 
SAO telescope in November 23, 2001 (JD 2452237.1) is presented. There, we marked the positions of some 
relevant \ion{Fe}{2} multiplets (27,28,37,38,42,48,49) and other important emission lines. As it can be 
easily noticed, the \ion{Fe}{2} emission is rather strong, as it is usually the case in NLS1 galaxies. 

From the flux data listed in Table~\ref{tab4}, we obtained light curves for the blue and
red continua, and for H$\alpha$, H$\beta$, and \ion{Fe}{2} emission (Fig.~\ref{fig05}),
and their line-segments (blue wing, core, and red wing, Fig.~\ref{fig06}).
As one can see in Figs.~\ref{fig05} and \ref{fig06}, the fluxes declined 
slowly from the beginning to the end of the monitoring period.
For H$\alpha$ and H$\beta$ a decline of $\sim$20\% is present, while, for the \ion{Fe}{2} 
emission, a decline of $\sim$30\% and one of $\sim$40\% for the continuum flux are seen (Fig. \ref{fig05}).
In the upper panel of Fig. \ref{fig05}, the upper dashed line represents the flux at the beginning of 
the monitoring campaign, and the lower dashed line at the end of it. There is only one red point in 2010
{\bf at which point} the red flux increased (two lower panels in Fig. \ref{fig05}). Similarly as in \cite{co01}, the light curves
show several flare-like increments (see Fig.~\ref{fig05}). 
The light curves of line wings and core (Fig.~\ref{fig06}) show
practically simultaneous variations.
 There might be 
up to five flare-like events detected in our data (see Table~\ref{tab6}) when the flux increases
$\sim$10-20\% for a short period of time ($\sim$1-3 days, Table~\ref{tab6}),
out of which two flares were prominent: in December 2003 and August 2004.
As it can be seen from Table~\ref{tab6}, as a rule, flare-like events in the continuum 
(see dF(cnt) in \%) are stronger than in emission lines.

{\bf Long-term flux-monitoring programs have shown that the flux variations of
AGNs tend to be stochastic \citep[i.e., there are few cases of
periodicity or quasi-periodicity, see e.g.][]{sh10}. However, the AGN light
curves sometimes, as in the case of Ark 564, can show a flare-like 
characteristics whose spectral properties are consistent with a 
shot-noise process \citep[][]{cd85,hb92,hu92}. One way to reproduce shot noise is
through a superposition of a series of identical impulses, occurring at
intervals dictated by Poisson statistics. In a Poisson process, the
overall rate of events is statistically constant, yet the starting times
of individual events are independent of all previous ones. The time
intervals between events follow an exponential distribution. It is
possible to use such a process in Ark 564 variability investigation, by
assuming a constant flare rate $\rho$, and let $Tj$ be the occurrence time of
the $j$-th flare. Probability of no occurrence of flare in the interval
$[Tj,Tj+\tau]$ is $exp(-\rho\tau)$.

The probability that a second flare will occur within a time   $\tau$
after the first one is
$ p(\tau) = 1-exp(-\rho\tau)$.  Actually, we can say that
$p[Tn+1-Tn<\tau|T0,T1,...Tn]= 1-exp(-\rho\tau)$ means that at least one
flare does occur between $Tn$ and $Tn+\tau$.

As it can be seen from Table 6, in the continuum and H$\beta$ we have 4 events in
4 separate years which gives a density of events in 10-year long
monitoring period of 0.4 events per year. In such a way we could
estimate probability of time between flare events (which could be recorded
in the continuum flux and H$\beta$ line) as $p(\tau)=1-exp(-0.4\tau)$.
As for H$\alpha$, we have 3 events in 3 separate years over 10 years
period, which leads to $p(\tau)=1-exp(-0.3\tau)$.
Finally, in the case of \ion{Fe}{2} line we have 5 events in 4 separate years (2
events occurred at the end of October 2006), so we could take
$\rho=3/10+2/10=1/2$, which leads to $p(\tau)=1-exp(-0.5\tau)$.
 The exponential density is monoton decreasing; hence there is a high
probability of a short interval, and a small probability of a long
interval between flares. This means that typically we will have  flares
occurring close to each other and  spaced out by long, but rare intervals
with no occurrence of flares.}

In Table~\ref{tab7} we list several parameters characterizing the
variability of the continuum, total line, and line-segments fluxes. 
{\bf There are several methods to estimate variability, here we will use the method given by \cite{ob98}.} There $F$ denotes 
the mean flux {\bf over the whole observing period} and $\sigma(F)$ its standard deviation. $R$(max/min) is the
ratio of the maximal to minimal fluxes in the monitoring period.
$F$(var) is a inferred (uncertainty-corrected) estimate of the
variation amplitude with respect to the mean flux, defined as:
$$ F({\rm var})= [\sqrt{\sigma(F)^2 -e^2}]/F({\rm mean}) $$
$e^2$ being the mean square value of the individual measurement
uncertainty for N observations, i.e. $e^2=\frac{1}{N}\sum_i^N e(i)^2$ \citep{ob98}.

From Table~\ref{tab7}  one can see that the amplitude of
variability $F$(var) is $\sim$10\% for the continuum and \ion{Fe}{2} emission and $\sim$7.5\% for the total H$\beta$ flux.
The H$\beta$ blue wing shows slightly larger variability (F(var)$\sim$15\%) than the red one
(F(var)$\sim$11\%, see Table~\ref{tab7}). However, the H$\alpha$ line wings and core
show lower amplitude variability (F(var)$\sim$8\%) than the H$\beta$ wings (F(var)$\sim$(11-15)\%).

\subsection{Mean and Root-Mean-Square Spectra}

We calculated the mean H$\alpha$ and H$\beta$ line profiles and their root-mean-square (rms) profiles. 
{\bf To find the different portion of variability in different line parts, as much as it is possible, first 
we inspect the spectra and conclude that the spectra with spectral resolution $\leq$11 \AA\ for 
H$\alpha$ and $\leq$10 \AA\ for H$\beta$ are good enough for this purpose,}
thus having a sample of 23 red spectra and 61 blue spectra (Fig.~\ref{fig07}). 
For this purposes the spectra were calibrated to have the same spectral resolution 
(11 \AA\ for H$\alpha$ and 10 \AA\ for H$\beta$).

Fig.~\ref{fig07} shows that the H$\alpha$ and H$\beta$ line profiles in Ark 564 
are Lorentzian like (with broad wings), 
that is a characteristic of the NLS1 galaxies \citep[see e.g.][]{su09, su11}. The rms profile of H$\beta$
resembles a Lorentzian like one, while in case of H$\alpha$ there is practically no change in the profile.  
The FWHM of the H$\beta$ line from the observed mean and rms profiles is 960 km/s, and from the 
observed mean profile of H$\alpha$ is 800 km/s. The full width at Zero Intensity (FW0I) of H$\beta$ is much 
more difficult to measure since the \ion{Fe}{2} emission contributes to the red wing. Thus we only give 
estimates of FWOI of H$\beta$ mean profiles (or rms) to be $\sim$8000 and for H$\alpha$ is also $\sim$8000 km/s.
As it can be seen in Fig. \ref{fig07}, the rms is relatively weak 
($F_{\rm rms}({\rm H}\alpha)/F_{\rm H \alpha} \sim 0.01$ and $F_{\rm rms}({\rm H}\beta)/F_{\rm H \beta} \sim 0.07$), 
meaning that there are no significant changes in the line profiles of H$\alpha$ and H$\beta$ 
during the monitoring period. {\bf Note here that we did a recalibration of the H$\beta$ line taking that the [OIII] lines have the same profile during the monitoring period, therefore we have small rms in forbidden lines, but it is also interesting that the rms shape of H$\beta$ is practically the same as the total H$\beta$ (composed from the broad and narrow components, see Fig. \ref{fig03}). This may indicate that the whole (Lorentzian-like) line is emitted from a complex BLR and that the contribution of the narrow component, that is coming from the same region as the [OIII] lines, is negligible. On the other hand, the H$\alpha$ line rms shows that the variability is caused mainly by variations in the line wings.}

\section{The continuum vs. line flux correlations}

To determine whether there are any changes in the structure of
the broad-line region (BLR), we investigated both the relationships between the total line flux
of H$\alpha$, H$\beta$ and \ion{Fe}{2} and different line-segments (wings and core) of
H$\alpha$ and H$\beta$.

In Fig.~\ref{fig08} the correlations between the total line flux of H$\beta$, H$\alpha$ and \ion{Fe}{2}
are presented. It is interesting to note that the correlations between the flux variation of H$\beta$ and 
H$\alpha$ is significantly weaker (r$\sim$0.40, and it seems statistically insignificant with 
P=0.0053) than that with the \ion{Fe}{2} (r$\sim$0.58, and P$<10^{-8}$). The lack of correlations between the
H$\alpha$ and H$\beta$ fluxes may indicate a very complex BLR structure. 
On the other hand, the correlation between different line-segment fluxes (i.e. blue/red wing-core,
blue wing-red wing) are better, especially for H$\alpha$ (Fig.~\ref{fig09}). 

In Fig.~\ref{fig10} we present the relationships between the continuum flux at 
6200\ \AA\ (for H$\alpha$) and 5100\ \AA\ (for H$\beta$) and the total line  
and line-segment fluxes for H$\alpha$ and H$\beta$. As it can be seen in 
Fig.~\ref{fig10} the correlation between line and continuum fluxes are  weak.
Such  weak  linear correlations of the lines  with continuum may indicate
existence of different sources of ionization (AGN source - photo-ionization, 
shock-impact excitation and etc.). It is interesting to note that the \ion{Fe}{2} emission seems to
show a slightly better correlation with the continuum at 5100 \AA\ 
(r$\sim$0.76, and P$<10^{-16}$) than Balmer lines (Fig.~\ref{fig11}).

\subsection{Balmer decrement}

We have calculated the BD=F(H$\alpha$)/F(H$\beta$) flux ratio, i.e. the Balmer decrement 
(see Fig.~\ref{fig12}), using $\sim$50 blue and red spectra taken 
in the same night (or {\bf one night before or after}). We obtained a mean Balmer decrement value  
BD(mean)=4.396$\pm$0.369, and no significant changes in the monitoring period. 
In Fig.~\ref{fig12} the BD against the continuum flux is shown, and it can be seen there is 
no correlation between the BD and continuum (R$\sim$0.02). It is apparent that the BD was 
more or less constant during the 11-year monitoring period. 
{\bf The ratio of H$\alpha$ and H$\beta$ depends on the physics in the BLR, and  
in the low density regime, the H$\alpha$/H$\beta$ ratio is expected to be below 4 
and it has a slight dependence on temperature \citep[see discussion and Figs. 6 
and 7 in][]{il12}. In the high density regime, the H$\alpha$/H$\beta$ ratio starts 
to depend on the temperatrue \citep[see][]{il12}. The obtained mean BD for Ark 564 seems 
to be close to the high density regime, and changes in the BD from 3.5 to 5.5 might 
be caused by an inhomogeneous BLR, i.e. may indicate a stratified BLR in density, 
temperature and rate of ionization.}

\subsection{Lags between continuum and permitted lines}

In order to determine potential time lags between the continuum and 
{permitted} line changes, we calculated the
cross-correlation function (CCF) for the continuum light curve
with the emission-line light curves. There are several ways to
construct a CCF, and it is always advisable the use of two or more
methods to confirm the obtained results. Therefore, we cross-correlated 
the 5100 \AA\ continuum light curve with both the H$\beta$ and H$\alpha$ 
line (and \ion{Fe}{2} emission) light curves using two methods: (i)
the z-transformed discrete correlation function (ZDCF) method
introduced by \cite{al97}, and (ii) the interpolation cross-correlation 
function method (ICCF) described by \cite{bi99}.

The time lags calculated by ZDCF are given in Table \ref{tab8}, where it can be seen that 
inferred lags have large associated errors.  It is interesting to note that the \ion{Fe}{2} lines tend to 
have shorter lag values, while the longest one is that of the H$\alpha$ line. {\bf If one takes a direct conversion from the time lag to the BLR size, the expected BLR sizes are of an order of $10^{-3}$ parsecs (0.003 to 0.0055 pc) that indicate a compact BLR, but also a strong stratification in the emitting region of Ark 564, where the \ion{Fe}{2} emitting region tends to be very compact and the largest one is the H$\alpha$ emitting line region}. 
We also calculated the lags using the ICCF and found a delay between the continuum
and H$\beta$ line of $\sim$6.7 days, and between the continuum and \ion{Fe}{2} emission
of about 0 days, that is in agreement with the ZDCF method (see Table \ref{tab8}). The errors are around 10 lightdays.

The uncertainties in the delays inferred from the CCFs are difficult to estimate, 
especially the evaluation of a realistic error of the CCF. 
In our case, the main problem in the time delay determination, likely comes from the small variation detected
in lines and continuum fluxes (see Table~\ref{tab7}) and also their weak correlations (see Figs.~\ref{fig10}
and \ref{fig11}). Therefore, all obtained lag times should be taken with caution.

\section{Variation of the \ion{Fe}{2} lines}

As mentioned above, the strong \ion{Fe}{2} emission is one of the main characteristics for NLS1 galaxies.
Optical \ion{Fe}{2} ($\lambda\lambda$4400--5400) emission is one of the most interesting features in 
AGN spectra. The emission arises from numerous transitions of the complex 
\ion{Fe}{2} ion \citep[see][for more details]{ko10}. The iron emission is seen in almost all type-1 
AGN spectra and it is especially strong in the NLS1s. The origin of the optical \ion{Fe}{2} lines, 
their excitation mechanisms, and the spatial location of the \ion{Fe}{2} emission region in AGNs are still 
open questions \citep[see e.g.][]{po09,ko10,po11}. There are also many correlations between the Fe 
II emission and other AGN properties which require a physical explanation. As discussed above, 
we found that \ion{Fe}{2} lines show a slightly better correlation with the continuum at 5100 \AA\ than Balmer 
lines (Fig.~\ref{fig11}). On the other hand, it seems that the H$\beta$ line flux is better correlated with the  
\ion{Fe}{2} emission, than with H$\alpha$.  {\bf We should note here that this better correlation might 
be caused by the fact that the \ion{Fe}{2} fluxes are much more susceptible to the contamination from
the continuum emission than the H$\alpha$ and H$\beta$ lines.}

In order to explore the variability of \ion{Fe}{2} lines in more detail, we fitted the \ion{Fe}{2} emission complex
by the multi-gaussian fitting method that \cite{ko10} and \cite{po11} have described in detail. 
Our spectra cover a wider wavelength range ($\lambda\lambda$ 4100--5600 \AA). Hence, simultaneous fits of
the \ion{Fe}{2} template and H$\delta$, H$\gamma$, He II $\lambda$4686, and H$\beta$ line were carried out. 
Additionally to the \ion{Fe}{2} line template introduced in \cite{ko10}, we included here 17 other \ion{Fe}{2} lines, basically  the transition arising from two groups with lower levels $^4$P and $^2$H\footnote{The atomic data was taken from 
the NIST atomic database: http://www.nist.gov} (see Fig~\ref{fig13} and Table~\ref{tab9}).

An example of a best fit is presented in Fig~\ref{fig14}, where all the strong emission lines and \ion{Fe}{2} features 
are labeled. Since the gaussian best-fit includes large number of free parameters (see Fig~\ref{fig14}), we focused our attention in the fit required to reproduce as close as possible the \ion{Fe}{2} line emission. To this aim, first we fit the strong hydrogen and helium lines, and correct their contribution to the \ion{Fe}{2} lines, and then we apply the best-fit 
procedure to the \ion{Fe}{2} emission. With this scheme, we substract the emission of all other lines and deal only 
with the \ion{Fe}{2} spectrum (Fig~\ref{fig15}). We fixed as many as possible 
Gaussian parameters, e.g. the ratio of [OIII] lines, or widths of the Balmer line components NLR, ILR and BLR 
\citep[ILR - intermidiate-line region, see also][]{zhang11} 
The line parameters inferred from our fits (width, shift, intensity relative to total H$\beta$) are given in Table~\ref{tab10}.

\cite{ko10} divided the \ion{Fe}{2} emission into subgroups according to the lower level of the transition. We used the same
criteria, thus we considered here 6 line groups: $^4$P, $^4$F, $^6$S, $^4$G, $^2$H, and I Zw 1 line group.  
In Fig.~\ref{fig16} we plot the fluxes of all these line groups, and the total \ion{Fe}{2} emission in the 4100--5600 \AA, against the continuum flux at 5100 \AA. We omitted the $^4$P group, since below 4200 \AA\ the points are missing for more than 50\% of considered spectra, and thus the flux measurements for this group are systematically lower.
The total \ion{Fe}{2} emission is correlating well with the continuum (r$\sim$0.63, P$<10^{-10}$). This correlation
is slightly smaller than the one obtained for the measured \ion{Fe}{2} in the wavelength region 5100--5470 \AA\ (see Fig. \ref{fig11}).
The Fe $^4$G line group consists of the transitions that contribute the most to the \ion{Fe}{2} emission in 5100--5470 \AA\ range, 
and for this line group we obtained practically the same correlation with the continuum variation as it was measured and presented in Fig.~\ref{fig11} (r$\sim$0.74, P$<10^{-16}$). A relatively good correlations (r$\sim$0.50, P$<10^{-6}$) is obtained for 
$^4$G group (lines located in the blue part of the \ion{Fe}{2} shelf) and for the high-excitation energy group
of I Zw 1 (r$\sim$0.56, P$<10^{-8}$), while the other two groups have no correlation at all: Fe $^2$H (r$\sim$-0.01 and P$=0.82$) and Fe $^6$S (r$\sim$0.14 and P$=0.18$).  In the case of Fe $^2$H this could probably be due the very weak emission coming from 
these transitions, while in Fe $^6$S it seems to be a real effect.

We compared the width of the \ion{Fe}{2} lines to the widths of different H$\beta$ components (Table~\ref{tab10}). 
The average value of the \ion{Fe}{2} lines is the same (within the error bars) to the average width value of the ILR component of 
the H$\beta$ line (Fig.~\ref{fig17}). This clearly supports the idea that the origin of the \ion{Fe}{2} emission is more within 
the ILR than the BLR as stated before \citep[see e.g.][]{ms93, po04, po09, ko10}.

The quantity R$_{\rm Fe}$, defined as the flux ratio of optical \ion{Fe}{2} emission to H$\beta$ line,
 is an important one in describing the EV1 parameter space \citep[see][]{bo92}. Here the H$\beta$ flux 
includes the contributions of all three components (narrow, intermidiate and broad), but still
represents the behavior of the broad H$\beta$ since the flux of the narrow component is 
expected to be constant. The variations of R$_{\rm Fe}$ as a function of the blue continuum flux are plotted in Fig.~\ref{fig18}. 
 This plot shows a weak (statistically insignificant), but positive correlation between R$_{\rm Fe}$ 
and continuum flux (the correlation coefficient r$\sim$0.36, and P$<10^{-3}$).

\section{Discussion}

\subsection{The structure of the line emitting region in Ark 564}

The {\bf permitted-line} profiles of Ark 564 are Lorentzian like, 
that can be found in a group of AGNs with FWHM of broad 
lines smaller than 4000 km s$^{-1}$ \citep{su09,mar10,su11}. The problem is, that, such line profiles are not expected 
in the classical BLR. The Lorentzian-like line profile can be caused by the composition of the three Gaussian 
profiles \citep[as it was shown in][]{ko10}, where the contribution of the narrow component is significant. 
As e.g. \cite{co03} roughly estimated the contribution of the BLR to the total H$\beta$ line and to the permitted 
lines in the UV and found that H$\beta_{\rm broad}$/H$\beta_{\rm narrow}$ should range between 1 and 2, that is not 
far from our estimation, that narrow component contributes to the total line flux with $\sim$20\%. Also, \cite{rod00} 
showed that the flux carried out by the narrow component of H$\beta$ in a sample of seven NLS1s is, on average, 50\% 
of the total line flux. Therefore, such high contribution of the narrow component to the total line flux can bring a 
small rate of variation in broad lines, and that the emission of the very broad component is very weak, i.e. 
that a relatively small fraction of the total flux in lines is coming from the BLR.

{\bf  The observed weak variation in the {\bf permitted} lines of Ark 564 during a 10-year period is in agreement with a short term monitoring covering 2-year period  given by \cite{sh01}, they found no significant optical line variations.} It is interesting that there is a weak correlation between 
the {\bf permitted} lines and continuum variation. This, as well as the lack of correlation between   H$\alpha$ and H$\beta$, may indicate different sources of ionization, as e.g. the shock wave ionization in addition to the photoionization.  Moreover, five flare-like events (two prominent and three possible ones) are registered during the monitoring period, {\bf that confirm flare-like variability reported in \cite{co01} and indicate burst events in the emission line regions.} These may indicate some kind of explosions (in starburst regions) which can
additionally affect the line and continuum emission. The small [OIII]/H$\beta_{\rm narrow}$ ratio may also indicate
a presence of starbursts in the center of Ark 564 \cite[as it was noted for Mrk 493, see][]{po09}.

Taking into account that it is very hard to properly decompose the narrow component \citep[see discussion in][in more details]{po11} from the broad one, it is hard to discuss about the geometry of the BLR of Ark 564. However, a lack of significant correlation between the H$\alpha$ and H$\beta$ flux variation, may indicate that there is a very stratified {\bf (in physical parameters, see Sec. 4.1) emitting region, where the H$\beta$ emitting region tends to be more compact than the H$\alpha$ one. This idea is supported by the detected differences in the FWHM and FW0I of H$\alpha$ and H$\beta$, and 
the absent correlation between the BD and continuum (Fig. \ref{fig08}). On the other hand, the quasi-simultaneous variations 
of H$\alpha$ and H$\beta$ blue and red wing (Fig. \ref{fig06}) and their good correlations (Fig. \ref{fig09}) indicate a 
predominantly circular motions in the BLR.}

We calculated the CCF and found the delay of only a couple of days. {\bf Taking the H$\beta$ width and lag (assuming that the H$\beta$ lag corresponds to the dimension of the BLR) one can estimate the mass of the black hole of $\sim 1\cdot 10^6$ Solar masses, that is in agreement with previous estimates by \cite{sh01,co01,po01}, and also well fit the hypothesis that NLS1s have lower black hole masses than typical Sy1s.}  However,  one should take with caution this estimates,
since there is no large correlation between the {\bf permitted} lines and continuum variability. The CCF may indicate that the variability (perturbation) is coming from relatively small region, but it is interesting that it causes amplification  of total line flux without {\bf significant} change in line profiles {\bf (even during flare-like events)}. The {\bf permitted} line profiles stay practically the same during the whole monitoring period (see Fig.~\ref{fig07}).

\subsection{\ion{Fe}{2} emission variability in Ark 564}

The variability behavior of the \ion{Fe}{2} complex in Seyfert galaxies has been poorly understood \citep[see][]{col00}. 
As e.g. \cite{kol01}  reported that in Mrk 110 the permitted optical \ion{Fe}{2} complex remained constant within 10\% error 
over 10 years, while the forbidden [Fe X]$\lambda $6375 line was variable. Similarly, in the Seyfert 1 galaxy NGC 5548 no 
significant variations of the optical \ion{Fe}{2} blends (less than 20\%) were detected \citep[][]{di93}. However, the opposite 
result was reported in a long term optical variability watch program on Seyfert 1 galaxy NGC 7603 over a period of nearly 20 
years \citep[][]{kol00}. This object displayed remarkable variability in the \ion{Fe}{2} feature, with amplitudes of the same 
order as for the H$\alpha$ and He I lines. \cite{gi96} found that, out of 12 NLS1s, at least 4 of them presented a significant 
variability of the \ion{Fe}{2} complex with percentage variations larger than 30\%. 
In addition, considerable variations of the \ion{Fe}{2} emission (larger than 50\%) were reported in two Seyfert 1 galaxies: 
Akn 120 and Fairall 9 \citep[][]{kol81,kol85}. 
On the other hand \cite{ku08} performed  a reverberation analysis of the strong, variable optical \ion{Fe}{2} emission 
bands in the spectrum of Ark 120 and they were unable to measure a clear reverberation 
lag for these \ion{Fe}{2} lines on any timescale. They concluded that the optical \ion{Fe}{2} emission does not 
come from a photoionization-powered region similar in size to the H$\beta$ emitting region. Our results confirm this since for
different groups there are different correlation with the continuum and in some groups (as e.g. $^6S$) there is no correlation 
at all (see discussion below).

The most interesting is that the \ion{Fe}{2} variation (at least in the red part of \ion{Fe}{2} shelfs) in 
Ark 564 is closely following the variations in the continuum. 
Similar result was obtained in the case of  NGC 4051, an NLS1 galaxy, where the variability of the optical \ion{Fe}{2} emission
also followed the continuum variability \citep[][]{wa05}. 

We investigate the time variability of several \ion{Fe}{2} multiplets in Ark 564.  An interesting result is that there
are different levels of correlations between the emission of \ion{Fe}{2} line groups and continuum flux. It seems that the level of \ion{Fe}{2} flux variability depends on the type of transition. For example, we found the good correlation for $^4$G and $^4$F
groups which mainly contribute to the blue and red \ion{Fe}{2} features around H$\beta$, and practically
no significant correlation between $^2$H and $^6$S group\footnote{\bf Note that the $^6$S FeII group may be affected by the [OIII]5007 line, therefore
the lack of a correlation for this component might be simply due to measurement bias.} and continuum (Fig.~\ref{fig13}). The emission from these 
two groups seem to be variable, but there is no respons to the continuum variability. This also may indicate 
that the \ion{Fe}{2} emission region in
Ark 564 is stratified. On the other hand, the width of \ion{Fe}{2} lines follow the width of the ILR component,
that is in a good agreement with results reported in previous work \citep{ms93, po04, po09, ko10}

We found a positive (yet, statistically not significant) trend of R$_{\rm Fe}$ with the blue continuum.
\cite{wa05} found the similar positive trend for the galaxy NGC 4051, in opposition to
the negative trend observed in NGC 7603 \citep[reported by][on the data of Kollatschny et al. 2000]{wa05}. They 
argued, by comparing the variability behaviors of different objects, that the 
objects with positive correlations have narrow H$\beta$ lines and consequently are classified as NLS1s. 
While the remaining two sources with negative correlations have relatively broad H$\beta$ profiles.
They interpreted that the dichotomy in variability behavior of R$_{\rm Fe}$ is due to the different 
physical conditions governing the variability of the optical \ion{Fe}{2} emission. Our result is 
consistent with their findings supporting their idea that 
in case of NLS1 we have that the bulk excitation of the optical \ion{Fe}{2} lines is due to collisional 
excitation in a high density optically thick cloud illuminated and heated mainly by X-rays photons 
\cite[see][and references therein]{wa05}.

\section{Conclusion}

\begin{figure}
\includegraphics[width=15cm]{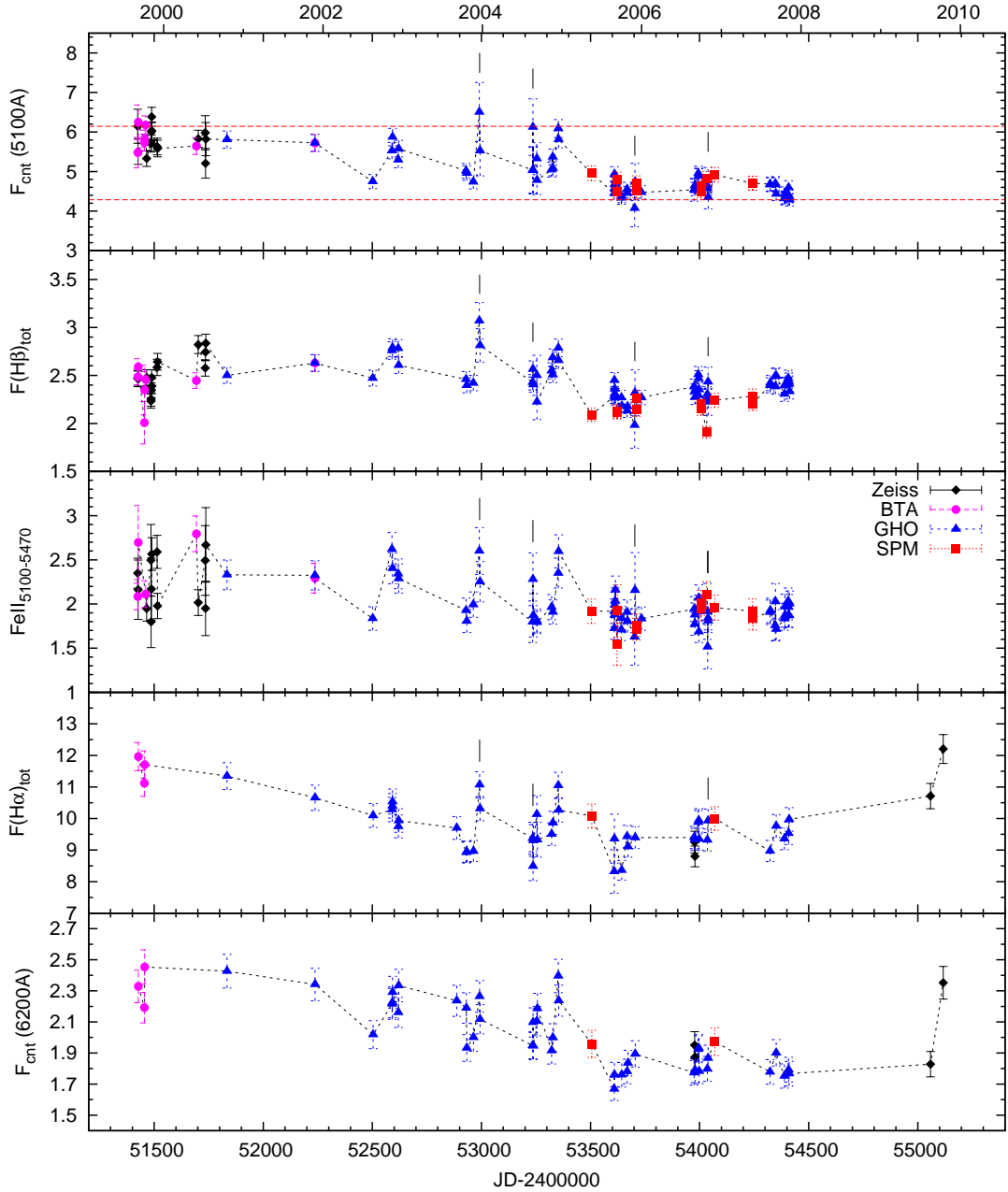}
\caption{Light curves (from top to the bottom) of the continuum at 5100 \AA\, H$\beta$, \ion{Fe}{2}, H$\alpha$ and 
the continuum at 6200 \AA. {\bf Data obtained with different telescopes are marked with different symbols:
diamonds - 6m BTA, circles - 1m Zeiss, triangles - 2.1m GHO, squares - 2.1m SPM}. 
The flares are marked on the upper four plot (see Table~\ref{tab6}), while
in the blue continuum plot (first upper plot) the dashed lines represent the first and last observed
continuum flux to show the decrease of the continuum flux during the monitoring campaign. 
Continuum fluxes are given in units 10$^{-15} \rm erg \ cm^{-2} s^{-1}$ \AA$^{-1}$ and line fluxes 
in 10$^{-13} \rm erg \ cm^{-2} s^{-1}$. \label{fig05}} 
\end{figure}

\begin{figure*}
\includegraphics[width=8cm]{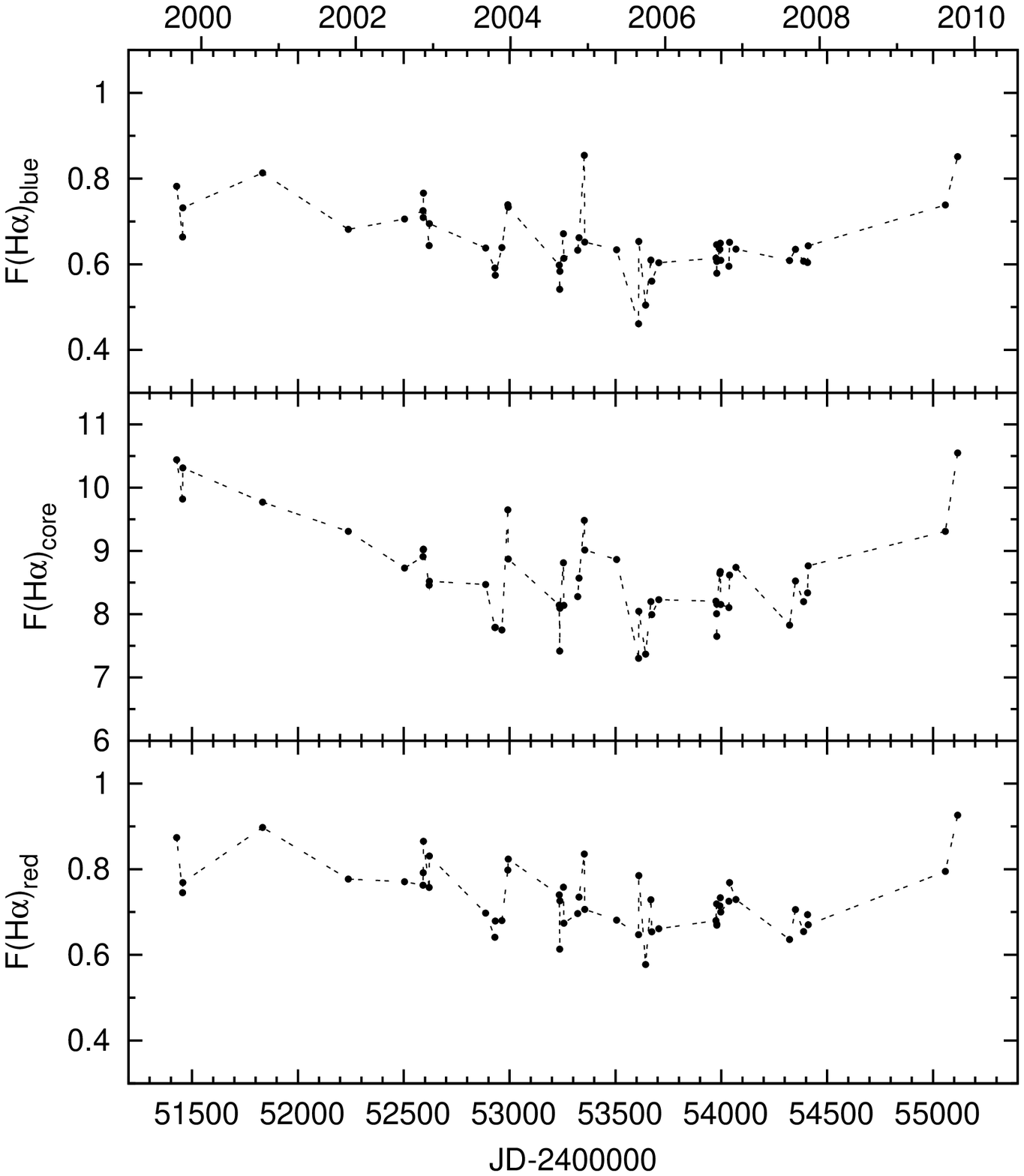}
\includegraphics[width=8cm]{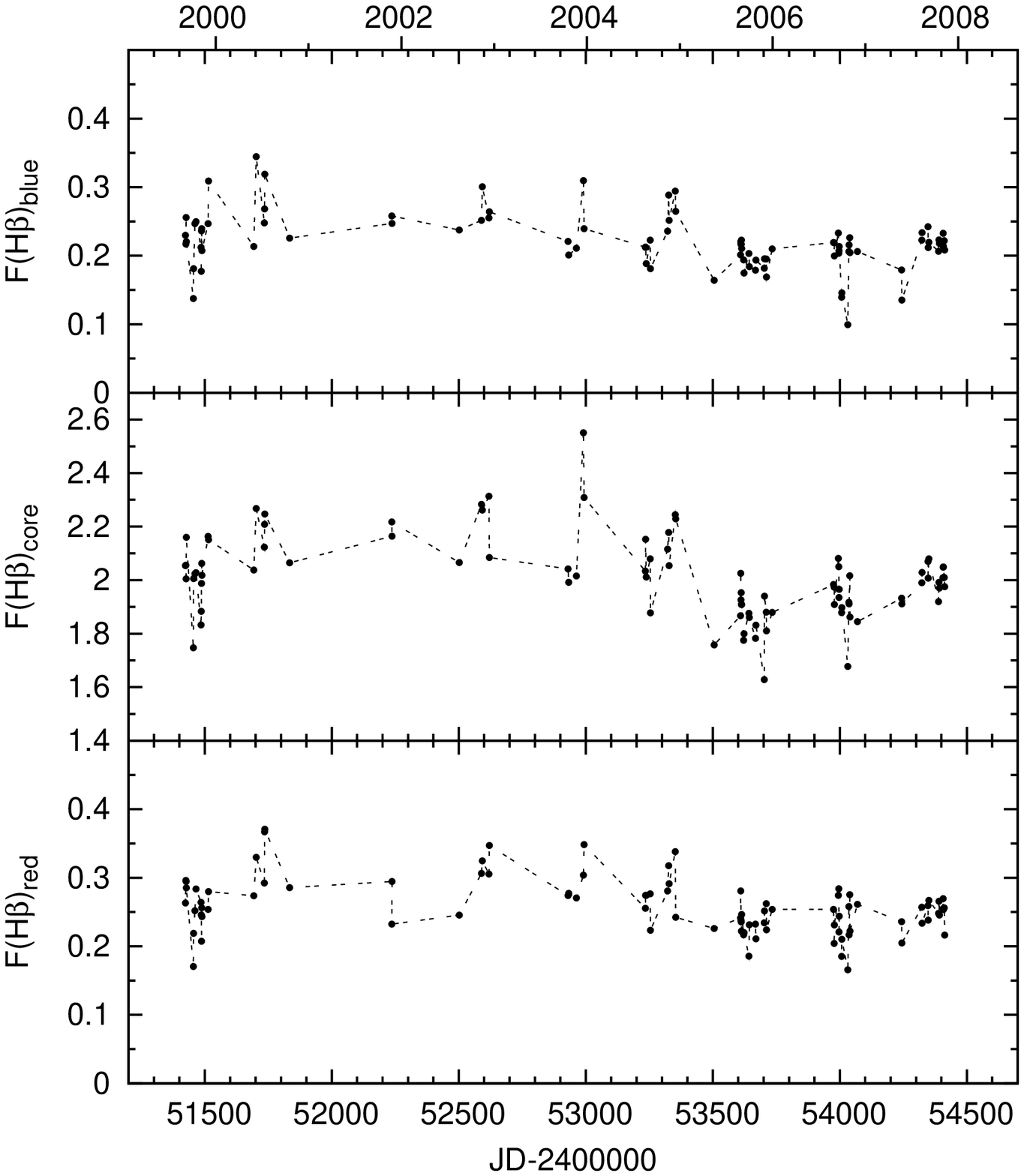}
\caption{Light curves of the H$\alpha$ (left) and H$\beta$ (right) line-segments (from top 
to bottom: blue, core, and red line parts).\label{fig06}} 
\end{figure*}

\begin{figure*}
\includegraphics[width=8cm]{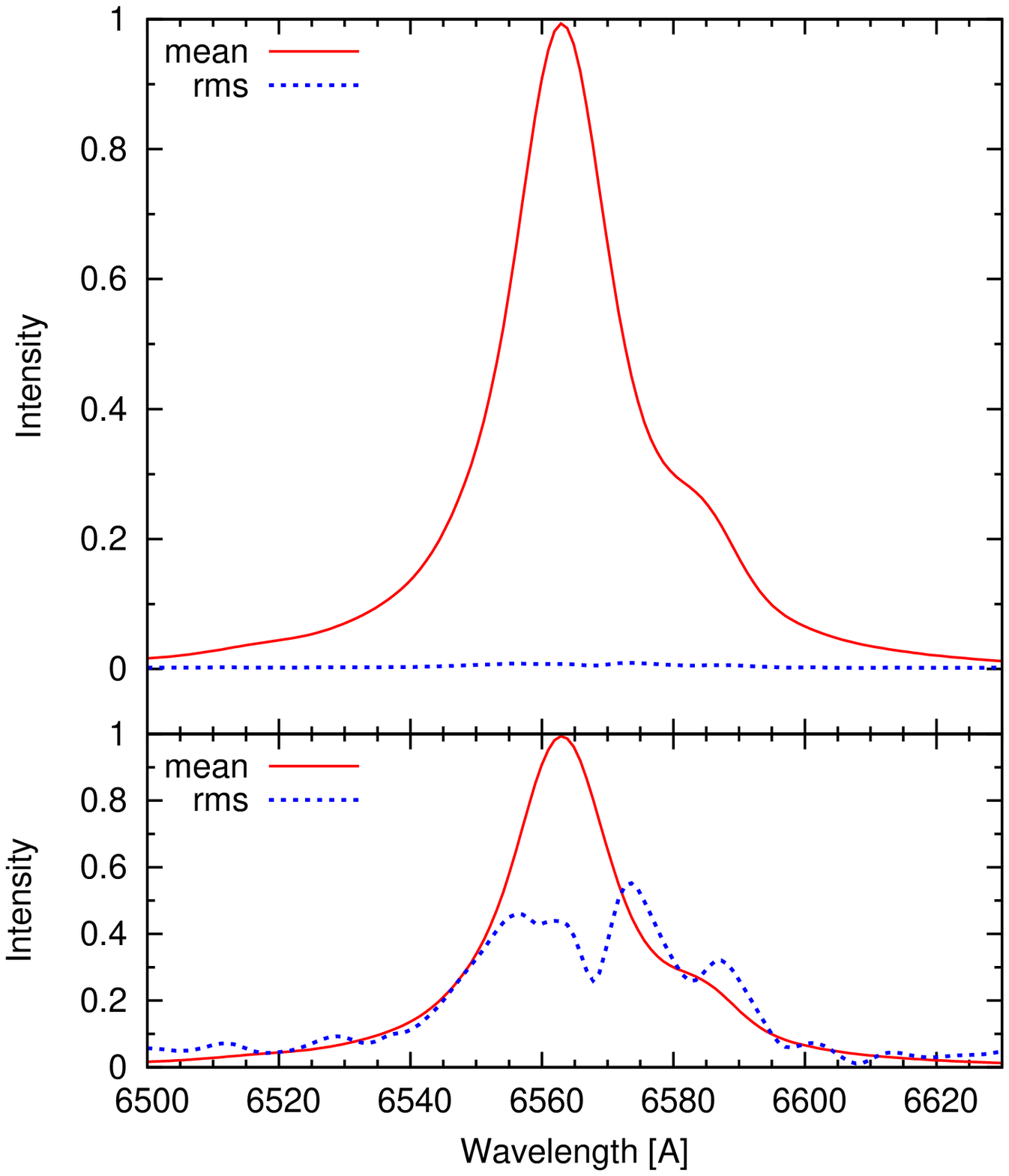}
\includegraphics[width=8cm]{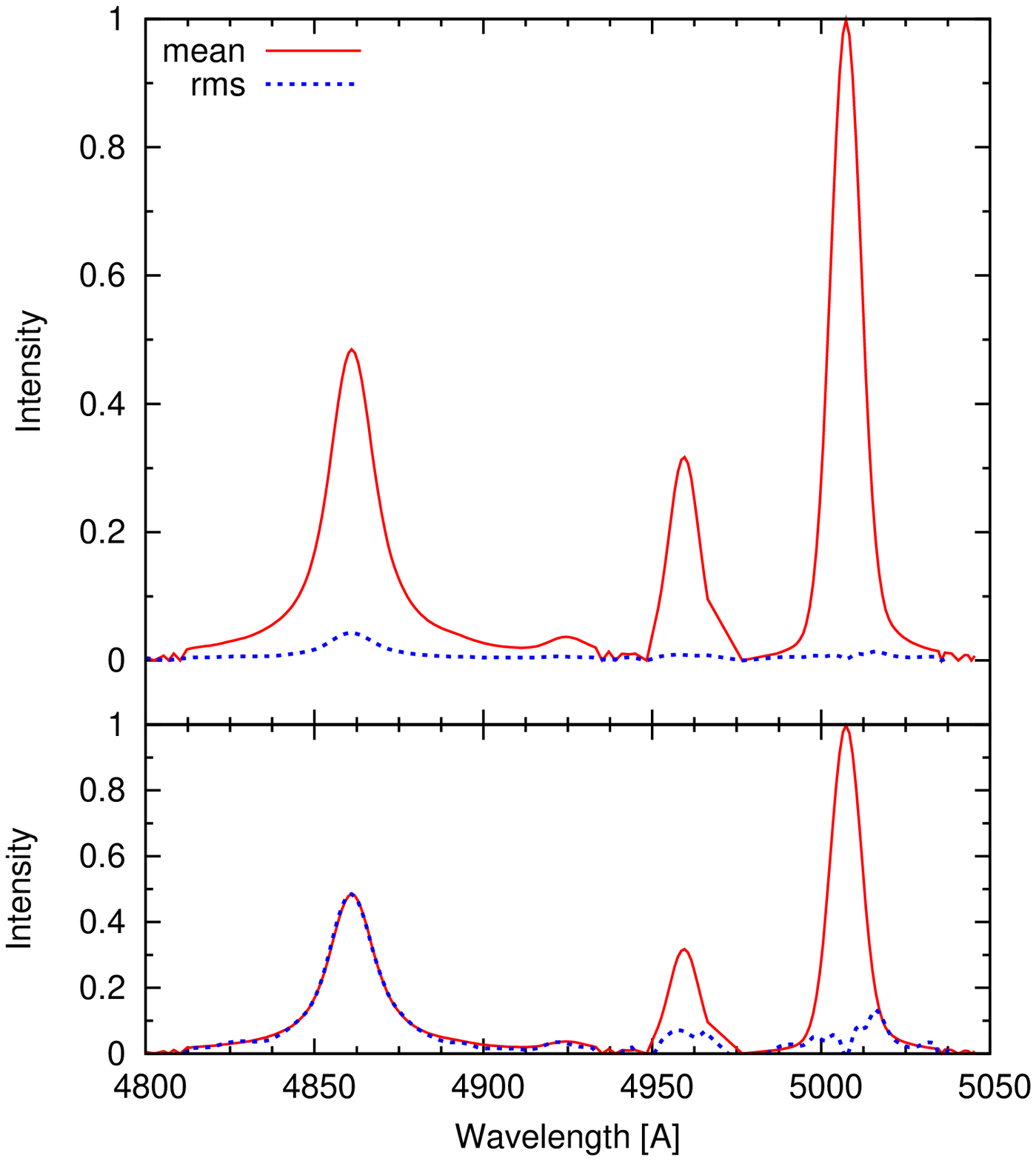}
\caption{The mean and rms spectra of H$\alpha$ (left) and H$\beta$ (right) after calibrating the spectra to
the same spectral resolution. Below plots show the normalized mean and rms spectra arbitrary scaled
for comparison.
\label{fig07}} 
\end{figure*}

\begin{figure*}
\includegraphics[width=8cm]{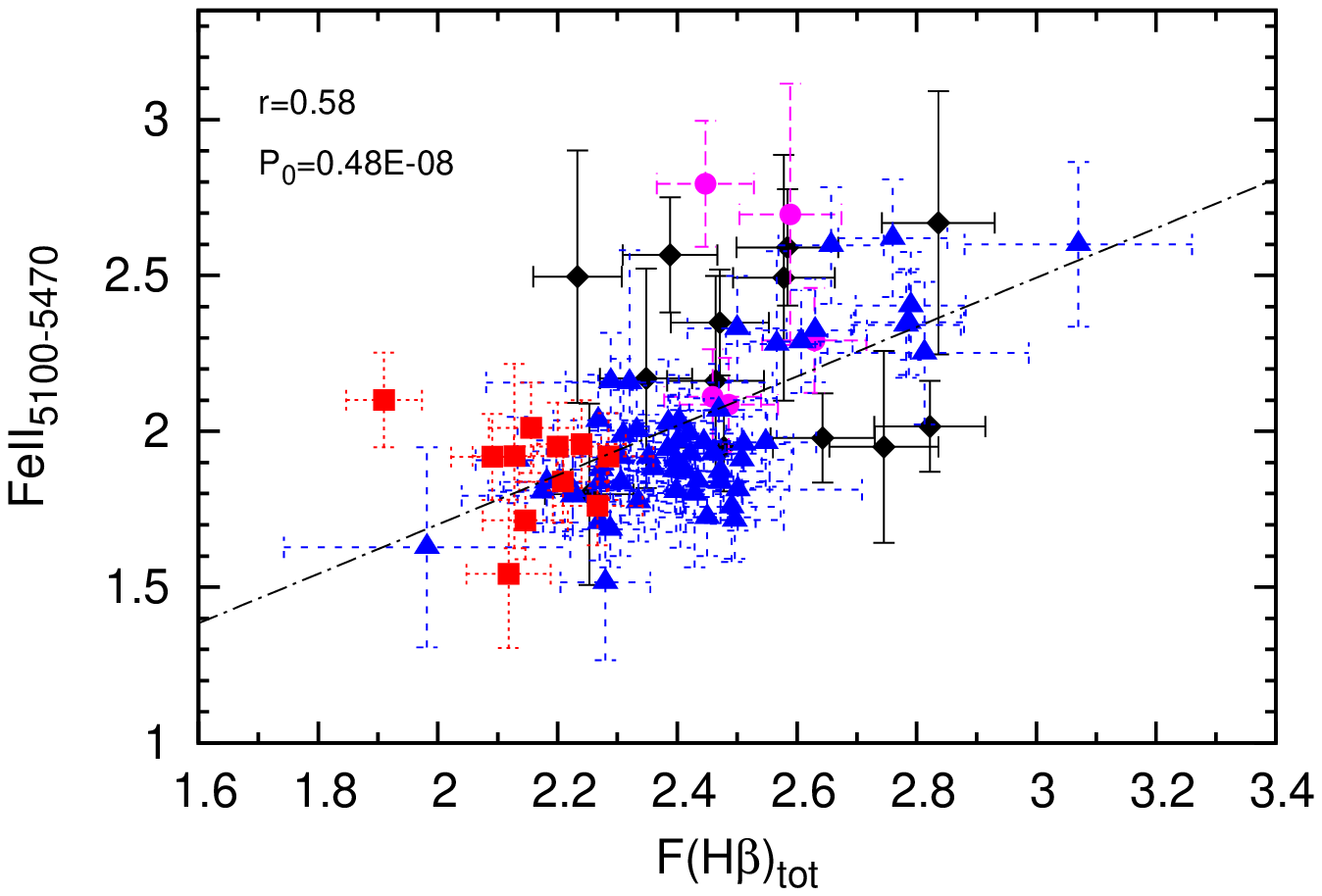}
\includegraphics[width=8cm]{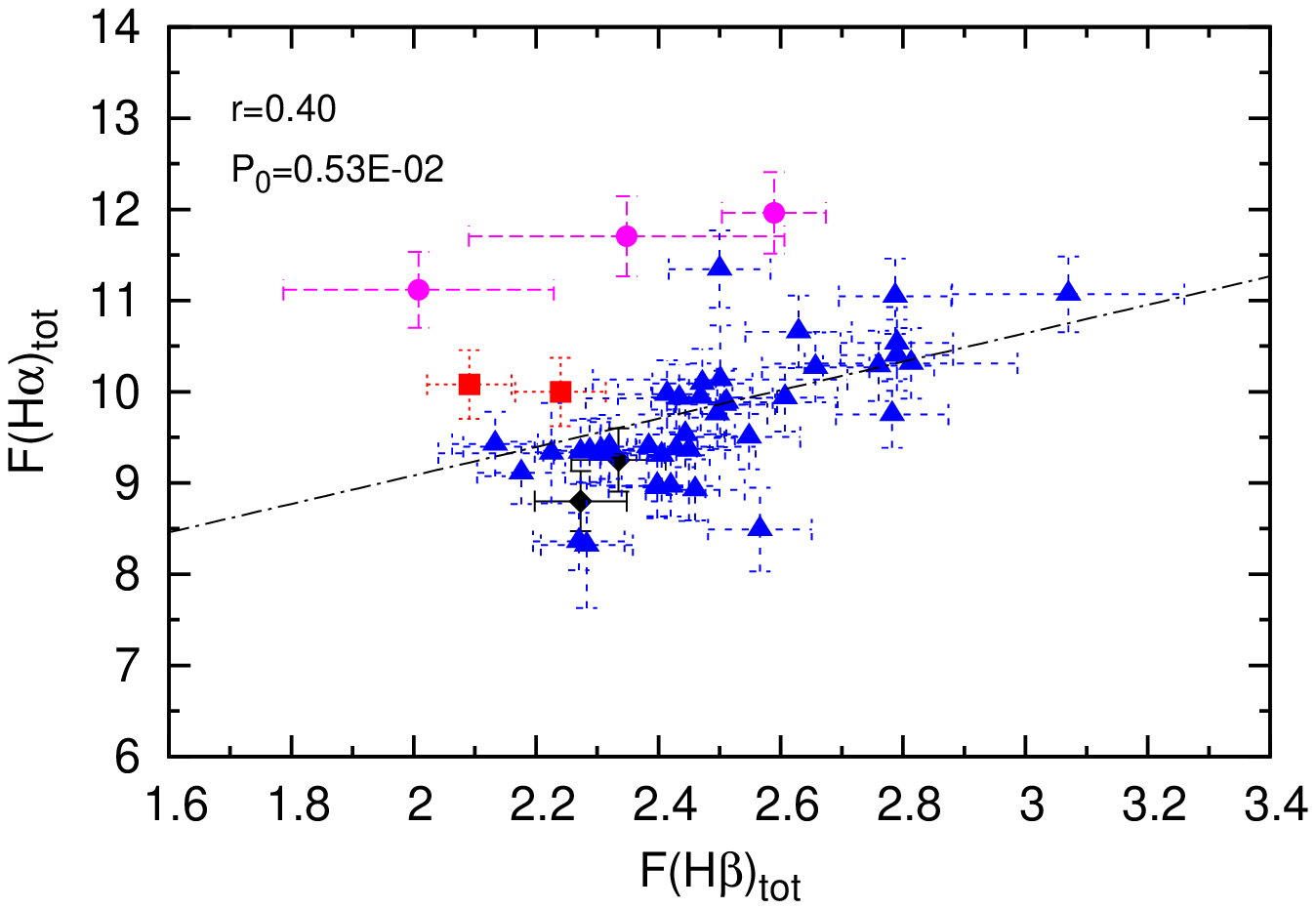}
\caption{The \ion{Fe}{2} red shelf (left) and H$\alpha$ (right) line fluxes as a function of the H$\beta$ line flux.
 The correlation coefficient and the corresponding P-value are given in the upper left corner.
{\bf The notation is the same as in Fig. \ref{fig05}}.\label{fig08}} 
\end{figure*}

\begin{figure*}
\includegraphics[width=8cm]{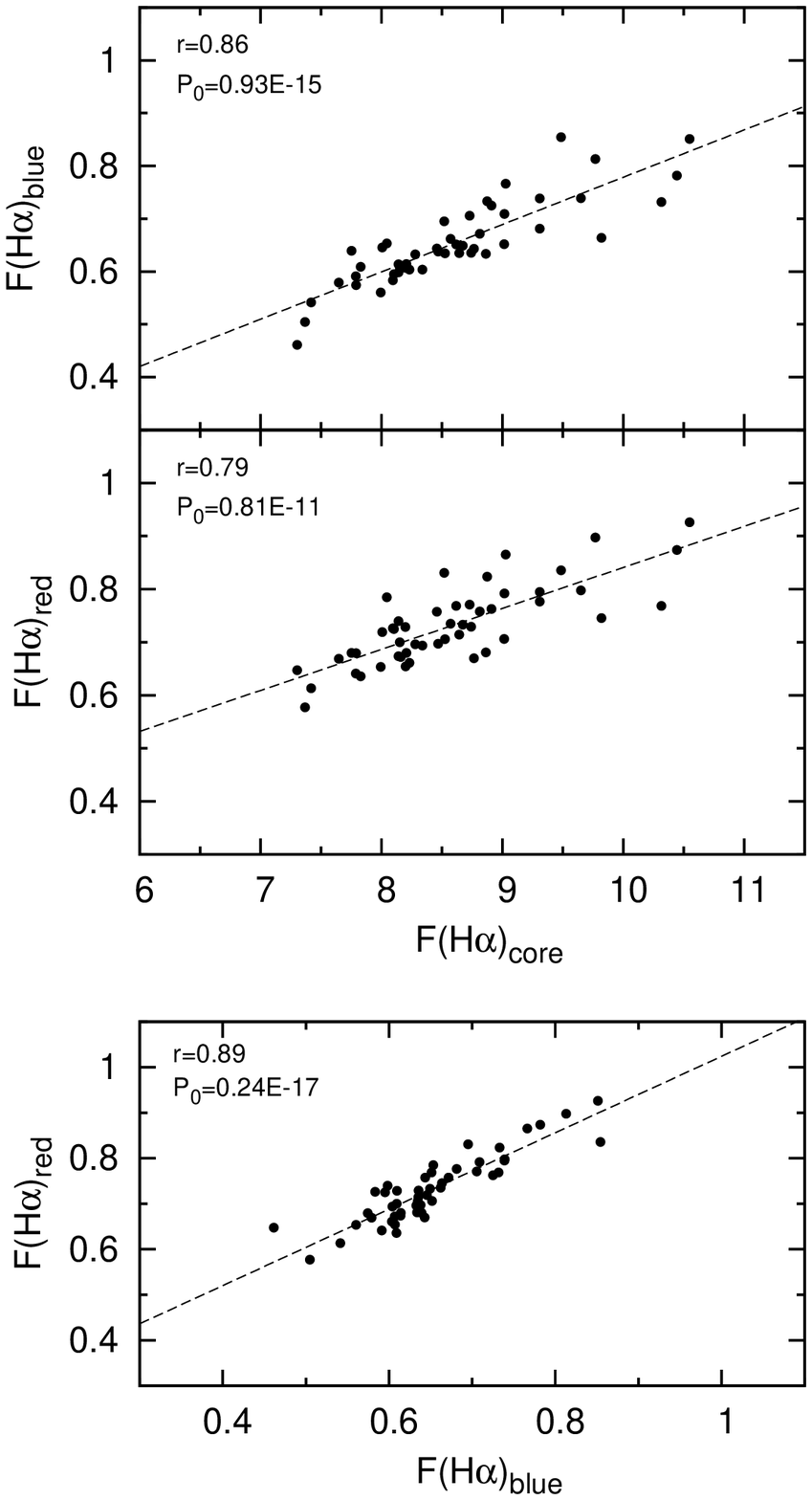}
\includegraphics[width=8cm]{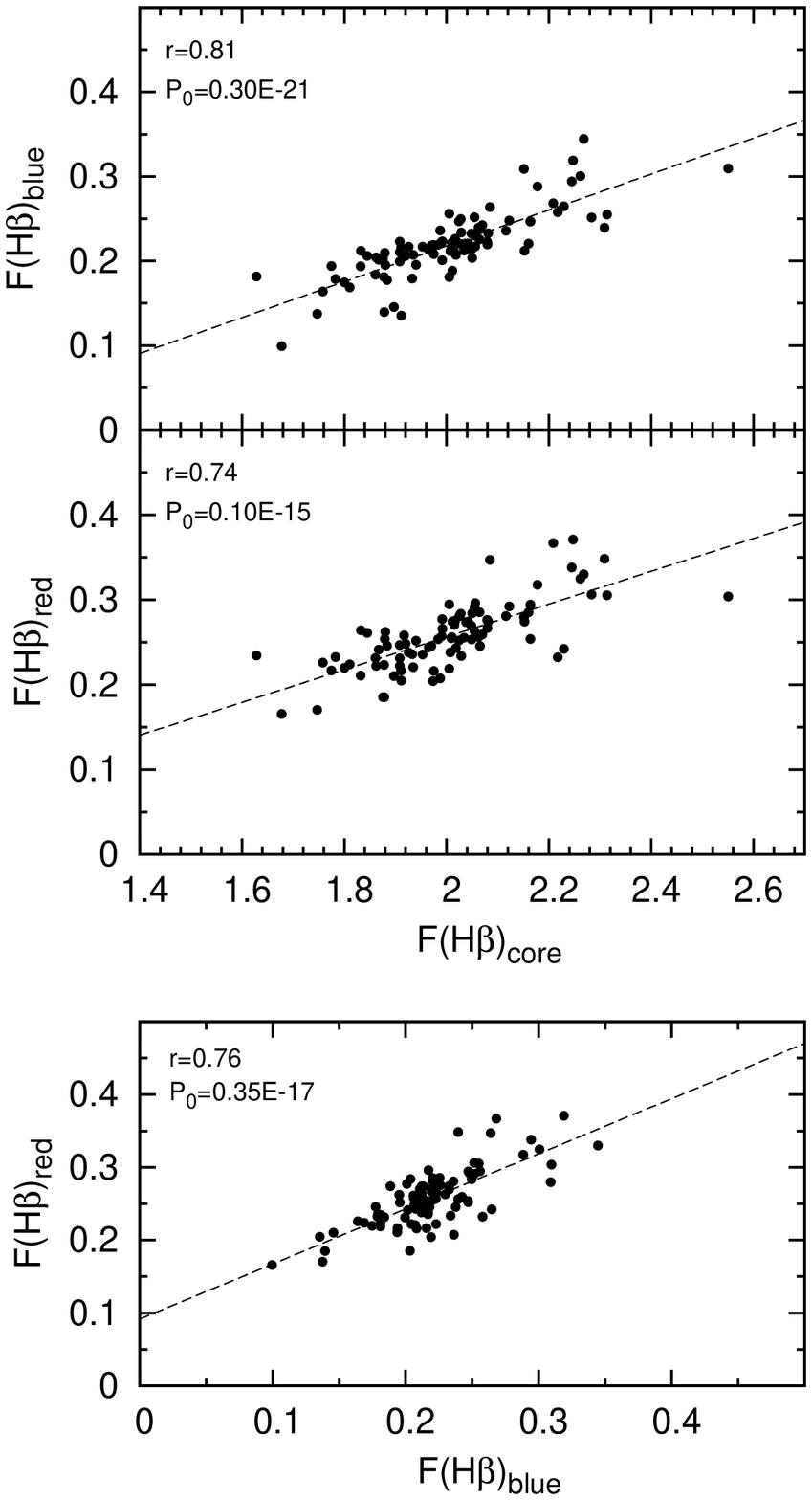}
\caption{H$\alpha$ and H$\beta$ line-wing fluxes (blue, red) vs. line-core flux 
(upper panels), and red vs. blue-wing (bottom panel). The correlation coefficient and the 
corresponding P-value are given in the upper left corner.\label{fig09}} 
\end{figure*}

\begin{figure*}
\includegraphics[width=8cm]{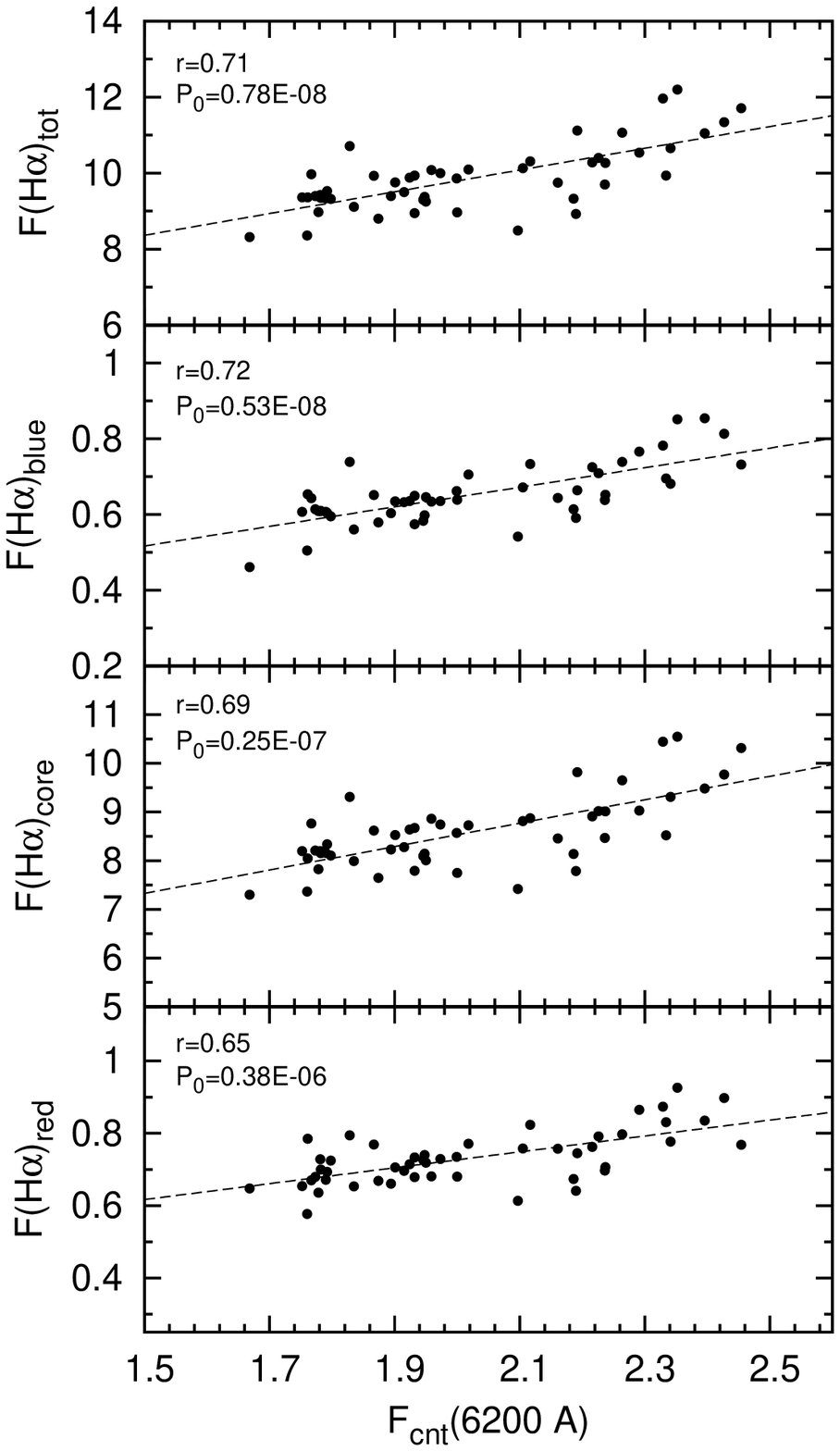}
\includegraphics[width=8cm]{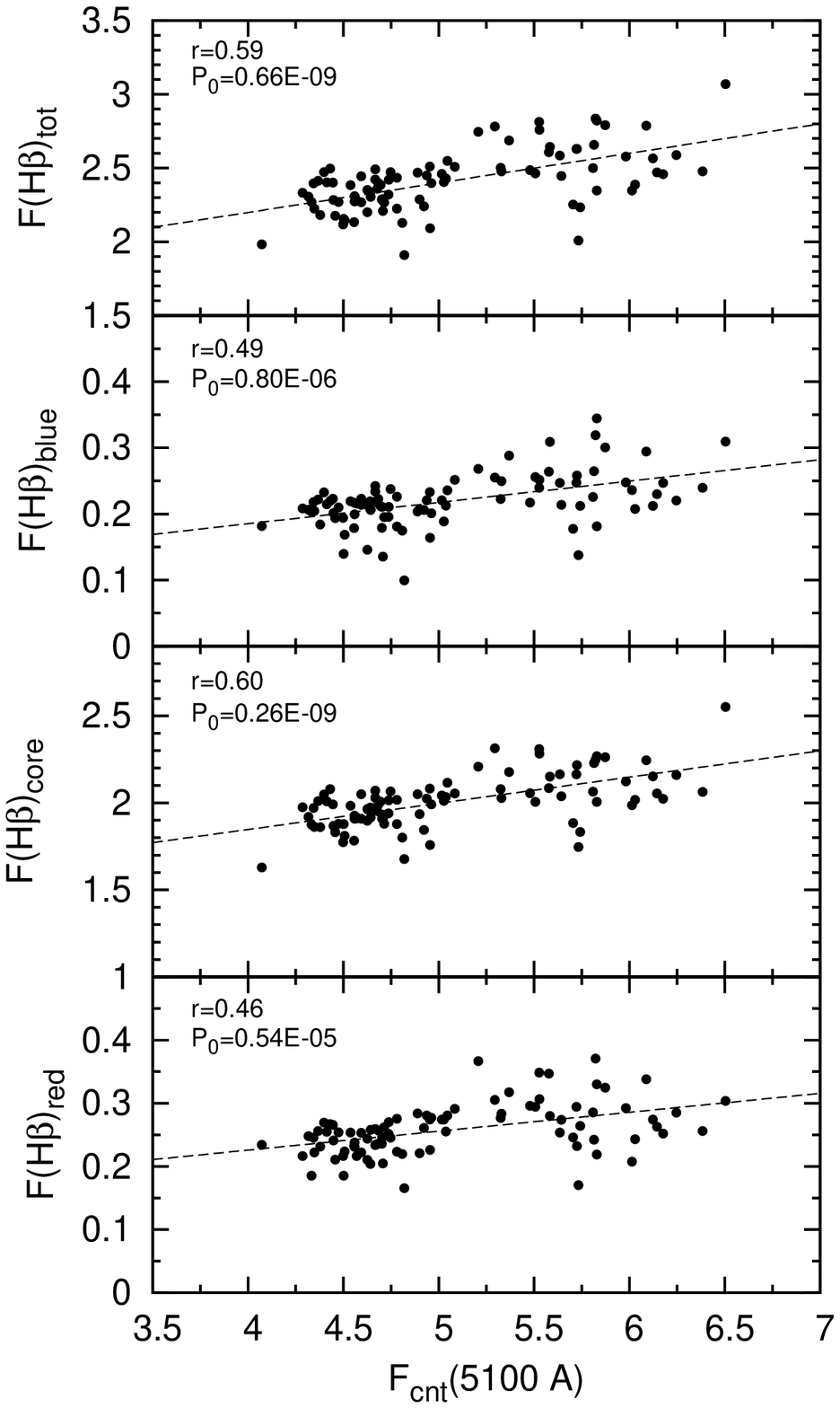}
\caption{H$\alpha$ and H$\beta$ line and line-segment fluxes (blue, core and red)
vs. continuum flux at 6300 and 5100, respectively. The correlation coefficient and the 
corresponding P-value are given in the upper left corner.\label{fig10}} 
\end{figure*}

\begin{figure}
\includegraphics[width=12cm]{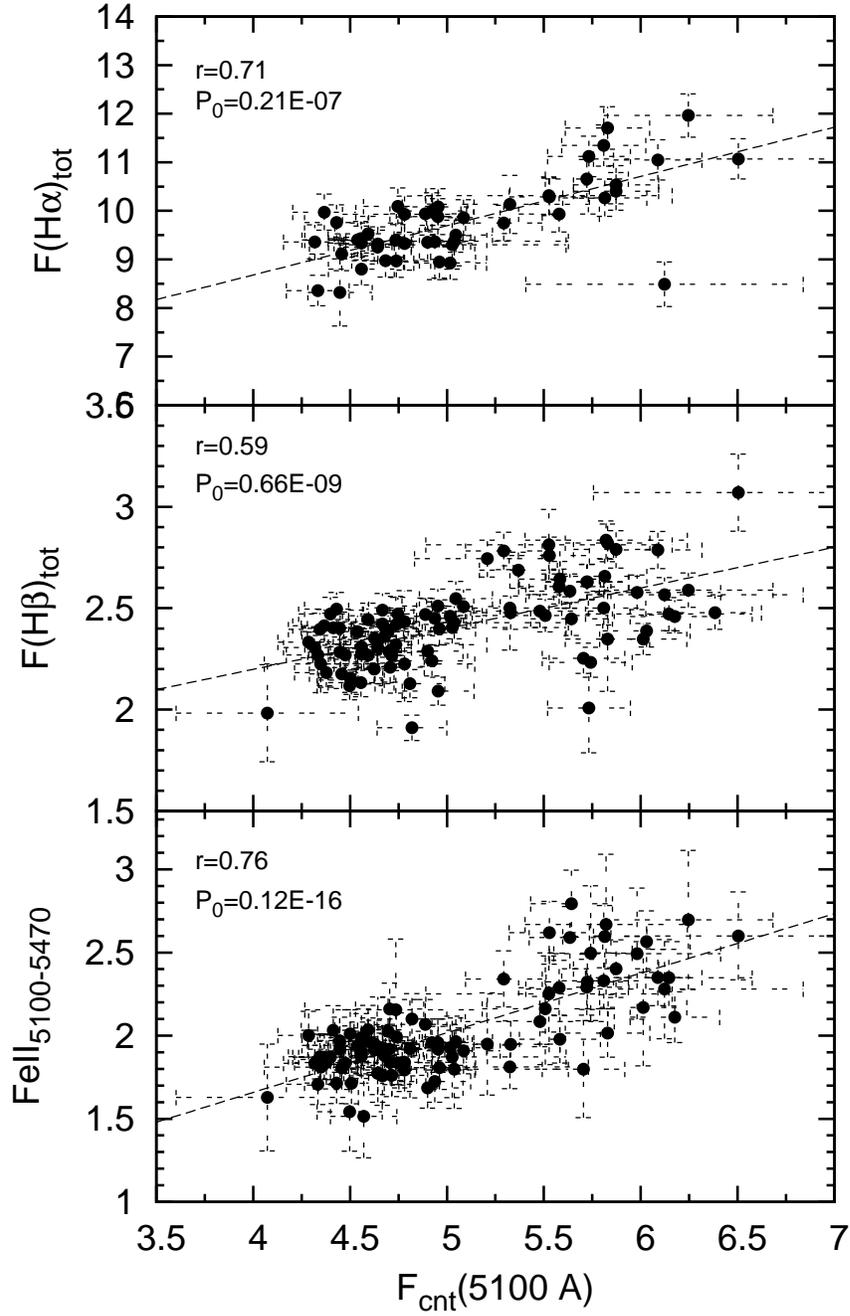}
\caption{H$\alpha$ (upper), H$\beta$ (middle), \ion{Fe}{2} (bottom) emission 
against the continuum flux at 5100. The correlation coefficient and the 
corresponding P-value are given in the upper left corner.\label{fig11}} 
\end{figure}

\begin{figure}
\includegraphics[width=13cm]{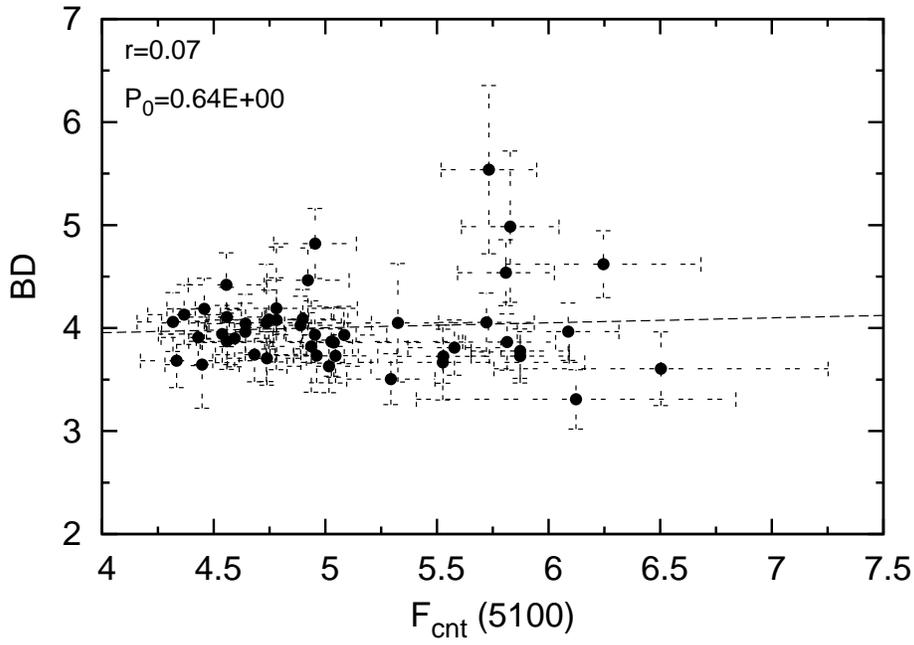}
\caption{Balmer decrement vs. continuum flux at 5100 \AA. The correlation coefficient and the 
corresponding P-value are given in the upper left corner.\label{fig12}} 
\end{figure}

\begin{figure}
\includegraphics[width=13.cm]{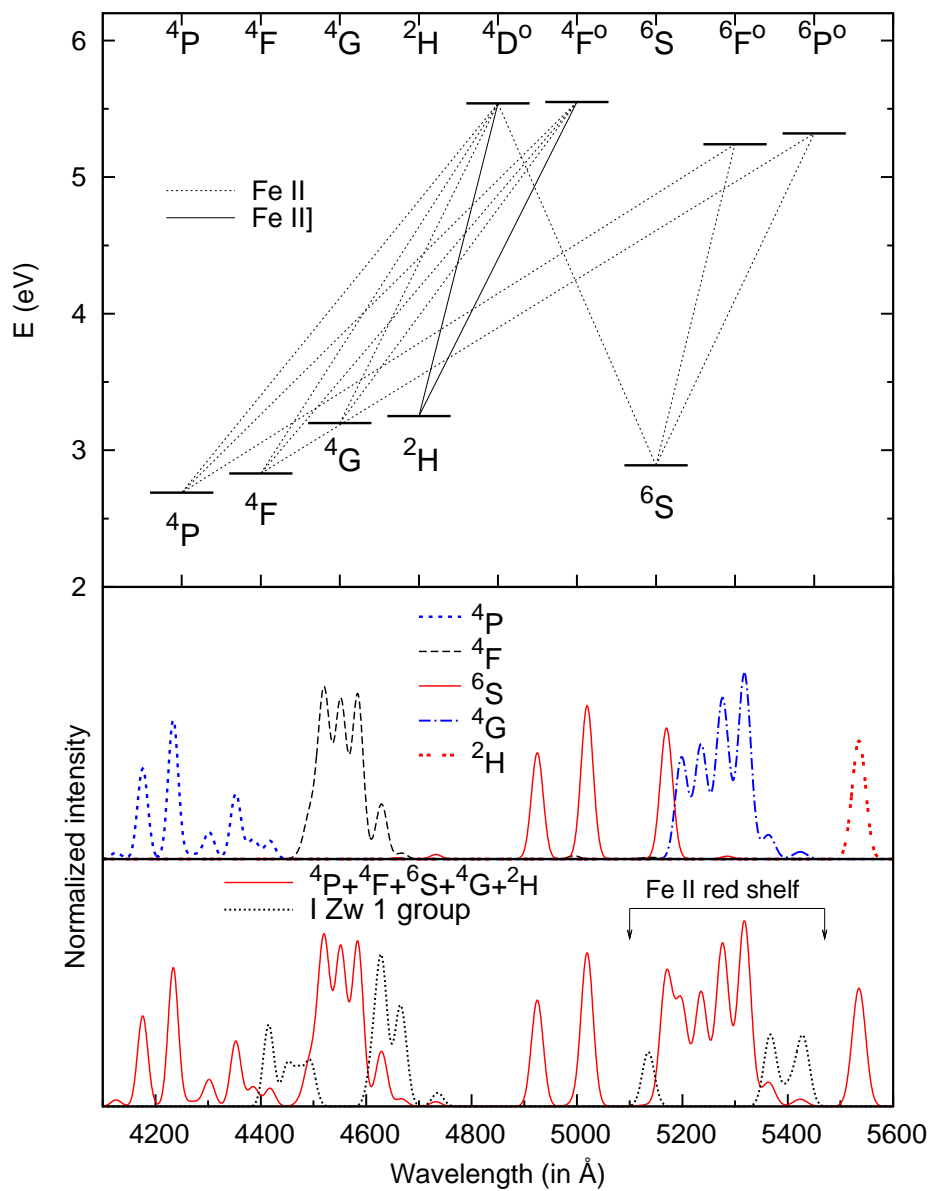}
\caption{The simplified Grotrian diagram showing the strongest \ion{Fe}{2} transitions in the 
$\lambda\lambda$ 4100-5600 \AA\ region (top). Lines are separated into five groups according to 
the lower level of transition (middle): P (dotted line), F (dashed line), 
S (solid line), G (dash - dotted line), and H (two-dashed line). Bottom: the lines from the three 
line groups (solid line) and additional lines taken from I Zw 1 \citep[see][]{ko10}, represented 
with dots. The measured red shelf region is also noted (Table \ref{tab4}). \label{fig13}}
\end{figure}

\begin{figure}
\includegraphics[width=13.cm]{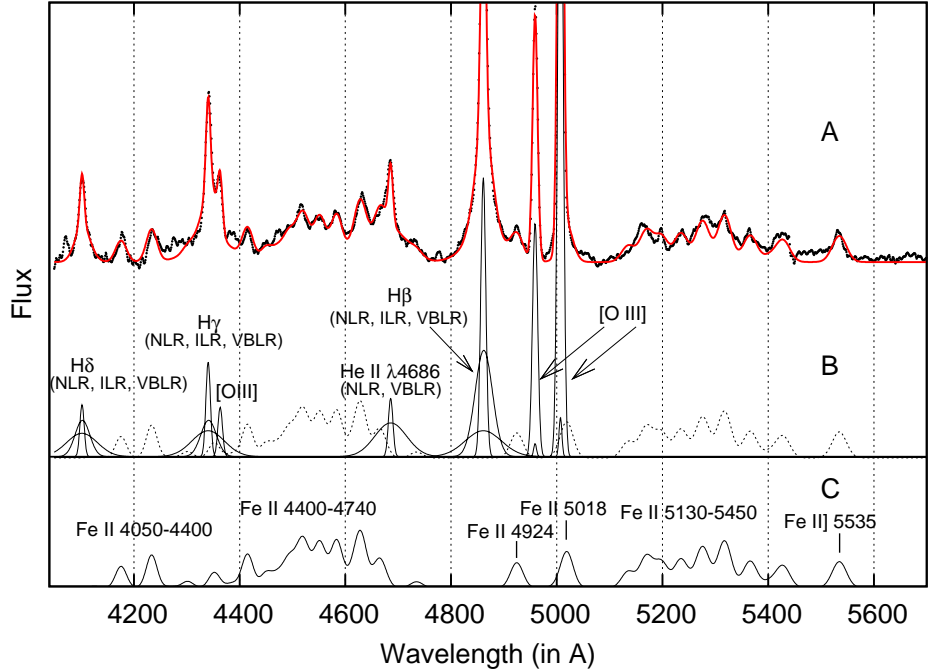}
\caption{An example of the best fit (date Nov 23, 2001) of the $\lambda\lambda$4000-5600 \AA\, 
region: (A) the observed spectra (dots) and the best fit (solid line). (B) H$\beta$, H$\gamma$, and H$\delta$ fit 
with the sum of three Gaussians representing emission from the NLR, ILR and BLR. 
The [O III] $\lambda\lambda$4959, 5007 \AA\ lines are fit with two Gaussians for each line of
the doublet and He II $\lambda$4686 \AA\ is fit with one broad and one narrow Gaussian. 
The \ion{Fe}{2} template is denoted with a dotted line, and also represented separately in panel (C).\label{fig14}} 
\end{figure}

\begin{figure}
\includegraphics[width=13.cm]{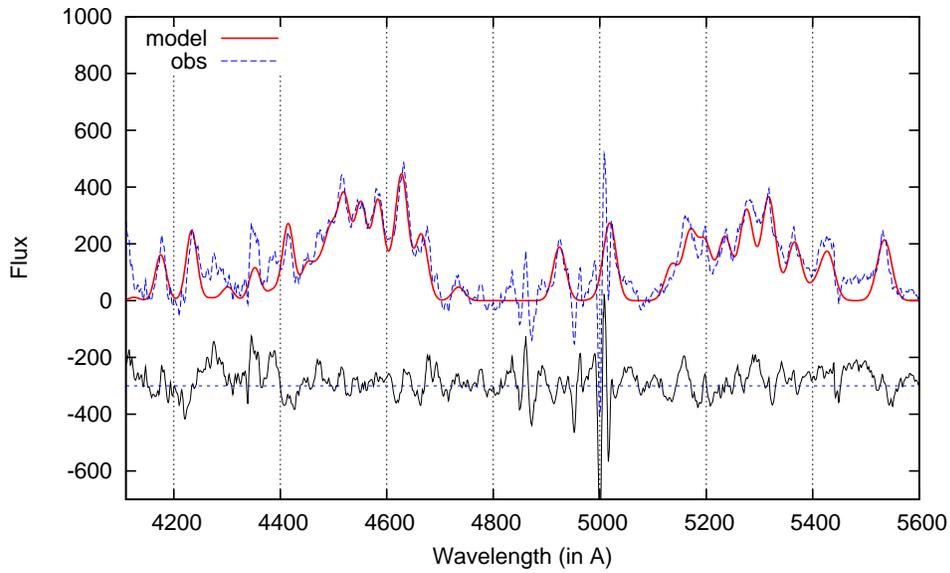}
\caption{An example of the \ion{Fe}{2} line emission (date Nov 23, 2001) in $\lambda\lambda$ 4100-5600 \AA\, 
cleared from the contamination of other strong lines in the field. The observed (dashed line) and fitted spectra
(solid line) are shown. The bottom line represents the residuals.\label{fig15}} 
\end{figure}

\begin{figure}
\includegraphics[width=8.cm]{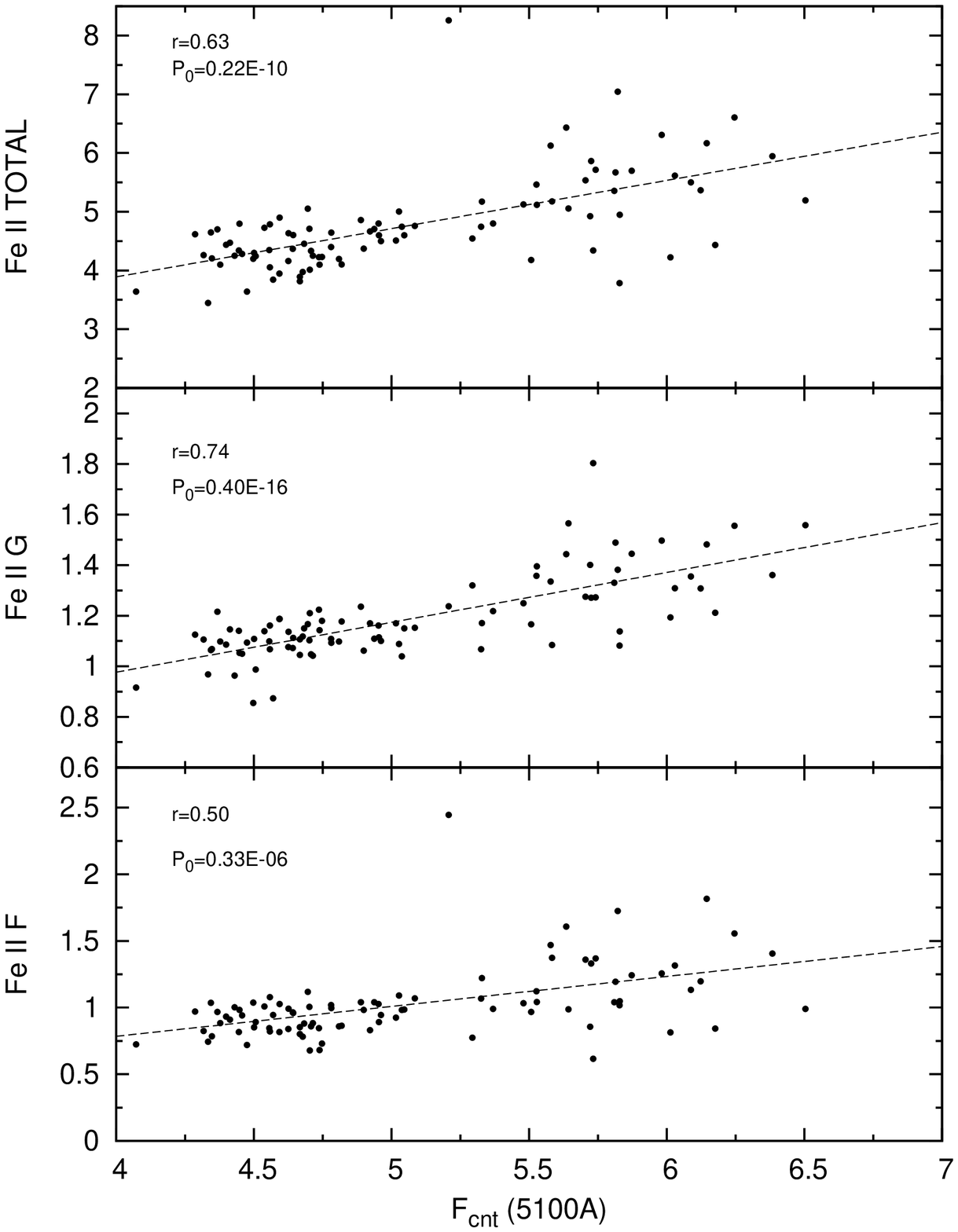}
\includegraphics[width=8.cm]{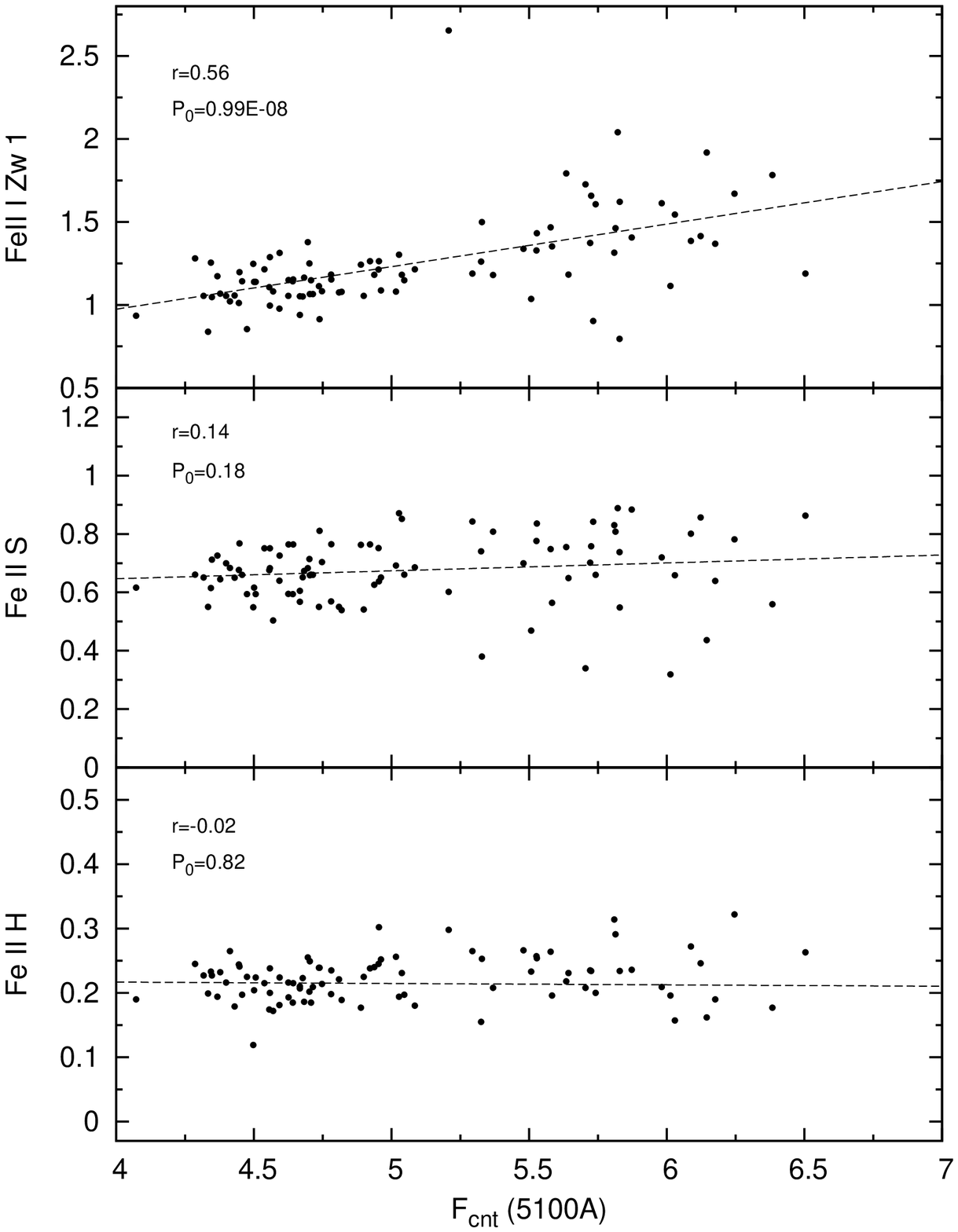}
\caption{\ion{Fe}{2} line fluxes (different groups and total fluxes) vs. continuum flux at 5100 \AA. 
The correlation coefficient and the corresponding p-value are given in the upper left corner.\label{fig16}} 
\end{figure}

\begin{figure}
\includegraphics[width=13.cm]{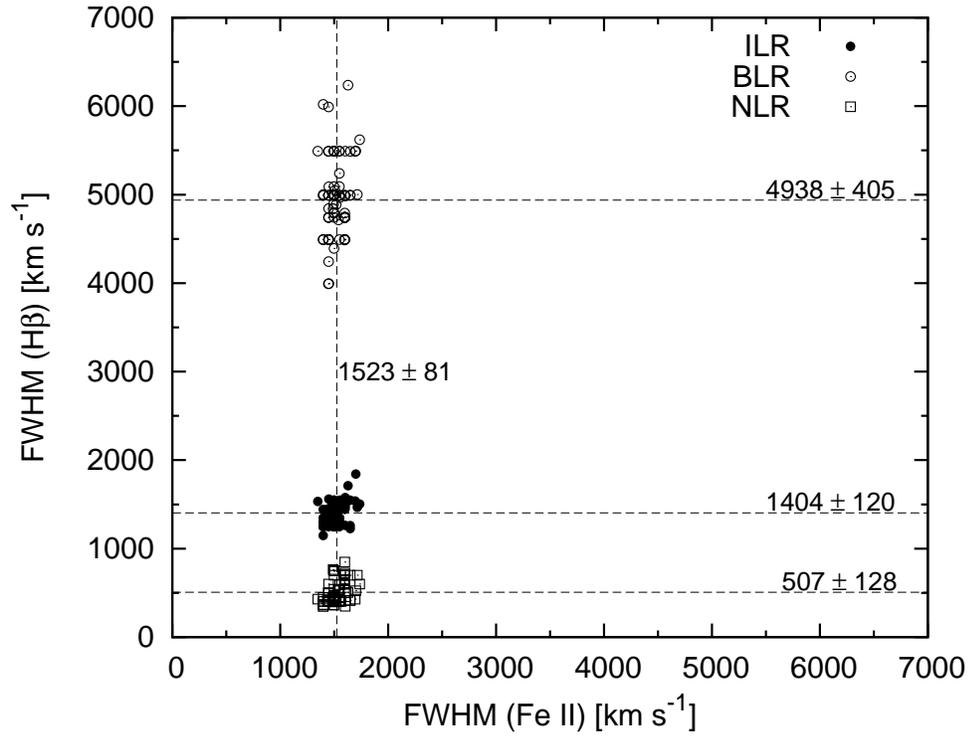}
\caption{Gaussian widths of the \ion{Fe}{2} lines compared with the widths of the H$\beta$ ILR component.
The vertical line shows the average value of \ion{Fe}{2} widths, while the
horizontal lines show the average values of the  H$\beta$ BLR, ILR and NLR  components.\label{fig17}} 
\end{figure}

\begin{figure}
\includegraphics[width=13.cm]{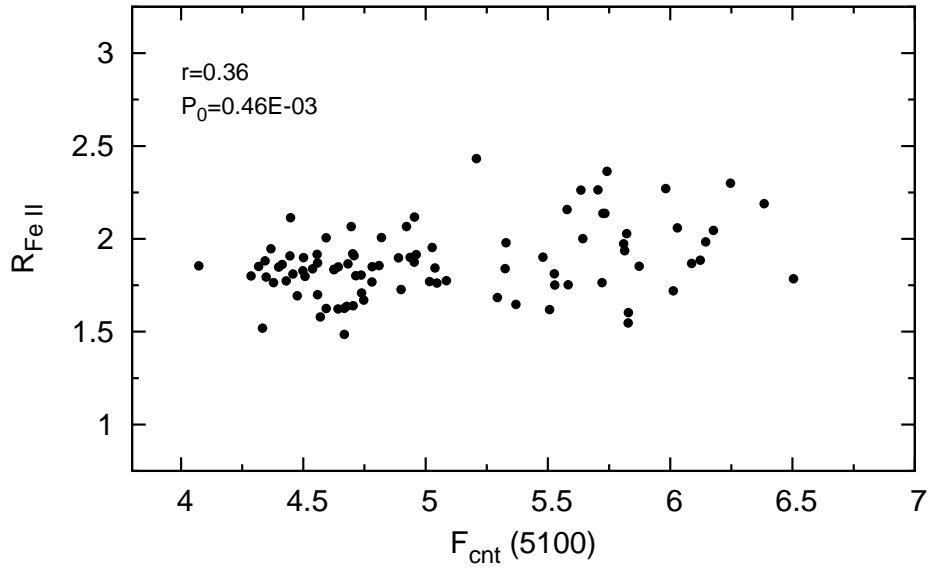}
\caption{R$_{\rm Fe II}$ plotted against the blue continuum flux (EV1 parameter plane).\label{fig18}} 
\end{figure}

In this paper, we analyzed the long spectral variability of NLS1 galaxy Ark 564, observed in the 11-year period from 1999 to 2010. 
We performed a detail analysis of optical spectra covering the continuum flux at 5100 \AA\ and 6200 \AA\ and H$\beta$, H$\alpha$ 
and \ion{Fe}{2} lines. Here we briefly outline our conclusion:

1) In Ark 564 during the monitoring period (1999--2010) the mean continuum
and lines fluxes decreased for $\sim$20\%-30\% (see Fig.~\ref{fig05}) from the beginning 
(1999) to the end of the monitoring (2010). The total flux of \ion{Fe}{2}  evidently
 increases with the continuum flux.

2) We registered five flare-like events (two prominent and three possible) lasting $\sim$1-3 days, when
fluxes in continuum and lines changed for $\sim$20\% (continuum and \ion{Fe}{2} emission)
and $\sim$10\% for Balmer lines.

3) The flux-flux correlations between the continuum and lines are weak,
where the correlation between the \ion{Fe}{2} lines (in the red shelf of the \ion{Fe}{2}) and continuum is
slightly higher (and more significant) than between the Balmer lines and
continuum. There is almost lack of correlation between the H$\alpha$ and H$\beta$
line fluxes. Such behavior indicate very complex physical processes in the
line forming region, i.e. beside the photoionization some additional physical
processes may be present.

4) We roughly estimated a lag of 2--6 days, but with large errorbars. Taking
that the photoionization is probably not the only source of line excitation,
the obtained results should be taken with caution.

5) We investigated in detail, the \ion{Fe}{2} emission variability. We divided 
the \ion{Fe}{2} emission in six groups according to the atomic transitions.
We found that correlation between the continuum flux and emission
of groups depends on the type of transition, i.e. in some case
there is relatively good correlation level between the \ion{Fe}{2} group emission ($^4$G, $^4$F group),
but for $^2$H and $^6$S there is no correlation at all. 

6) The Gaussian multicomponent analysis indicates that the emission
of the \ion{Fe}{2} lines is probably coming from the intermidiate line region,
having velocities around 1500 km/s.

The spectral variability of Ark 564 seems to be complex and different from the one 
observed in BLS1 \citep[see e.g.][]{sh09}. The observed flare-like events, the small (or even lack of) correlation 
between H$\alpha$ and H$\beta$ fluxes, the different correlation degree for \ion{Fe}{2} group emission
and continuum light level, may indicate complex physics in the emitting regions, as e.g. there may
be, beside the AGN, contribution of star explosions and internal shock waves.

\acknowledgments

We thank the anonymous referee for very useful comments and suggestions.
This work was supported by RFBR research grants N00-02-16272, N03-02-17123, 
06-02-16843, N09-02-01136, N12-02-01237a (Russia), CONACYT research grants 
39560-F, 54480, 151494 and PAPIIT-UNAM research grant
IN111610  (M\'exico), and the projects ``Astrophysical Spectroscopy
of Extragalactic Objects'' (176001)  supported by the Ministry of
Education and Science of the Republic of Serbia. L. \v C. P, W. K., D. I. and
J. K. are grateful to the Alexander von Humboldt foundation
for support in the frame of program ``Research Group Linkage''.

\clearpage

\clearpage

\clearpage

\clearpage
 
\clearpage

\clearpage

\clearpage

\clearpage

\clearpage

\clearpage

\clearpage

\clearpage

\clearpage

\clearpage

\clearpage

\clearpage

\clearpage

\clearpage

\begin{deluxetable}{lcccccc}
\tabletypesize{\scriptsize}
\tablecaption{Sources of spectroscopic observations.\label{tab1}}
\tablewidth{0pt}
\tablehead{
\colhead{Observatory} & \colhead{Code} & \colhead{Tel.aperture + equipment} & \colhead{Aperture} & \colhead{Focus}
& \colhead{No.} & \colhead{Period}\\
\colhead{1} & \colhead{2} & \colhead{3} & \colhead{4} & \colhead{5} & \colhead{6} & \colhead{7}
}
\startdata
SAO(Russia)             & L(U) & 6 m + UAGS          &  2.0$\times$6.0   & Prime       &  9 & 1999--2001\\
SAO(Russia)             & L(N) & 6 m + UAGS          &  2.0$\times$6.0   & Nasmith     &  1 & 1999 Oct 09\\
SAO(Russia)             & Z1K  & 1 m + UAGS+CCD1K    &  4.0$\times$19.8  & Cassegrain  & 19 & 1999--2001\\
SAO(Russia)             & Z2K  & 1 m + UAGS+CCD2K    &  4.0$\times$4.0   & Cassegrain  &  5 & 2006--2009\\
Gullermo Haro (M\'exico)& GHO  & 2.1 m + B\&C        &  2.5$\times$6.0   & Cassegrain  & 74 & 2000--2007\\
San–Pedro Martir (M\'exico)    & SPM  & 2.1 m + B\&C &  2.5$\times$6.0   & Cassegrain  & 12 & 2005--2007\\
\enddata
\tablecomments{Col.(1): Observatory. Col.(2): Code assigned to each combination of 
telescope + equipment used throughout this paper. Col.(3): Telescope aperture and spectrograph. 
Col.(4): Projected spectrograph entrance apertures (slit width$\times$slit length in
arcsec). Col.(5): Focus of the telescope. Col.(6): Number or spectra obtained.
Col.(7): Observation period.\\}
\end{deluxetable}

\begin{deluxetable}{clcccccc}
\tablecaption{The log of spectroscopic observations.\label{tab2}}
\tablewidth{0pt}
\tablehead{
\colhead{N} & \colhead{UT-date} & \colhead{JD+} & \colhead{CODE\tablenotemark{a}} & \colhead{Aperture} &
\colhead{Sp.range} & \colhead{Res\tablenotemark{b}} & \colhead{Seeing} \\
  &    & \colhead{2400000+} &  & \colhead{arcsec} & \colhead{\AA} & \colhead{\AA}   & \colhead{arcsec}   \\
\colhead{1} & \colhead{2} & \colhead{3} & \colhead{4} & \colhead{5} & \colhead{6} & \colhead{7} & \colhead{8}
}
\startdata
  1 &  1999Sep02 &  51424.4 &  Z1K  &   4.0$\times$19.8 &  4025-5825  &    7  &   2.0    \\
  2 &  1999Sep03 &  51425.4 &  L(U) &   2.0$\times$6.0  &  3620-6044  &    9  &   1.5    \\
  3 &  1999Sep04 &  51426.4 &  Z1K  &   4.0$\times$19.8 &  4025-5825  &    7  &   2.0    \\
  4 &  1999Sep05 &  51427.3 &  L(U) &   2.0$\times$6.0  &  3650-6074  &    8  &   1.6    \\
  5 &  1999Sep05 &  51427.4 &  L(U) &   2.0$\times$6.0  &  4900-7324  &    9  &   1.6    \\
  6 &  1999Oct03 &  51455.2 &  L(U) &   2.0$\times$6.0  &  4320-5568  &    5  &   1.3    \\
  7 &  1999Oct03 &  51455.3 &  L(U) &   2.0$\times$6.0  &  6030-7278  &    5  &   1.3    \\
  8 &  1999Oct04 &  51456.2 &  L(U) &   2.0$\times$6.0  &  4320-5556  &    6  &   1.3    \\
  9 &  1999Oct04 &  51456.3 &  L(U) &   2.0$\times$6.0  &  6040-7276  &    5  &   1.3    \\
 10 &  1999Oct09 &  51461.3 &  L(N) &   2.0$\times$6.0  &  4240-6590  &    8  &   2.5    \\
 11 &  1999Oct13 &  51465.2 &  Z1K  &   4.0$\times$19.8 &  4050-5850  &    6  &   2.0    \\
 12 &  1999Nov02 &  51485.3 &  Z1K  &   4.0$\times$19.8 &  4025-5825  &    7  &   2.0    \\
 13 &  1999Nov03 &  51486.2 &  Z1K  &   4.0$\times$19.8 &  4025-5825  &    7  &   2.0    \\
 14 &  1999Nov04 &  51487.2 &  Z1K  &   4.0$\times$19.8 &  4025-5825  &    6  &   2.0    \\
 15 &  1999Nov05 &  51488.2 &  Z1K  &   4.0$\times$19.8 &  4025-5825  &    8  &   2.0    \\
 16 &  1999Nov06 &  51489.2 &  Z1K  &   4.0$\times$19.8 &  4025-5825  &    7  &   2.0    \\
 17 &  1999Nov30 &  51513.2 &  Z1K  &   4.0$\times$19.8 &  4025-5825  &    7  &   2.0    \\
 18 &  1999Dec02 &  51515.2 &  Z1K  &   4.0$\times$19.8 &  4050-5850  &    8  &   2.0    \\
 19 &  2000May28 &  51693.5 & L(U)  &   2.0$\times$6.0  &  3550-5974  &    8  &   1.6    \\
 20 &  2000Jun06 &  51702.4 &  Z1K  &   4.0$\times$19.8 &  4020-5820  &    8  &   3.0    \\
 21 &  2000Jul08 &  51734.4 &  Z1K  &   4.0$\times$19.8 &  4050-5850  &    8  &   3.0    \\
 22 &  2000Jul09 &  51735.4 &  Z1K  &   4.0$\times$19.8 &  4050-5850  &    7  &   3.0    \\
 23 &  2000Jul10 &  51736.4 &  Z1K  &   4.0$\times$19.8 &  4030-5830  &    7  &   3.0    \\
 24 &  2000Oct16 &  51833.7 &  GHO  &   2.5$\times$6.0  &  4000-7300  &   12  &   2.3    \\
 25 &  2001Aug29 &  52151.5 &  Z1K  &   4.0$\times$19.8 &  4040-5840  &    7  &   2.0    \\
 26 &  2001Aug29 &  52151.5 &  Z1K  &   4.0$\times$19.8 &  5600-7290  &    8  &   2.0    \\
 27 &  2001Oct08 &  52191.2 &  Z1K  &   4.0$\times$19.8 &  4050-5850  &    8  &   2.0    \\
 28 &  2001Oct09 &  52192.3 &  Z1K  &   4.0$\times$19.8 &  4040-5840  &    7  &   2.0    \\
 29 &  2001Oct09 &  52192.4 &  Z1K  &   4.0$\times$19.8 &  5640-7290  &   11  &   2.0    \\
 30 &  2001Nov23 &  52237.1 & L(U)  &   2.0$\times$6.0  &  3600-6024  &   10  &   3.5    \\
 31 &  2001Nov23 &  52236.6 &  GHO  &   2.5$\times$6.0  &  4200-5960  &    8  &   2.5    \\
 32 &  2001Nov24 &  52237.6 &  GHO  &   2.5$\times$6.0  &  6000-7360  &    9  &   1.5    \\
 33 &  2002Aug15 &  52501.8 &  GHO  &   2.5$\times$6.0  &  4270-5840  &   10  &   2.5    \\
 34 &  2002Aug17 &  52503.9 &  GHO  &   2.5$\times$6.0  &  5700-7460  &   10  &   2.0    \\
 35 &  2002Nov11 &  52589.7 &  GHO  &   2.5$\times$6.0  &  4300-6060  &    8  &   4.5    \\
 36 &  2002Nov12 &  52590.7 &  GHO  &   2.5$\times$6.0  &  5700-7460  &   10  &   2.7    \\
 37 &  2002Nov13 &  52591.7 &  GHO  &   2.5$\times$6.0  &  5700-7460  &    9  &   2.7    \\
 38 &  2002Nov14 &  52592.7 &  GHO  &   2.5$\times$6.0  &  3800-7100  &   10  &   2.7    \\
 39 &  2002Dec10 &  52618.6 &  GHO  &   2.5$\times$6.0  &  4300-6060  &    8  &   1.5    \\
 40 &  2002Dec11 &  52619.6 &  GHO  &   2.5$\times$6.0  &  5700-7460  &    9  &   1.8    \\
 41 &  2002Dec12 &  52620.6 &  GHO  &   2.5$\times$6.0  &  3800-7100  &   13  &   1.8    \\
 42 &  2003Sep04 &  52886.9 &  GHO  &   2.5$\times$6.0  &  5700-7460  &   11  &   2.3    \\
 43 &  2003Oct17 &  52929.7 &  GHO  &   2.5$\times$6.0  &  4300-6060  &   10  &   2.3    \\
 44 &  2003Oct18 &  52930.7 &  GHO  &   2.5$\times$6.0  &  5700-7460  &   11  &   1.8    \\
 45 &  2003Oct20 &  52932.7 &  GHO  &   2.5$\times$6.0  &  3800-7100  &   12  &   1.8    \\
 46 &  2003Nov19 &  52962.6 &  GHO  &   2.5$\times$6.0  &  4300-6060  &   10  &   2.3    \\
 47 &  2003Nov20 &  52963.7 &  GHO  &   2.5$\times$6.0  &  5700-7460  &   12  &   2.6    \\
 48 &  2003Dec17 &  52990.6 &  GHO  &   2.5$\times$6.0  &  4300-6060  &    9  &   3.1    \\
 49 &  2003Dec18 &  52991.6 &  GHO  &   2.5$\times$6.0  &  5300-7460  &   12  &   2.7    \\
 50 &  2003Dec20 &  52993.6 &  GHO  &   2.5$\times$6.0  &  3800-7100  &   15  &   2.3    \\
 51 &  2004Aug17 &  53234.9 &  GHO  &   2.5$\times$6.0  &  3800-7100  &   15  &   2.5    \\
 52 &  2004Aug18 &  53235.8 &  GHO  &   2.5$\times$6.0  &  4300-6060  &    9  &   3.1    \\
 53 &  2004Aug19 &  53236.8 &  GHO  &   2.5$\times$6.0  &  5700-7460  &   12  &   3.1    \\
 54 &  2004Aug20 &  53237.9 &  GHO  &   2.5$\times$6.0  &  3800-7100  &   15  &   2.7    \\
 55 &  2004Sep05 &  53253.9 &  GHO  &   2.5$\times$6.0  &  3800-7100  &   15  &   2.7    \\
 56 &  2004Sep06 &  53254.8 &  GHO  &   2.5$\times$6.0  &  4300-6060  &    8  &   2.7    \\
 57 &  2004Sep08 &  53256.8 &  GHO  &   2.5$\times$6.0  &  5700-7460  &   10  &   3.6    \\
 58 &  2004Nov12 &  53321.6 &  GHO  &   2.5$\times$6.0  &  3800-7100  &   14  &   2.3    \\
 59 &  2004Nov17 &  53326.6 &  GHO  &   2.5$\times$6.0  &  4300-6060  &   11  &   2.7    \\
 60 &  2004Nov18 &  53327.6 &  GHO  &   2.5$\times$6.0  &  3800-7100  &   14  &   2.3    \\
 61 &  2004Dec13 &  53352.6 &  GHO  &   2.5$\times$6.0  &  3800-7100  &   12  &   3.6    \\
 62 &  2004Dec14 &  53353.6 &  GHO  &   2.5$\times$6.0  &  4300-6060  &    7  &   3.6    \\
 63 &  2004Dec15 &  53354.6 &  GHO  &   2.5$\times$6.0  &  5700-7460  &    8  &   2.3    \\
 64 &  2005May14 &  53505.0 &  SPM  &   2.5$\times$6.0  &  3880-5960  &    7  &   4.9    \\
 65 &  2005May15 &  53506.0 &  SPM  &   2.5$\times$6.0  &  5720-7580  &    7  &   3.3    \\
 66 &  2005Aug26 &  53608.9 &  GHO  &   2.5$\times$6.0  &  3800-7100  &   13  &   2.9    \\
 67 &  2005Aug27 &  53609.9 &  GHO  &   2.5$\times$6.0  &  4150-7460  &   12  &   3.4    \\
 68 &  2005Aug28 &  53610.8 &  GHO  &   2.5$\times$6.0  &  4330-6000  &    7  &   2.8    \\
 69 &  2005Aug29 &  53611.8 &  GHO  &   2.5$\times$6.0  &  4330-6000  &    8  &   3.9    \\
 70 &  2005Aug30 &  53612.8 &  GHO  &   2.5$\times$6.0  &  4330-6000  &    8  &   3.3    \\
 71 &  2005Aug31 &  53613.8 &  GHO  &   2.5$\times$6.0  &  4330-6000  &    7  &   2.7    \\
 72 &  2005Sep08 &  53621.9 &  SPM  &   2.5$\times$6.0  &  3700-5790  &   10  &   3.0    \\
 73 &  2005Sep09 &  53622.9 &  SPM  &   2.5$\times$6.0  &  3700-5780  &    8  &   -      \\
 74 &  2005Sep28 &  53641.8 &  GHO  &   2.5$\times$6.0  &  4320-5980  &    7  &   2.7    \\
 75 &  2005Sep29 &  53642.6 &  GHO  &   2.5$\times$6.0  &  5740-7400  &    8  &   3.1    \\
 76 &  2005Sep30 &  53643.8 &  GHO  &   2.5$\times$6.0  &  4290-5960  &    7  &   2.8    \\
 77 &  2005Oct24 &  53667.6 &  GHO  &   2.5$\times$6.0  &  3750-7050  &   12  &   2.3    \\
 78 &  2005Oct26 &  53669.7 &  GHO  &   2.5$\times$6.0  &  4260-5920  &    8  &   3.0    \\
 79 &  2005Oct28 &  53671.7 &  GHO  &   2.5$\times$6.0  &  5740-7400  &    8  &   3.1     \\
 80 &  2005Nov28 &  53702.6 &  GHO  &   2.5$\times$6.0  &  3800-6908  &   14  &   3.0    \\
 81 &  2005Nov29 &  53703.6 &  GHO  &   2.5$\times$6.0  &  4300-5917  &    8  &   5.0    \\
 82 &  2005Nov30 &  53704.6 &  GHO  &   2.5$\times$6.0  &  5740-7400  &    8  &   2.0     \\
 83 &  2005Dec05 &  53710.6 &  SPM  &   2.5$\times$6.0  &  3700-5770  &    7  &   -      \\
 84 &  2005Dec07 &  53711.6 &  SPM  &   2.5$\times$6.0  &  3700-5770  &    7  &   2.8    \\
 85 &  2005Dec29 &  53733.6 &  GHO  &   2.5$\times$6.0  &  4300-6010  &   10  &   2.3    \\
 86 &  2006Jun28 &  53915.5 &  Z2K  &   4.0$\times$4.0 &  3740-7400  &    9  &   2.5    \\
 87 &  2006Aug27 &  53974.9 &  GHO  &   2.5$\times$6.0  &  3600-7050  &   12  &   2.2    \\
 88 &  2006Aug29 &  53977.5 &  Z2K  &   4.0$\times$4.0 &  3750-7400  &    9  &   2.0    \\
 89 &  2006Aug30 &  53978.5 &  Z2K  &   4.0$\times$4.0 &  3750-7400  &    8  &   2.0    \\
 90 &  2006Aug30 &  53977.8 &  GHO  &   2.5$\times$6.0  &  4120-5920  &    8  &   2.5    \\
 91 &  2006Aug31 &  53978.8 &  GHO  &   2.5$\times$6.0  &  3600-7050  &   12  &   2.2    \\
 92 &  2006Sep15 &  53993.8 &  GHO  &   2.5$\times$6.0  &  3600-7050  &   13  &   3.4    \\
 93 &  2006Sep17 &  53995.7 &  GHO  &   2.5$\times$6.0  &  3600-7050  &   13  &   2.4    \\
 94 &  2006Sep18 &  53996.8 &  GHO  &   2.5$\times$6.0  &  4130-5030  &    7  &   2.5    \\
 95 &  2006Sep19 &  53997.8 &  GHO  &   2.5$\times$6.0  &  3600-7000  &   12  &   2.8    \\
 96 &  2006Sep28 &  54006.7 &  SPM  &   2.5$\times$6.0  &  3740-5810  &    7  &   3.3    \\
 97 &  2006Sep29 &  54007.7 &  SPM  &   2.5$\times$6.0  &  3740-5810  &    7  &   3.3    \\
 98 &  2006Oct23 &  54031.7 &  SPM  &   2.5$\times$6.0  &  3700-5900  &    8  &   2.6    \\
 99 &  2006Oct27 &  54035.7 &  GHO  &   2.5$\times$6.0  &  3700-7280  &   12  &   2.8    \\
100 &  2006Oct28 &  54036.7 &  GHO  &   2.5$\times$6.0  &  4230-6040  &    8  &   2.4    \\
101 &  2006Oct30 &  54038.7 &  GHO  &   2.5$\times$6.0  &  3700-7270  &   14  &   2.3    \\
102 &  2006Oct31 &  54039.7 &  GHO  &   2.5$\times$6.0  &  4160-5960  &    8  &   2.3    \\
103 &  2006Nov30 &  54069.6 &  SPM  &   2.5$\times$6.0  &  3680-7560  &   12  &   4.6    \\
104 &  2007May22 &  54242.9 &  SPM  &   2.5$\times$6.0  &  3730-5810  &    8  &   3.0    \\
105 &  2007May23 &  54244.0 &  SPM  &   2.5$\times$6.0  &  3730-5810  &    8  &   3.2    \\
106 &  2007Aug10 &  54322.9 &  GHO  &   2.5$\times$6.0  &  3870-7430  &   11  &   3.0    \\
107 &  2007Aug11 &  54323.8 &  GHO  &   2.5$\times$6.0  &  4340-6140  &    7  &   3.2    \\
108 &  2007Sep03 &  54346.8 &  GHO  &   2.5$\times$6.0  &  4330-6130  &    7  &   3.6    \\
109 &  2007Sep04 &  54347.8 &  GHO  &   2.5$\times$6.0  &  4150-5950  &    7  &   2.6    \\
110 &  2007Sep07 &  54350.9 &  GHO  &   2.5$\times$6.0  &  3860-7420  &   12  &   3.0    \\
111 &  2007Oct15 &  54388.7 &  GHO  &   2.5$\times$6.0  &  3870-7440  &   12  &   1.8    \\
112 &  2007Oct17 &  54390.6 &  GHO  &   2.5$\times$6.0  &  4190-6000  &    8  &   2.5    \\
113 &  2007Oct18 &  54391.7 &  GHO  &   2.5$\times$6.0  &  4190-6000  &    8  &   2.2    \\
114 &  2007Nov01 &  54405.7 &  GHO  &   2.5$\times$6.0  &  4190-6000  &    8  &   3.0    \\
115 &  2007Nov02 &  54406.6 &  GHO  &   2.5$\times$6.0  &  4190-6000  &    8  &   3.3    \\
116 &  2007Nov03 &  54407.6 &  GHO  &   2.5$\times$6.0  &  3820-7390  &   12  &   2.9    \\
117 &  2007Nov06 &  54410.7 &  GHO  &   2.5$\times$6.0  &  3830-7400  &   12  &   2.4    \\
118 &  2007Nov08 &  54412.6 &  GHO  &   2.5$\times$6.0  &  4290-6100  &    8  &   2.4    \\
119 &  2009Aug14 &  55058.5 &  Z2K  &   4.0$\times$4.0 &  3750-7390  &    8  &   2.0    \\
120 &  2009Oct11 &  55116.4 &  Z2K  &   4.0$\times$4.0 &  3750-7390  &    8  &   1.5    \\
\enddata
\tablecomments{Col.(1): Number. Col.(2): UT date. Col.(3): Julian
date (JD). Col.(4): CODE\tablenotemark{a}. Col.(5): Projected
spectrograph entrance apertures. Col.(6): Wavelength range covered.
Col.(7): Spectral resolution\tablenotemark{b}. Col.(8): Mean seeing in arcsec.}
\tablenotetext{a}{Code given according to Table~\ref{tab1}.}
\tablenotetext{b}{Resolution determined from [OIII]5007 line, and from
[OI]6300 when only red part of the spectrum present.}
\end{deluxetable}

\clearpage

\begin{deluxetable}{lcccc}
\tablecaption{Flux scale factors for optical spectra.\label{tab3}}
\tablewidth{0pt}
\tablehead{
\colhead{Sample} & \colhead{Years} & \colhead{Aperture} & \colhead{Scale factor} & \colhead{Extended source  correction} \\
  &  & \colhead{(arcsec)} & \colhead{($\varphi$)} & \colhead{G(g)\tablenotemark{a}}
}
\startdata
L(U,N)    & 1999--2010 &  2.0$\times$6.0   & 1.089  & -0.130 \\
GHO       & 1999--2007 &  2.5$\times$6.0   & 1.000  & 0.000 \\
SPM       & 2005--2007 &  2.5$\times$6.0   & 1.000  & 0.000 \\
Z1K       & 1999--2001 &  4.0$\times$19.8  & 1.152$\pm$0.013  & 0.998$\pm$0.368\\
Z2K       & 2005--2007 &  4.0$\times$4.0  & 0.893$\pm$0.052  & 1.005$\pm$0.548\\
GHO\tablenotemark{*}& 1999--2007 &  2.5$\times$6.0   & 1.067$\pm$0.048  & 0.000   \\
\enddata
\tablenotetext{*}{Resolution 15\AA}
\tablenotetext{a}{In units 10$^{-15} \rm erg s^{−1} cm^{−2}$\AA$^{−1}$}
\end{deluxetable}

\begin{deluxetable}{clccccc}
\tabletypesize{\scriptsize}
\tablecaption{The measured line and continuum fluxes.\label{tab4}}
\tablewidth{0pt}
\tablehead{
\colhead{N} & \colhead{UT-date} & \colhead{JD+} & \colhead{F$_{\rm cnt}\pm \sigma$} & \colhead{F(H$\alpha$)$\pm \sigma$} &
\colhead{F(H$\beta$)$\pm \sigma$} & \colhead{\ion{Fe}{2}$_{5100-5470}\pm \sigma$}  \\
  &    & \colhead{2400000+} &  \colhead{$10^{-15} \rm erg \, cm^{-2} s^{-1}$\AA$^{-1}$} & 
\colhead{$ 10^{-13} \rm erg \, cm^{-2} s^{-1}$} & \colhead{$ 10^{-13} \rm erg \, cm^{-2} s^{-1}$}   
& \colhead{$ 10^{-13} \rm erg \, cm^{-2} s^{-1}$}   \\
\colhead{1} & \colhead{2} & \colhead{3} & \colhead{4} & \colhead{5} & \colhead{6} & \colhead{7} 
}
\startdata
1 &  1999Sep02 & 51424.4  & 6.145$\pm$ 0.430 &         -        & 2.471 $\pm$ 0.082 &  2.349$\pm$  0.170  \\
2 &  1999Sep03 & 51425.4  & 5.479$\pm$ 0.384 &         -        & 2.486 $\pm$ 0.082 &  2.085$\pm$  0.151  \\
3 &  1999Sep04 & 51426.4  & 5.507$\pm$ 0.325 &         -        & 2.464 $\pm$ 0.081 &  2.162$\pm$  0.336  \\
4 &  1999Sep05 & 51427.3  & 6.246$\pm$ 0.437 &11.963$\pm$ 0.447 & 2.589 $\pm$ 0.085 &  2.696$\pm$  0.419  \\
5 &  1999Oct03 & 51455.2  & 5.733$\pm$ 0.214 &11.119$\pm$ 0.416 & 2.008 $\pm$ 0.221 &  -  \\
6 &  1999Oct04 & 51456.2  & 5.828$\pm$ 0.218 &11.707$\pm$ 0.438 & 2.348 $\pm$ 0.258 &  -  \\
7 &  1999Oct09 & 51461.3  & 6.176$\pm$ 0.231 &        -         & 2.459 $\pm$ 0.081 &  2.111$\pm$  0.152  \\
8 &  1999Oct13 & 51465.2  & 5.329$\pm$ 0.199 &        -         & 2.478 $\pm$ 0.082 &  1.948$\pm$  0.141  \\
9 &  1999Nov02 & 51485.3  & 5.742$\pm$ 0.215 &        -         & 2.233 $\pm$ 0.074 &  2.496$\pm$  0.405  \\
10 &  1999Nov03 & 51486.2  & 5.705$\pm$ 0.213 &        -         & 2.253 $\pm$ 0.074 &  1.798$\pm$  0.291  \\
11 &  1999Nov04 & 51487.2  & 6.013$\pm$ 0.225 &        -         & 2.348 $\pm$ 0.077 &  2.170$\pm$  0.352  \\
12 &  1999Nov05 & 51488.2  & 6.384$\pm$ 0.239 &        -         & 2.478 $\pm$ 0.082 &  -  \\
13 &  1999Nov06 & 51489.2  & 6.030$\pm$ 0.226 &        -         & 2.388 $\pm$ 0.079 &  2.566$\pm$  0.185  \\
14 &  1999Nov30 & 51513.2  & 5.635$\pm$ 0.211 &        -         & 2.584 $\pm$ 0.085 &  2.590$\pm$  0.187  \\
15 &  1999Dec02 & 51515.2  & 5.583$\pm$ 0.209 &        -         & 2.643 $\pm$ 0.087 &  1.978$\pm$  0.143  \\
16 &  2000May28 & 51693.5  & 5.643$\pm$ 0.211 &        -         & 2.447 $\pm$ 0.081 &  2.794$\pm$  0.202  \\
17 &  2000Jun06 & 51702.4  & 5.829$\pm$ 0.218 &        -         & 2.822 $\pm$ 0.093 &  2.016$\pm$  0.146  \\
18 &  2000Jul08 & 51734.4  & 5.982$\pm$ 0.431 &        -         & 2.578 $\pm$ 0.085 &  2.493$\pm$  0.394  \\
19 &  2000Jul09 & 51735.4  & 5.208$\pm$ 0.375 &        -         & 2.745 $\pm$ 0.091 &  1.950$\pm$  0.308  \\
20 &  2000Jul10 & 51736.4  & 5.822$\pm$ 0.419 &        -         & 2.836 $\pm$ 0.094 &  2.669$\pm$  0.422  \\
21 &  2000Oct16 & 51833.7  & 5.809$\pm$ 0.217 &11.346$\pm$ 0.424 & 2.500 $\pm$ 0.083 &  2.330$\pm$  0.168  \\
22 &  2001Nov23 & 52236.6  & 5.722$\pm$ 0.214 &        -         & 2.630 $\pm$ 0.087 &  2.292$\pm$  0.168  \\
23 &  2001Nov23 & 52237.1  & 5.725$\pm$ 0.214 &        -         & 2.629 $\pm$ 0.087 &  2.324$\pm$  0.166  \\
24 &  2001Nov24 & 52237.6  &     -            &10.658$\pm$ 0.399 & - & -\\
25 &  2002Aug15 & 52501.8  & 4.747$\pm$ 0.178 &        -         & 2.472 $\pm$ 0.082 &  1.836$\pm$  0.133  \\
26 &  2002Aug17 & 52503.9  &     -            &10.094$\pm$ 0.378 & - & -\\
27 &  2002Nov11 & 52589.7  & 5.528$\pm$ 0.207 &        -         & 2.760 $\pm$ 0.091 &  2.619$\pm$  0.189  \\
28 &  2002Nov12 & 52590.7  &     -            &10.285$\pm$ 0.385 & - & -\\
29 &  2002Nov13 & 52591.7  &     -            &10.402$\pm$ 0.389 & - & -\\
30 &  2002Nov14 & 52592.7  & 5.873$\pm$ 0.220 &10.537$\pm$ 0.394 & 2.790 $\pm$ 0.092 &  2.403$\pm$  0.173  \\
31 &  2002Dec10 & 52618.6  & 5.294$\pm$ 0.198 &        -         & 2.782 $\pm$ 0.092 &  2.341$\pm$  0.169  \\
32 &  2002Dec11 & 52619.6  &     -            & 9.750$\pm$ 0.365 & - & -\\
33 &  2002Dec12 & 52620.6  & 5.578$\pm$ 0.209 & 9.936$\pm$ 0.372 & 2.607 $\pm$ 0.086 &  2.288$\pm$  0.165  \\
34 &  2003Sep04 & 52886.9  &     -            & 9.700$\pm$ 0.363 & - & -\\
35 &  2003Oct17 & 52929.7  & 5.016$\pm$ 0.188 &         -        & 2.460 $\pm$ 0.081 &  1.928$\pm$  0.139  \\
36 &  2003Oct18 & 52930.7  &     -            & 8.925$\pm$ 0.334 & - & -\\
37 &  2003Oct20 & 52932.7  & 4.961$\pm$ 0.186 & 8.950$\pm$ 0.335 & 2.398 $\pm$ 0.079 &  1.807$\pm$  0.130  \\
38 &  2003Nov19 & 52962.6  & 4.738$\pm$ 0.177 &         -        & 2.420 $\pm$ 0.080 &  1.993$\pm$  0.144  \\
39 &  2003Nov20 & 52963.7  &     -            & 8.968$\pm$ 0.335 & - & -\\
40 &  2003Dec17 & 52990.6  & 6.504$\pm$ 0.748 &         -        & 3.070 $\pm$ 0.190 &  2.600$\pm$  0.264  \\
41 &  2003Dec18 & 52991.6  &     -            &11.069$\pm$ 0.414 & - & -\\
42 &  2003Dec20 & 52993.6  & 5.526$\pm$ 0.636 &10.313$\pm$ 0.386 & 2.813 $\pm$ 0.174 &  2.251$\pm$  0.229  \\
43 &  2004Aug17 & 53234.9  & 5.038$\pm$ 0.589 & 9.375$\pm$ 0.506 & 2.429 $\pm$ 0.080 &  1.798$\pm$  0.235  \\
44 &  2004Aug18 & 53235.8  & 6.123$\pm$ 0.716 &       -          & 2.566 $\pm$ 0.085 &  2.280$\pm$  0.298  \\
45 &  2004Aug19 & 53236.8  &     -            & 8.489$\pm$ 0.458 & - & -\\
46 &  2004Aug20 & 53237.9  & 5.027$\pm$ 0.588 & 9.303$\pm$ 0.502 & 2.405 $\pm$ 0.079 &  1.872$\pm$  0.245  \\
47 &  2004Sep05 & 53253.9  & 5.325$\pm$ 0.405 &10.132$\pm$ 0.598 & 2.501 $\pm$ 0.208 &  1.812$\pm$  0.131  \\
48 &  2004Sep06 & 53254.8  & 4.781$\pm$ 0.363 &       -          & 2.225 $\pm$ 0.185 &  1.793$\pm$  0.129  \\
49 &  2004Sep08 & 53256.8  &     -            & 9.327$\pm$ 0.550 & - & -\\
50 &  2004Nov12 & 53321.6  & 5.046$\pm$ 0.189 & 9.504$\pm$ 0.355 & 2.548 $\pm$ 0.084 &  1.964$\pm$  0.142  \\
51 &  2004Nov17 & 53326.6  & 5.369$\pm$ 0.201 &       -          & 2.687 $\pm$ 0.089 &  -  \\
52 &  2004Nov18 & 53327.6  & 5.085$\pm$ 0.190 & 9.862$\pm$ 0.369 & 2.508 $\pm$ 0.083 &  1.909$\pm$  0.138  \\
53 &  2004Dec13 & 53352.6  & 6.088$\pm$ 0.228 &11.049$\pm$ 0.413 & 2.787 $\pm$ 0.092 &  2.350$\pm$  0.170  \\
54 &  2004Dec14 & 53353.6  & 5.814$\pm$ 0.217 &       -          & 2.657 $\pm$ 0.088 &  2.596$\pm$  0.187  \\
55 &  2004Dec15 & 53354.6  &     -            &10.268$\pm$ 0.384 & - & -\\
56 &  2005May14 & 53505.0  & 4.954$\pm$ 0.185 &       -          & 2.091 $\pm$ 0.069 &  1.918$\pm$  0.138  \\
57 &  2005May14 & 53506.0  &     -            &10.081$\pm$ 0.377 & - & -\\
58 &  2005Aug26 & 53608.9  & 4.448$\pm$ 0.166 & 8.321$\pm$ 0.691 & 2.283 $\pm$ 0.075 &  1.919$\pm$  0.139  \\
59 &  2005Aug27 & 53609.9  & 4.937$\pm$ 0.185 & 9.362$\pm$ 0.777 & 2.450 $\pm$ 0.081 &  1.724$\pm$  0.125  \\
60 &  2005Aug28 & 53610.8  & 4.559$\pm$ 0.170 &      -           & 2.310 $\pm$ 0.076 &  1.988$\pm$  0.144  \\
61 &  2005Aug29 & 53611.8  & 4.678$\pm$ 0.175 &      -           & 2.360 $\pm$ 0.078 &  1.880$\pm$  0.136  \\
62 &  2005Aug30 & 53612.8  & 4.593$\pm$ 0.172 &      -           & 2.268 $\pm$ 0.075 &  2.035$\pm$  0.147  \\
63 &  2005Aug31 & 53613.8  & 4.704$\pm$ 0.176 &      -           & 2.289 $\pm$ 0.076 &  2.160$\pm$  0.156  \\
64 &  2005Sep08 & 53621.9  & 4.498$\pm$ 0.168 &      -           & 2.118 $\pm$ 0.070 &  1.543$\pm$  0.238  \\
65 &  2005Sep09 & 53622.9  & 4.809$\pm$ 0.180 &      -           & 2.128 $\pm$ 0.070 &  1.921$\pm$  0.296  \\
66 &  2005Sep28 & 53641.8  & 4.333$\pm$ 0.162 &      -           & 2.270 $\pm$ 0.075 &  1.708$\pm$  0.123  \\
67 &  2005Sep29 & 53642.6  &     -            & 8.359$\pm$ 0.313 & - & -\\
68 &  2005Sep30 & 53643.8  & 4.378$\pm$ 0.164 &      -           & 2.182 $\pm$ 0.072 &  1.838$\pm$  0.133  \\
69 &  2005Oct24 & 53667.6  & 4.556$\pm$ 0.170 & 9.429$\pm$ 0.353 & 2.133 $\pm$ 0.070 &  1.908$\pm$  0.138  \\
70 &  2005Oct26 & 53669.7  & 4.457$\pm$ 0.167 &      -           & 2.176 $\pm$ 0.072 &  1.806$\pm$  0.130  \\
71 &  2005Oct28 & 53671.7  &     -            & 9.112$\pm$ 0.341 & - & -\\
72 &  2005Nov28 & 53702.6  & 4.072$\pm$ 0.470 &       -          & 1.982 $\pm$ 0.239 &  1.628$\pm$  0.321  \\
73 &  2005Nov29 & 53703.6  & 4.736$\pm$ 0.470 &       -          & 2.320 $\pm$ 0.239 &  2.156$\pm$  0.425  \\
74 &  2005Nov30 & 53704.6  &     -            & 9.397$\pm$0.351  & - & -\\
75 &  2005Dec05 & 53710.6  & 4.714$\pm$ 0.176 &       -          & 2.267 $\pm$ 0.075 &  1.761$\pm$  0.127  \\
76 &  2005Dec07 & 53711.6  & 4.507$\pm$ 0.169 &       -          & 2.146 $\pm$ 0.071 &  1.714$\pm$  0.124  \\
77 &  2005Dec29 & 53733.6  & 4.475$\pm$ 0.167 &       -          & 2.270 $\pm$ 0.075 &  1.834$\pm$  0.132  \\
78 &  2006Aug27 & 53974.9  & 4.538$\pm$ 0.286 & 9.399$\pm$ 0.352 & 2.384 $\pm$ 0.079 &  1.942$\pm$  0.140  \\
79 &  2006Aug29 & 53977.5  &     -            & 9.254$\pm$ 0.346 & - & -\\
80 &  2006Aug30 & 53977.8  & 4.642$\pm$ 0.174 &       -          & 2.335 $\pm$ 0.077 &  1.774$\pm$  0.128  \\
81 &  2006Aug30 & 53978.5  &     -            & 8.801$\pm$ 0.329 & - & -\\
82 &  2006Aug31 & 53978.8  & 4.558$\pm$ 0.170 & 9.337$\pm$ 0.349 & 2.273 $\pm$ 0.075 &  1.877$\pm$  0.135  \\
83 &  2006Sep15 & 53993.8  & 4.953$\pm$ 0.185 & 9.883$\pm$ 0.370 & 2.511 $\pm$ 0.083 &  1.959$\pm$  0.141  \\
84 &  2006Sep17 & 53995.7  & 4.889$\pm$ 0.183 & 9.941$\pm$ 0.372 & 2.469 $\pm$ 0.081 &  2.068$\pm$  0.149  \\
85 &  2006Sep18 & 53996.8  & 4.899$\pm$ 0.183 &      -           & 2.288 $\pm$ 0.076 &  1.686$\pm$  0.122  \\
86 &  2006Sep19 & 53997.8  & 4.626$\pm$ 0.173 & 9.352$\pm$ 0.350 & 2.352 $\pm$ 0.078 &  1.916$\pm$  0.138  \\
87 &  2006Sep28 & 54006.7  & 4.501$\pm$ 0.168 &       -          & 2.156 $\pm$ 0.071 &  2.011$\pm$  0.145  \\
88 &  2006Sep29 & 54007.7  & 4.625$\pm$ 0.173 &       -          & 2.200 $\pm$ 0.073 &  1.951$\pm$  0.141  \\
89 &  2006Oct23 & 54031.7  & 4.819$\pm$ 0.180 &       -          & 1.910 $\pm$ 0.063 &  2.101$\pm$  0.152  \\
90 &  2006Oct27 & 54035.7  & 4.643$\pm$ 0.174 & 9.319$\pm$ 0.349 & 2.305 $\pm$ 0.076 &  1.916$\pm$  0.317  \\
91 &  2006Oct28 & 54036.7  & 4.570$\pm$ 0.171 &       -          & 2.280 $\pm$ 0.075 &  1.515$\pm$  0.250  \\
92 &  2006Oct30 & 54038.7  & 4.781$\pm$ 0.320 & 9.930$\pm$ 0.371 & 2.434 $\pm$ 0.153 &  1.840$\pm$  0.133  \\
93 &  2006Oct31 & 54039.7  & 4.348$\pm$ 0.291 &       -          & 2.225 $\pm$ 0.141 &  1.810$\pm$  0.131  \\
94 &  2006Nov30 & 54069.6  & 4.922$\pm$ 0.184 & 9.999$\pm$ 0.374 & 2.240 $\pm$ 0.074 &  1.959$\pm$  0.141  \\
95 &  2007May22 & 54242.9  & 4.702$\pm$ 0.176 &       -          & 2.285 $\pm$ 0.075 &  1.919$\pm$  0.139  \\
96 &  2007May23 & 54244.0  & 4.707$\pm$ 0.176 &       -          & 2.209 $\pm$ 0.073 &  1.839$\pm$  0.133  \\
97 &  2007Aug10 & 54322.9  & 4.683$\pm$ 0.175 & 8.973$\pm$ 0.336 & 2.398 $\pm$ 0.079 &  1.907$\pm$  0.138  \\
98 &  2007Aug11 & 54323.8  & 4.667$\pm$ 0.175 &       -          & 2.422 $\pm$ 0.080 &  1.927$\pm$  0.139  \\
99 &  2007Sep03 & 54346.8  & 4.668$\pm$ 0.175 &       -          & 2.491 $\pm$ 0.082 &  1.758$\pm$  0.177  \\
100&  2007Sep04 & 54347.8  & 4.696$\pm$ 0.176 &       -          & 2.385 $\pm$ 0.079 &  2.027$\pm$  0.204  \\
101&  2007Sep07 & 54350.9  & 4.430$\pm$ 0.166 & 9.760$\pm$ 0.365 & 2.496 $\pm$ 0.082 &  1.715$\pm$  0.124  \\
102&  2007Oct15 & 54388.7  & 4.318$\pm$ 0.161 & 9.360$\pm$ 0.350 & 2.306 $\pm$ 0.076 &  1.834$\pm$  0.132  \\
103&  2007Oct17 & 54390.6  & 4.446$\pm$ 0.166 &       -          & 2.401 $\pm$ 0.079 &  1.969$\pm$  0.142  \\
104&  2007Oct18 & 54391.7  & 4.344$\pm$ 0.162 &       -          & 2.396 $\pm$ 0.079 &  1.874$\pm$  0.135  \\
105&  2007Nov01 & 54405.7  & 4.413$\pm$ 0.165 &       -          & 2.403 $\pm$ 0.079 &  2.033$\pm$  0.147  \\
106&  2007Nov02 & 54406.6  & 4.399$\pm$ 0.165 &       -          & 2.472 $\pm$ 0.082 &  1.871$\pm$  0.135  \\
107&  2007Nov03 & 54407.6  & 4.594$\pm$ 0.172 & 9.530$\pm$ 0.356 & 2.444 $\pm$ 0.081 &  1.964$\pm$  0.142  \\
108&  2007Oct06 & 54410.7  & 4.367$\pm$ 0.163 & 9.973$\pm$ 0.373 & 2.414 $\pm$ 0.080 &  1.870$\pm$  0.135  \\
109&  2007Nov08 & 54412.6  & 4.287$\pm$ 0.160 &       -          & 2.332 $\pm$ 0.077 &  2.002$\pm$  0.145  \\
110&  2009Aug14 & 55058.5  &     -            &10.712$\pm$ 0.401 & - & -\\
111&  2009Oct11 & 55116.4  &     -            &12.204$\pm$ 0.456 & - & -\\
\enddata
\end{deluxetable}

\begin{deluxetable}{lcccc}
\tablecaption{Estimates of the errors for line and line-segment fluxes.\label{tab5}}
\tablewidth{0pt}
\tablehead{
 \colhead{Line}    &  \colhead{Spectral Region}   &        & \colhead{$\sigma \pm$e} & \colhead{$V_r$ region} \\
                   &  \colhead{[\AA] (obs)} & \colhead{[\AA] (rest)} &  \colhead{[\%]}     & \colhead{km s$^{-1}$}
}
\startdata
 cont 5100    &   5220--5250  & 5094--5123  &  3.7 $\pm$3.3     &  -   \\
 cont 6200    &   6320--6370  & 6168--6216  &  4.5 $\pm$2.2     &  -   \\
 H$\alpha$ - total &   6640--6810  & 6480--6646  &  3.7 $\pm$2.2     & (-3792;+3792)  \\
 H$\beta$ - total  &   4936--5030  & 4817--4909  &  3.3 $\pm$2.5     & (-2710;+2950)  \\
 \ion{Fe}{2}             &   5226--5605  & 5100--5470  &  7.2 $\pm$5.6     &  - \\
 H$\alpha$ - blue  &   6635--6698  & 6475--6537  &  6.8 $\pm$6.6     & (-4015;-1204)  \\
 H$\alpha$ - core  &   6698--6752  & 6537--6589  &  3.5 $\pm$2.1     & (-1204;+1204)  \\
 H$\alpha$ - red   &   6752--6816  & 6589--6652  &  5.6 $\pm$4.0     & (+1204;+4015)  \\
 H$\beta$ - blue   &   4928--4960  & 4809--4840  &  8.0 $\pm$4.9     & (-3200;-1200) \\
 H$\beta$ - core   &   4960--5001  & 4840--4880  &  3.0 $\pm$2.9     & (-1200;+1200)  \\
 H$\beta$ - red    &   5001--5034  & 4880--4913  &  8.7 $\pm$5.3     & (+1200;+3200)  \\
\enddata
\end{deluxetable}

\begin{deluxetable}{lcccccccccc}
\tabletypesize{\scriptsize}
\tablecaption{Flares in light curves of the blue continuum, H$\beta$, H$\alpha$, and \ion{Fe}{2} emission.\label{tab6}}
\tablewidth{0pt}
\tablehead{
 \colhead{N} & \colhead{DATE} &  \colhead{JD+} & \colhead{F(cnt)\tablenotemark{a}} & \colhead{dF(cnt)} 
&\colhead{F(H$\beta$)\tablenotemark{b}}  & \colhead{dF(H$\beta$)} &  \colhead{F(H$\alpha$)\tablenotemark{b}} & 
 \colhead{dF(H$\alpha$)} &  \colhead{F(\ion{Fe}{2})\tablenotemark{b}}& \colhead{dF(\ion{Fe}{2})} \\
  &      &  \colhead{2400000}  & \colhead{(5235)\AA} &  \colhead{\%}  &    &  \colhead{\%} &     &    \colhead{\%}  & & \colhead{\%}
}
\startdata
1 &2003Dec17 & 52990.6 &  6.5038  &   18\%  &   3.070 &  9\%   &  11.069   & 7\%  &  2.600 &  10\%   \\
  &2003Dec20 & 52993.6 &  5.5264  &         &   2.813 &        &  10.313   &      &  2.251 &         \\
\hline
2 &2004Aug17 & 53234.9 &  5.0378  &   22\%   &  2.429  & 6\%   &  9.375  &  10\%  &  1.798 &  13\%   \\
  &2004Aug18 & 53235.8 &  6.1232  &          &  2.566  &       &  8.489  &        &  2.280 &          \\
  &2004Aug20 & 53237.9 &  5.0271  &          &  2.405  &       &  9.303  &        &  1.872  &       \\
\hline
3 &2005Nov28 & 53702.6 &  4.0724  &   16\%   &  1.982  & 17\%  &   -    &   -   &  1.628 &  20\%   \\
  &2005Nov29 & 53703.6 &  4.7362  &          &  2.320  &       &   -    &       &  2.156 &          \\
\hline
4 &2006Oct27 & 54035.7  &          &         &          &       &         &        & 1.916 &  17\%     \\
  &2006Oct28 & 54036.7  &          &         &          &       &         &        & 1.515 &        \\
\hline
5 &2006Oct30 & 54038.7 &  4.7813   &  10\%   &  2.434  &  9\%  &  9.319   &  7\%  &  1.840 &  7\%     \\
  &2006Oct31 & 54039.7 &  4.3478   &         &  2.225  &       &  9.930   &       &  1.810 &        \\
\enddata
\tablenotetext{a}{Continuum flux is in units $10^{-15} \rm erg \ cm^{-2} s^{-1} A^{-1}$.} 
\tablenotetext{b}{Line fluxes are in units $10^{-13} \rm erg \ cm^{-2}s^{-1}$.} 
\end{deluxetable}

\begin{deluxetable}{lcccccc}
\tablecaption{Parameters of the continuum and line variabilities.\label{tab7}}
\tablewidth{0pt}
\tablehead{
 \colhead{Feature}  &  \colhead{N}   &  \colhead{Region [\AA]}  &  \colhead{$F$(mean)\tablenotemark{a}}  &  
\colhead{$\sigma$($F$)\tablenotemark{a}} & \colhead{$R$(max/min)} & \colhead{$F$(var)}\\
\colhead{1} & \colhead{2} & \colhead{3} & \colhead{4} & \colhead{5} & \colhead{6} & \colhead{7} 
}
\startdata
 continuum 5100    & 91 &   5094--5123   &   5.068    & 0.608   &   1.597   &  0.107 \\
 continuum 6300    & 50 &   6168--6216   &   2.021    & 0.218   &   1.471   &  0.098 \\
 H$\alpha$ - total & 50 &   6480--6646   &   9.856    & 0.878   &   1.467   &  0.079 \\
 H$\beta$ - total  & 91 &   4817--4909   &   2.413    & 0.206   &   1.607   &  0.075 \\
 \ion{Fe}{2}             & 87 &   5100--5470   &   2.029    & 0.280   &   1.844   &  0.096 \\
 H$\alpha$ - blue  & 50 &   6475--6537   &   0.652    & 0.079   &   1.852   &  0.081 \\
 H$\alpha$ - core  & 50 &   6537--6589   &   8.581    & 0.757   &   1.445   &  0.080 \\
 H$\alpha$ - red   & 50 &   6589--6652   &   0.731    & 0.074   &  1.605    &  0.077 \\
 H$\beta$ - blue   & 91 &   4809--4840   &   0.220    & 0.040   &  3.465    &  0.149 \\
 H$\beta$ - core   & 91 &   4840--4880   &   2.008    & 0.152   &  1.566    &  0.064 \\
 H$\beta$ - red    & 91 &   4880--4913   &   0.258    & 0.040   &  2.238    &  0.110 \\
\enddata
\tablecomments{Col.(1): Analyzed feature of the spectrum. Col.(2): Total number of spectra. 
Col.(3): Wavelength region (in the rest frame). Col.(4): Mean flux.\tablenotemark{a} Col.(5): Standard
deviation\tablenotemark{a}. Col.(6): Ratio of the maximal to minimal flux . Col.(7): Variation amplitude (see text).}
\tablenotetext{a}{Continuum flux is in units $10^{-15} \rm erg \ cm^{-2} s^{-1} A^{-1}$.
and line fluxes and line-segment fluxes are in $10^{-13} \rm erg \ cm^{-2}s^{-1}$.} 
\end{deluxetable}

\begin{deluxetable}{lcc}
\tablecaption{Lags and CCF between the continuum and lines.\label{tab8}}
\tablewidth{0pt}
\tablehead{
 \colhead{LC1-LC2}    &  \colhead{lag (days)}   &  \colhead{CCF} 
}
\startdata
cnt-H{$\beta$}$_{\rm tot}$&    3.56$^{+27.44}_{3.56}$ & 0.49$^{+0.08}_{-0.09}$\\
cnt-\ion{Fe}{2} &  0.02$^{+2.02}_{2.08}$ & 0.52$^{+0.08}_{-0.08}$\\
cnt-H{$\alpha$}$_{\rm tot}$ &  4.54$^{+5.54}_{14.46}$ & 0.49$^{+0.01}_{-0.01}$\\
\enddata
\end{deluxetable}

\clearpage

\begin{deluxetable}{lcc}
\tablecaption{Line transitions added to the \ion{Fe}{2} template given in Tables 1 and 2 of \citet{ko10}. 
The atomic data are taken from the NIST atomic database.\label{tab9}}
\tablewidth{0pt}
\tablehead{
 \colhead{\ion{Fe}{2} multiplet}    &  \colhead{Transition}   &    \colhead{Wavelength [\AA]}  \\
}
\startdata
\ion{Fe}{2} 27-28 & b$^4$P$_{5/2}$ -- z$^4$F$_{3/2}^o$  & 4087.284    \\
             & b$^4$P$_{5/2}$ -- z$^4$F$_{5/2}^o$  & 4122.668    \\
             & b$^4$P$_{5/2}$ -- z$^4$D$_{3/2}^o$  & 4128.748    \\
             & b$^4$P$_{5/2}$ -- z$^4$D$_{5/2}^o$  & 4173.461    \\
             & b$^4$P$_{5/2}$ -- z$^4$F$_{7/2}^o$  & 4178.862    \\
             & b$^4$P$_{5/2}$ -- z$^4$D$_{7/2}^o$  & 4233.172    \\
             & b$^4$P$_{3/2}$ -- z$^4$F$_{3/2}^o$  & 4258.154    \\
             & b$^4$P$_{3/2}$ -- z$^4$D$_{1/2}^o$  & 4273.326    \\
             & b$^4$P$_{3/2}$ -- z$^4$F$_{5/2}^o$  & 4296.572    \\
             & b$^4$P$_{3/2}$ -- z$^4$D$_{3/2}^o$  & 4303.176    \\
             & b$^4$P$_{3/2}$ -- z$^4$D$_{5/2}^o$  & 4351.769    \\
             & b$^4$P$_{1/2}$ -- z$^4$F$_{3/2}^o$  & 4369.411    \\
             & b$^4$P$_{1/2}$ -- z$^4$D$_{1/2}^o$  & 4385.387    \\
             & b$^4$P$_{1/2}$ -- z$^4$D$_{3/2}^o$  & 4416.830    \\
             & b$^4$P$_{5/2}$ -- z$^6$F$_{7/2}^o$  & 4670.182    \\
 \ion{Fe}{2}] 55   & b$^2$H$_{9/2}$ -- z$^4$D$_{7/2}^o$  & 5525.125    \\
                   & b$^2$H$_{11/2}$ -- z$^4$F$_{9/2}^o$  & 5534.847    \\
\enddata
\end{deluxetable}

\clearpage

\begin{deluxetable}{ccccccccccccc}
\tabletypesize{\scriptsize}
\rotate
\tablecaption{The line parameters (w-widths, s-shifts, i-intensity\tablenotemark{*}) from the gaussian best-fitting 
of H$\beta$ and \ion{Fe}{2} lines for 91 good-resolution spectra.\label{tab10}}
\tablewidth{0pt}
\tablehead{
\colhead{JD+} & \colhead{w NLR} & \colhead{s NLR} & \colhead{i NLR} & \colhead{w ILR} &  \colhead{s ILR}  &  \colhead{i ILR}  & \colhead{w BLR} &  \colhead{s BLR}  & \colhead{i BLR} & \colhead{w \ion{Fe}{2}} &  \colhead{s \ion{Fe}{2}} & \colhead{i \ion{Fe}{2}}\\
\colhead{2400000+} & \colhead{km/s}& \colhead{km/s} &  \colhead{[\%]} & \colhead{km/s}  & \colhead{km/s}    & \colhead{[\%]} &   \colhead{km/s}  & \colhead{km/s}  &    \colhead{[\%]} &   \colhead{km/s}  & \colhead{km/s} & \colhead{[\%]} \\
\colhead{1} & \colhead{2} & \colhead{3} & \colhead{4} & \colhead{5} & \colhead{6} & \colhead{7} 
& \colhead{8} & \colhead{9} & \colhead{10} & \colhead{11} & \colhead{12} & \colhead{13} 
}
\startdata
51424.4  &    525    &   36   & 0.21 & 1843    &    0   & 0.35 &   5493  &     0   &0.44 & 1699    &   77 & 1.98  \\
51425.4  &    504    &   20   & 0.20 & 1712    &    0   & 0.53 &   6238  &   -90   &0.27 & 1627    &    6 & 1.90  \\
51426.4  &    429    &   21   & 0.18 & 1534    &    0   & 0.48 &   5491  &   390   &0.33 & 1347    &  125 & 1.62  \\
51427.3  &    429    &   21   & 0.15 & 1538    &    0   & 0.48 &   5488  &   150   &0.37 & 1694    &   19 & 2.30  \\
51455.2  &    349    &   24   & 0.19 & 1578    &    0   & 0.60 &   5491  &   150   &0.21 & 1602    &    3 & 2.14  \\
51456.2  &    346    &   24   & 0.17 & 1443    &   30   & 0.54 &   6021  &   148   &0.29 & 1398    &   56 & 1.55  \\
51461.3  &    429    &   21   & 0.21 & 1560    &    0   & 0.57 &   3994  &   240   &0.22 & 1450    &  -24 & 2.04  \\
51465.2  &    409    &   15   & 0.17 & 1446    &    0   & 0.45 &   4995  &  -240   &0.38 & 1497    &  -25 & 1.98  \\
51485.3  &    409    &   15   & 0.16 & 1498    &  150   & 0.43 &   4992  &     0   &0.41 & 1498    &   93 & 2.36  \\
51486.2  &    409    &   15   & 0.15 & 1248    &   30   & 0.42 &   5241  &    90   &0.43 & 1547    &    0 & 2.26  \\
51487.2  &    409    &   15   & 0.16 & 1248    &   30   & 0.43 &   4992  &     0   &0.41 & 1448    &    0 & 1.72  \\
51488.2  &    409    &   15   & 0.16 & 1447    &   30   & 0.43 &   5491  &     0   &0.41 & 1448    &  174 & 2.19  \\
51489.2  &    409    &   15   & 0.15 & 1348    &   30   & 0.45 &   5491  &     0   &0.40 & 1498    &    0 & 2.06  \\
51513.2  &    409    &   15   & 0.15 & 1348    &   30   & 0.43 &   5491  &     0   &0.42 & 1497    &  -22 & 2.26  \\
51515.2  &    409    &   15   & 0.14 & 1348    &   30   & 0.41 &   5990  &  -450   &0.44 & 1448    &    0 & 1.75  \\
51693.5  &    410    &  101   & 0.16 & 1349    &   30   & 0.46 &   5489  &   300   &0.39 & 1446    &   79 & 2.00  \\
51702.4  &    449    &   30   & 0.15 & 1348    &   30   & 0.39 &   4992  &     0   &0.46 & 1398    &  267 & 1.60  \\
51734.4  &    450    &   30   & 0.17 & 1349    &   30   & 0.44 &   4998  &    -1   &0.39 & 1499    &  173 & 2.27  \\
51735.4  &    444    &   30   & 0.14 & 1348    &   30   & 0.36 &   5491  &     0   &0.51 & 1547    &    0 & 2.43  \\
51736.4  &    444    &   30   & 0.13 & 1298    &   30   & 0.33 &   5492  &     0   &0.54 & 1547    &  180 & 2.03  \\
51833.7  &    593    &    5   & 0.16 & 1548    &   -6   & 0.46 &   4968  &     0   &0.37 & 1552    &  -66 & 1.97  \\
52237.1  &    699    &    0   & 0.20 & 1467    &   60   & 0.45 &   5001  &     0   &0.35 & 1714    &   21 & 2.14  \\
52236.6  &    499    &    0   & 0.19 & 1548    &    0   & 0.42 &   4991  &   -30   &0.39 & 1598    &   26 & 1.76  \\
52501.8  &    540    &   15   & 0.19 & 1548    &    0   & 0.48 &   4992  &   -30   &0.33 & 1497    & -133 & 1.67  \\
52589.7  &    474    &    0   & 0.19 & 1548    &    0   & 0.42 &   5492  &   -30   &0.39 & 1497    &   17 & 1.75  \\
52592.7  &    599    &  -42   & 0.14 & 1503    & -101   & 0.45 &   5621  &   -98   &0.40 & 1736    &    0 & 1.85  \\
52618.6  &    444    &   18   & 0.20 & 1529    &  -41   & 0.48 &   4893  &   -33   &0.33 & 1494    &   48 & 1.68  \\
52620.6  &    700    &    1   & 0.17 & 1548    &    0   & 0.38 &   5490  &   -30   &0.45 & 1647    &    1 & 2.16  \\
52929.7  &    534    &    9   & 0.19 & 1454    &   31   & 0.40 &   4714  &    28   &0.40 & 1539    &   12 & 1.77  \\
52932.7  &    703    &    1   & 0.21 & 1549    &    0   & 0.46 &   4492  &     0   &0.33 & 1597    &   41 & 1.92  \\
52962.6  &    524    &    0   & 0.20 & 1448    &   30   & 0.41 &   4493  &    30   &0.39 & 1547    &  -59 & 1.71  \\
52990.6  &    500    &   15   & 0.21 & 1448    &   30   & 0.47 &   3993  &    30   &0.32 & 1446    &   56 & 1.78  \\
52993.6  &    849    &   30   & 0.21 & 1547    &   30   & 0.40 &   4993  &    60   &0.39 & 1597    &   46 & 1.81  \\
53234.9  &    749    &   27   & 0.21 & 1448    &    0   & 0.41 &   4743  &     0   &0.37 & 1498    &  -49 & 1.84  \\
53235.8  &    499    &   15   & 0.18 & 1447    &   30   & 0.43 &   4742  &     0   &0.39 & 1448    &   -2 & 1.88  \\
53237.9  &    749    &   27   & 0.20 & 1447    &    0   & 0.42 &   4795  &    15   &0.38 & 1497    &  -29 & 1.95  \\
53253.9  &    764    &   27   & 0.21 & 1443    &   -6   & 0.41 &   4840  &    20   &0.38 & 1490    &  -34 & 1.84  \\
53254.8  &    414    &   18   & 0.15 & 1260    &   30   & 0.44 &   4993  &     0   &0.41 & 1647    &   14 & 1.77  \\
53321.6  &    749    &   27   & 0.20 & 1448    &    0   & 0.43 &   4791  &    15   &0.37 & 1498    &   -7 & 1.76  \\
53326.6  &    599    &   15   & 0.20 & 1448    &   30   & 0.42 &   4742  &     0   &0.38 & 1448    &  135 & 1.65  \\
53327.6  &    749    &   27   & 0.18 & 1448    &    0   & 0.42 &   4793  &    15   &0.40 & 1498    &   23 & 1.77  \\
53352.6  &    649    &   27   & 0.18 & 1548    &  -60   & 0.41 &   4793  &    15   &0.42 & 1597    &  -57 & 1.87  \\
53353.6  &    499    &   15   & 0.20 & 1448    &    0   & 0.42 &   4741  &     0   &0.38 & 1448    &   62 & 1.94  \\
53495.0  &    397    &   15   & 0.18 & 1398    &   15   & 0.46 &   4986  &    16   &0.36 & 1498    &   52 & 2.12  \\
53608.9  &    699    &    0   & 0.21 & 1497    &    0   & 0.42 &   4742  &    15   &0.36 & 1597    &  -44 & 2.11  \\
53609.9  &    699    &    0   & 0.21 & 1498    &    0   & 0.42 &   4742  &    15   &0.37 & 1597    &    3 & 1.90  \\
53610.8  &    400    &   17   & 0.16 & 1347    &    0   & 0.46 &   4991  &    15   &0.38 & 1548    &  -22 & 1.70  \\
53611.8  &    404    &   24   & 0.16 & 1353    &   -1   & 0.45 &   4886  &    16   &0.38 & 1518    &  -59 & 1.64  \\
53612.8  &    433    &   19   & 0.17 & 1325    &   -6   & 0.43 &   5048  &   -34   &0.40 & 1505    &  -92 & 1.62  \\
53613.8  &    399    &    0   & 0.16 & 1332    &  -27   & 0.44 &   5091  &    15   &0.40 & 1548    &  -90 & 1.64  \\
53620.9  &    465    &  -14   & 0.17 & 1337    &  -60   & 0.43 &   5091  &    15   &0.40 & 1497    & -134 & 1.83  \\
53622.9  &    464    &  -15   & 0.18 & 1278    &    0   & 0.42 &   4992  &    15   &0.40 & 1498    &    2 & 1.86  \\
53641.8  &    419    &   15   & 0.19 & 1278    &    0   & 0.42 &   4992  &    15   &0.39 & 1498    &  -71 & 1.52  \\
53643.8  &    414    &   22   & 0.16 & 1225    &   30   & 0.46 &   4992  &     0   &0.39 & 1648    &  -19 & 1.76  \\
53667.6  &    699    &    0   & 0.21 & 1497    &    0   & 0.42 &   4742  &    15   &0.36 & 1597    &  -87 & 1.92  \\
53669.7  &    419    &    0   & 0.17 & 1278    &    0   & 0.40 &   4991  &    15   &0.42 & 1497    &   -5 & 1.81  \\
53702.6  &    699    &    3   & 0.19 & 1298    &    0   & 0.40 &   4991  &     0   &0.41 & 1498    &  -67 & 1.85  \\
53703.6  &    449    &   12   & 0.19 & 1298    &    0   & 0.44 &   4992  &     0   &0.37 & 1498    &   61 & 1.81  \\
53710.6  &    364    &   12   & 0.15 & 1248    &    0   & 0.42 &   4991  &    15   &0.43 & 1498    &  -20 & 1.80  \\
53711.6  &    364    &   12   & 0.15 & 1248    &    0   & 0.42 &   4992  &    15   &0.43 & 1497    &    0 & 1.80  \\
53733.6  &    449    &   12   & 0.16 & 1248    &    0   & 0.47 &   4393  &    60   &0.37 & 1498    & -119 & 1.69  \\
53974.9  &    649    &    0   & 0.19 & 1498    &    0   & 0.46 &   4742  &    15   &0.35 & 1598    &  -67 & 1.84  \\
53977.8  &    399    &   12   & 0.15 & 1248    &    0   & 0.40 &   5491  &  -300   &0.44 & 1498    &   -9 & 1.62  \\
53978.8  &    649    &    0   & 0.19 & 1498    &    0   & 0.46 &   4743  &    15   &0.35 & 1597    &  -37 & 1.87  \\
53993.8  &    650    &    0   & 0.19 & 1497    &    0   & 0.46 &   4746  &    15   &0.36 & 1598    &  -67 & 1.87  \\
53995.7  &    650    &   15   & 0.19 & 1497    &    0   & 0.46 &   4741  &    15   &0.35 & 1598    &  -92 & 1.90  \\
53996.8  &    415    &   18   & 0.16 & 1263    &   30   & 0.43 &   4985  &     0   &0.41 & 1601    &   -7 & 1.73  \\
53997.8  &    639    &   15   & 0.19 & 1472    &    0   & 0.45 &   4742  &    15   &0.36 & 1597    &  -75 & 1.84  \\
54006.7  &    364    &   12   & 0.17 & 1248    &    0   & 0.48 &   4492  &   300   &0.35 & 1397    &   19 & 1.90  \\
54007.7  &    364    &   12   & 0.18 & 1247    &    0   & 0.48 &   4493  &   300   &0.34 & 1397    &    6 & 1.83  \\
54031.7  &    369    &   60   & 0.19 & 1148    &   90   & 0.43 &   4492  &   300   &0.38 & 1398    &  248 & 2.01  \\
54035.7  &    639    &   15   & 0.19 & 1473    &    0   & 0.45 &   4742  &    15   &0.36 & 1597    &  -90 & 1.85  \\
54036.7  &    405    &   20   & 0.18 & 1298    &    0   & 0.42 &   4992  &     0   &0.40 & 1398    &   -8 & 1.58  \\
54038.7  &    725    &    1   & 0.20 & 1470    &    0   & 0.44 &   4740  &    15   &0.36 & 1600    &  -85 & 1.85  \\
54039.7  &    401    &    3   & 0.16 & 1296    &    0   & 0.42 &   5001  &    -1   &0.42 & 1399    &  -84 & 1.79  \\
54069.6  &    599    &    0   & 0.19 & 1448    &   30   & 0.45 &   4743  &    15   &0.37 & 1597    &  -57 & 2.07  \\
54242.9  &    400    &    3   & 0.18 & 1297    &    0   & 0.42 &   5088  &     0   &0.41 & 1450    &    8 & 1.92  \\
54244.0  &    399    &    3   & 0.20 & 1298    &    0   & 0.45 &   4842  &     0   &0.35 & 1447    &  -16 & 1.91  \\
54322.9  &    604    &    0   & 0.20 & 1446    &   30   & 0.46 &   4740  &    15   &0.35 & 1598    &  -26 & 1.86  \\
54323.8  &    414    &   18   & 0.17 & 1306    &   30   & 0.43 &   4992  &     0   &0.39 & 1547    &   -1 & 1.49  \\
54346.8  &    399    &    3   & 0.20 & 1298    &    0   & 0.44 &   4493  &     0   &0.36 & 1448    &   -6 & 1.63  \\
54347.8  &    399    &    3   & 0.19 & 1298    &    0   & 0.41 &   4493  &     0   &0.40 & 1447    &   -2 & 2.07  \\
54350.9  &    606    &    0   & 0.21 & 1460    &   29   & 0.46 &   4488  &    14   &0.33 & 1596    &  -65 & 1.77  \\
54388.7  &    599    &   15   & 0.19 & 1448    &   30   & 0.47 &   4493  &    15   &0.34 & 1598    & -100 & 1.85  \\
54390.6  &    400    &    3   & 0.20 & 1298    &    0   & 0.43 &   4243  &   150   &0.37 & 1448    &   15 & 1.91  \\
54391.7  &    399    &    3   & 0.18 & 1298    &    0   & 0.43 &   4493  &   150   &0.40 & 1448    &   25 & 1.88  \\
54405.7  &    399    &    3   & 0.18 & 1298    &    0   & 0.43 &   4493  &   150   &0.39 & 1447    &  -15 & 1.86  \\
54406.6  &    399    &    3   & 0.18 & 1298    &    0   & 0.43 &   4493  &   150   &0.39 & 1448    &  -22 & 1.85  \\
54407.6  &    599    &   15   & 0.21 & 1448    &   30   & 0.45 &   4493  &    15   &0.34 & 1597    &  -49 & 2.01  \\
54410.7  &    607    &   14   & 0.20 & 1451    &   30   & 0.45 &   4495  &    16   &0.35 & 1595    &  -71 & 1.95  \\
54412.6  &    399    &    3   & 0.17 & 1297    &    0   & 0.40 &   4993  &     0   &0.42 & 1448    &    3 & 1.80  \\
\hline
mean & 507$\pm$128 & 13$\pm$16 & 0.18$\pm$0.02 & 1404$\pm$120 & 9$\pm$28 & 0.44$\pm$0.04 & 4938$\pm$405 & 25$\pm$109 & 0.38$\pm$0.05 
& 1523$\pm$81 & -2$\pm$74 & 1.87$\pm$0.19 \\
\enddata
\tablenotetext{*}{The intensity is given as a ratio to the total H$\beta$ [in \%], i.e. for \ion{Fe}{2} it is 
the parameter R$_{\rm Fe}$.}         
\end{deluxetable}
                                                                       	
\end{document}